\documentclass{aa}  

\usepackage{graphicx}
\usepackage{txfonts}
\usepackage{psfig}
\usepackage{amsmath}
\usepackage{amssymb}
\usepackage{upgreek}
\usepackage{longtable}
\usepackage{supertabular}
\usepackage{lscape}
\usepackage[mathscr]{eucal}
\usepackage{float}
\usepackage{afterpage}
\usepackage{subcaption}
\usepackage[export]{adjustbox}
\usepackage[usenames, dvipsnames]{color}
\usepackage{titlesec}
\usepackage{csvsimple}
\usepackage{colortbl}
\usepackage{caption}
\usepackage{booktabs}
\usepackage{dblfloatfix}
\begin{document}

   \title{Expansion kinematics of young clusters.} 

   \subtitle{III. The kiloparsec sample}

   \author{Joseph J. Armstrong
          \inst{1}
          \and
          Jonathan C. Tan\inst{1,2}
          }

   \institute{$^1$Department of Space, Earth \& Environment, Chalmers University of Technology, SE-412 96 Gothenburg, Sweden\\
   $^2$Department of Astronomy, University of Virginia, Charlottesville, VA, USA\\
              \email{jarmstrongastro@gmail.com}
             }

 
  \abstract
    {Most stars form in clusters, 
    but only a small fraction are thought to
    remain gravitationally bound for longer than $\sim10\:$Myr.
    }
   {Cluster
   formation and dispersal can be investigated by analysing the spatial and kinematic structure in nearby young systems, particularly expansion.
   }
   {We combine Gaia DR3 5-parameter astrometry with calibrated radial velocities for 
   23 nearby ($<1\:$kpc) young ($<60\:$Myr) clusters. 
   }
   {We characterise the plane-of-sky structure of the clusters 
   using $Q$-Parameter and Angular Dispersion Parameter (ADP) methods. We measure plane-of-sky expansion using several methods. We determine plane-of-sky orientations along which expansion is maximised. We estimate expansion timescales and traceback ages and compare to isochronal ages. We look for correlations between cluster properties and discuss sample-wide trends.}
   {We find that most young clusters are more smoothly structured in their centers where the rate of dynamical interactions is highest, while hierarchical structure can survive in the sparse outskirts for $>10\:$Myr. We find that the majority of nearby young clusters exhibit clear signatures of expansion in the plane-of-sky, which in many cases is significantly anisotropic, even at ages $>30$ Myr. We find evidence that older clusters tend to have directions of maximum expansion oriented closer to parallel with the Galactic plane. The high degree of spatial structure and significant expansion anisotropy imply that the majority of these young clusters have formed with significant spatial and kinematic substructure and not as dense, monolithic clusters. Kinematic ages estimated from expansion timescales and on-sky traceback are generally in good agreement with estimates inferred from stellar evolution models for clusters $<10\:$Myr old. However, many clusters with older isochronal ages appear to have significantly younger kinematic ages. We discuss potential reasons for this discrepancy, including a prolonged embedded and/or gravitationally bound phase in the early stages of the clusters.} 

   \keywords{Surveys: Gaia; methods: data analysis; Open Clusters:
               }

   \maketitle

\section{Introduction}

Most stars form in clusters \citep[e.g.,][]{lada03,gutermuth09,quintana25}, which are thus the building blocks of galaxies and the nurseries of most planetary systems. Hundreds to thousands of sibling stars are born from collapsing gas clumps inside giant molecular clouds (GMCs), creating `embedded' clusters with initial structure and kinematics inherited from clumpy and turbulent natal gas \citep[e.g.,][]{kuhn14,sills18}. Once a cluster forms massive stars, their stellar feedback begins to disperse the surrounding molecular gas, which then causes the loss of the majority of the cluster's binding mass \citep[e.g.,][]{kroupa01b,goodwin06}. For many young clusters, the mutual gravity of the stellar population is insufficient to keep the cluster bound and thus the cluster begins to expand and eventually disperse into the Galactic field, as many observational studies are finding \citep[e.g.,][]{kuhn19,wright19,armstrong20,guilherme23,dellacroce24,wright24,armstrong24,cheshire25}. 

Observable signatures of cluster expansion and dissolution have become measurable for many nearby clusters and stellar associations thanks to the ultra-precise astrometry available from the Gaia mission \citep{Gaiaedr3}. It has also become possible to measure the rates of expansion and trajectories of individual stars to sufficient precision to enable robust kinematic age estimates, i.e., constraints on cluster age based on the motion of cluster member stars \citep[e.g.,][]{miret-roig20,armstrong24,fajrin25}. Kinematic ages offer constraints independent of stellar evolution models, of which there are many flavours that differ in treatment of magnetic activity, location of the birth-line \citep[e.g.,][]{baraffe15,marigo17,somers20} and which suffer from many uncertainties when used to interpret observational data, such as binarity, extinction, variability, leading to large uncertainties in ages derived for young stellar populations \citep[e.g.,][]{ratzenbock23,cheshire25}. Thus, there is the potential to cross-check and validate timescales of star-formation, gas expulsion and cluster dissolution by comparing ages derived from these different approaches \citep[e.g.,][]{miret-roig24,armstrong24,cheshire25}.

Previously, in \citet{armstrong24} (i.e., Paper I, which we also refer to as AT24), we employed a variety of techniques to investigate the internal substructure and kinematics of the $\lambda$~Ori cluster, using the cluster membership list of \citet{cantat-gaudin20}. Utilizing both the $Q$-Parameter \citep{cartwright04} and Angular Dispersion Parameter (ADP) \citep{dario14}, we found evidence that the cluster contains significant substructure outside the smooth central cluster core. We also found strong evidence for asymmetric expansion
and determined the direction at which the rate of expansion is at a maximum. We inverted the maximum rate of expansion of $0.144^{+0.003}_{-0.003}$ km/s/pc to give an expansion timescale of 
$6.94^{+0.15}_{-0.14}\:$Myr, 
which we compared to other kinematic age methods applied to this cluster \citep{squicciarini21,quintana22,pelkonen24} and literature age estimates \citep{kounkel18,zari19,cao22}. We also found significant asymmetry in the velocity dispersions and signatures of cluster rotation in the plane-of-sky. Putting all of these results together, we concluded that the $\lambda$ Ori cluster likely formed in a sparse, substructured configuration, and is not simply the dispersing remnant of an initially bound, monolithic cluster which began to expand after the dispersal of its parent molecular cloud. The asymmetric kinematic signatures, and the discovery of a group of candidate ejected cluster members which we consider tentative evidence of a `sub-cluster ejection' \citep{polak24}, suggest a more complex dynamical history for the $\lambda$ Ori cluster. Scenarios for inducing complex dynamics in star-forming gas include triggering of gravitational instability via GMC-GMC collisions \citep[e.g.,][]{tan00,wu17} or feedback-driven shell collisions \citep[e.g.,][]{inutsuka15}.

In \citet{cheshire25} (i.e., Paper II, which we also refer to as CAT26), we investigated the structure and kinematics of the young clusters NGC 2264 and Collinder 95 using a sample of young stellar objects (YSOs) confirmed by spectra from HectoSpec (EW(Li), EW(H$\alpha$)) and Gaia DR3 variability flags \citep{marton23}. We found that NGC 2264 contains a significant level of substructure, including two dense likely bound sub-clusters, which we denote as NGC 2264 N \& S, while Collinder 95 is sparse and likely wholly unbound. All clusters show some evidence for expansion, but this is most significant in Collinder 95. We also calculated kinematic traceback ages for YSOs belonging to these clusters and compared them to ages inferred from several stellar evolution models \citep{baraffe15,marigo17,somers20} using either Gaia DR3 BP-RP or G-RP colour. We found that traceback ages for YSOs in NGC 2264 N \& S clusters are generally lower than model ages, implying that these YSOs formed within the bound cluster cores and have subsequently become unbound, while traceback ages for YSOs in Collinder 95 are often greater than model ages, which may imply that they formed in an initially sparse configuration, outside the cluster core, or otherwise that stellar evolution models systematically underestimate ages for these stars. 

In this work we apply the structural and kinematic analysis techniques described in AT24 and CAT26 to a larger sample of nearby ($<1\:$kpc) young ($\lesssim 60\:$Myr) clusters. We estimate cluster distances using Gaia DR3 parallaxes, investigate cluster substructure, search for evidence of expansion, measure kinematic anisotropy and estimate kinematic ages. We then compare the results across the sample and look for correlations between structural and kinematic measures, and with cluster age and mass. We discuss our results in the context of recent literature and draw conclusions about the formation and early evolution of star clusters.

\section{Data}
\label{data}
As in AT24, we begin with the open cluster catalog of \citet{cantat-gaudin20}, which provides mean positions, proper motions, distances and ages for 2,017 open clusters, as well as the list of individual members from the Gaia DR2 catalog and their cluster membership probabilities. The cluster parameters are based on $\geq$0.7 probability cluster members selected using the UPMASK method \citep{krone-martins14} applied to data from Gaia DR2. In order to have a clean sample of cluster members with high precision astrometry, we match cluster members with Gaia DR3, which has improved astrometric precision over DR2, and then filter out sources with Gaia DR3 RUWE$>1.4$, the recommended threshold for astrometric quality \citep{Gaiaedr3}. 

\subsection{Cluster sample}
\label{clusters}

In order to investigate how detailed structure and kinematic trends within clusters may vary with cluster age, mass and dynamical history, we need to select a sample of clusters with a range of ages and environments, but which each contain a sufficiently large number of cluster members with precise, quality astrometry (and complementary RVs) to enable us to to measure kinematic trends with the necessary precision. For this reason, we select only clusters with $>80$ cluster members with quality astrometry (RUWE $>1.4$), which have a median distance $<1\:$kpc, ensuring that proper motion uncertainties are kept small relative to internal cluster velocities and the on-sky area of the cluster, that have $>5$ RVs available in the \citet{tsantaki22} compilation with which to obtain a robust cluster average RV, and which have literature age estimates $\lesssim 60\:$Myr. This gives us a sample of 23 clusters in the \citet{cantat-gaudin20} catalog.

Our cluster sample includes Platais~8, $\sigma$~Ori, NGC~2541B, Melotte 20 ($\alpha$ Persei), IC~2391, Gulliver 6, Pozzo~1 ($\gamma$~Vel), IC~2602, NGC~2232, NGC~1977, NGC~2264, IC~2395, Alessi~20, Collinder~359, Gulliver~9, ASCC~16 (Briceno~1), ASCC~21, IC~348, NGC~1980, Collinder~135, NGC~2547, and ASCC~19. We also include the AT24 results for Collinder~69 ($\lambda$~Ori) in the sample. 

Notable nearby young clusters included in the \citet{cantat-gaudin20} catalog, but which do not have sufficient RVs in \textit{Survey of Surveys} \citep{tsantaki22} are IC~1396, Collinder~95, Collinder~197, Alessi~43, etc. Furthermore, notable nearby young clusters which are missing from the \citet{cantat-gaudin20} catalog include the Orion Nebula Cluster, NGC 2024 (Flame Nebula Cluster), 25 Ori, Rho Ophiuchus, Chamaeleon~I, ASCC 50, etc.

We calculate the mean sky positions (in R.A., Dec. and $l$, $b$) and proper motions from our sample of high-probability members, and we estimate uncertainties on these values with a bootstrapping approach, calculating these means and medians from 100,000 randomly selected (with replacement) samples of cluster members and taking the 16th and 84th percentiles of the posterior distributions as the uncertainties. These values are given in Table~\ref{kinematic_table}.

\subsection{Radial velocities}
\label{RVs}
We match cluster members to the \textit{Survey of Surveys} \citep{tsantaki22} compilation of radial velocities, which combines RVs from large-scale spectroscopic surveys including Gaia, APOGEE, GALAH, Gaia-ESO, RAVE and LAMOST into a single cross-calibrated sample, shifted to the Gaia reference frame. The catalog also includes flags for the quality of RVs for individual sources $flagRV$ and for likely binary systems with large discrepancies between multiple RV measurements $flagBinary$. We exclude from our sample all RVs with $flagBinary=binary$ and $flagRV=2$, leaving only sufficiently reliable single-source RVs for determining cluster-average RVs.

As in AT24, we only require that clusters have a sufficient number of RVs to determine an average RV for each cluster for the purpose of correcting for projection effects (see Sect. \ref{tangential velocities}). For each cluster we report the number of usable RVs for cluster members, the cluster median RV with its uncertainties and the cluster RV in the local standard of rest (LSR) in Table~\ref{kinematic_table}.  

\begin{figure} 
    {\includegraphics[width=245pt]{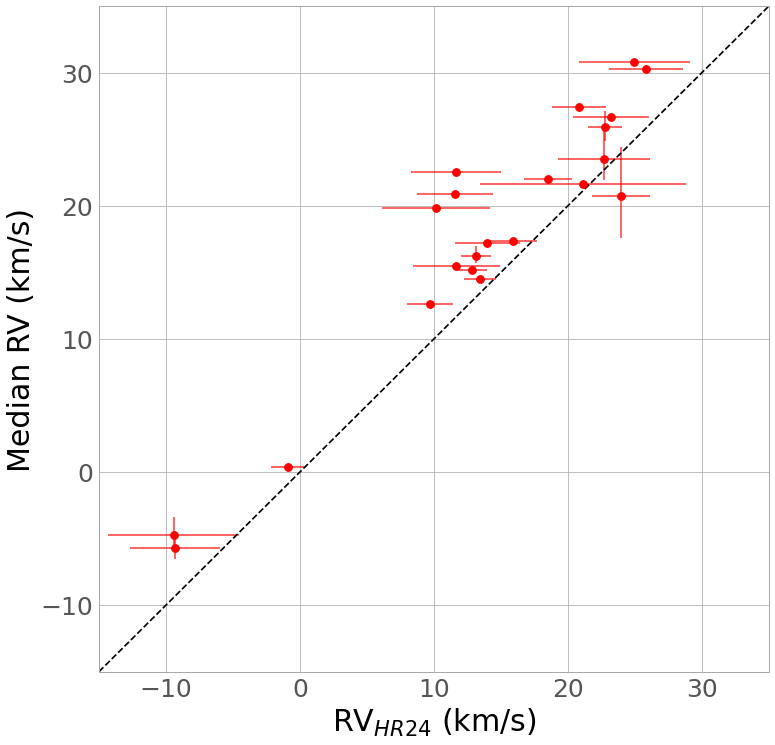} }%
    \setlength{\belowcaptionskip}{-10pt}
    \setlength{\textfloatsep}{0pt}
    \caption{Scatter plot of median cluster RVs which we calculate from RVs available in \protect\citet{tsantaki22} versus cluster average RVs from \protect\citet{hunt24}.} 
    \label{RVcomp}%
\end{figure}

Cluster-average RVs are available in the \citet{hunt24} catalog for the majority of our clusters (with the exception of NGC 1977, which is not included in their catalog). In Fig.~\ref{RVcomp} we plot their cluster RVs against those we calculate from RVs available in \citet{tsantaki22}. Generally, there is good agreement, with most cluster RVs consistent between catalogs within their uncertainties. However, there is a systematic trend that our median cluster RVs are typically $\sim5\:{\rm km\:s}^{-1}$ greater than the \citet{hunt24} catalog values. A possible reason for this discrepancy is that the \citet{hunt24} catalog RVs are calculated only using RVs from the Gaia DR3 catalog, and do not include RVs from other spectroscopic surveys which are included in the \citet{tsantaki22} survey. Gaia RVs are known to be unreliable for young stars in particular \citep[e.g., see][]{kounkel22,kounkel23} and have large uncertainties. Also, there is no filtering for likely binary systems in \citet{hunt24}. It is for these reasons that we continue our analysis with cluster RVs calculated using the \citet{tsantaki22} catalog.

\subsection{Literature ages}
\label{ages}
The \citet{cantat-gaudin20} cluster catalog provides age estimates for clusters based on the best fitting PARSEC v1.2 \citep{marigo17} isochrones to the cluster members selected by their algorithm. We report these (in Myr) for each cluster in our sample in Table ~\ref{kinematic_table}, along with age estimates from \citet{dias21} and \citet{hunt24} when available, which are other recent cluster catalogs compiled from Gaia data using different approaches and criteria to determine cluster membership compared to \citet{cantat-gaudin20}. We will denote these isochronal age estimates as $\tau_{\rm iso}$.

\begin{figure} 
    {\includegraphics[width=245pt]{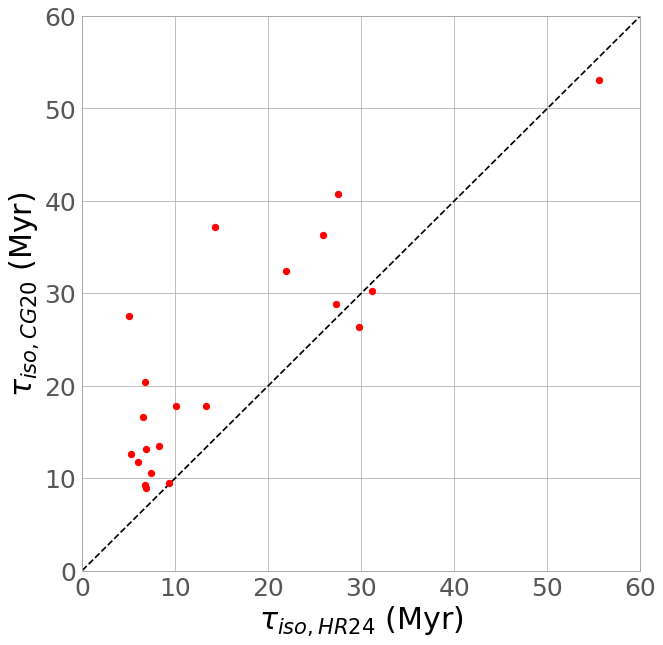} }%
    \setlength{\belowcaptionskip}{-10pt}
    \setlength{\textfloatsep}{0pt}
    \caption{Scatter plot of isochronal ages from \protect\citet{cantat-gaudin20} $\tau_{iso,CG20}$ versus ages from \protect\citet{hunt24} $\tau_{iso,HR24}$.} 
    \label{Agecomp}%
\end{figure}

For some clusters there is significant disagreement between these literature age estimates. For example, the age estimate given for NGC 1977 in \citet{cantat-gaudin20} is $97.7$ Myr, whereas in \citet{dias21} it is estimated to be $12$ Myr and in other works as young as $~2-3$ Myr \citep[e.g.,][]{getman19,zuniga23}. Generally, ages from \citet{hunt24} tend to be younger than those from \citet{cantat-gaudin20}, as illustrated in Fig.~\ref{Agecomp}.

\subsection{Distances}
\label{Distances}

We derive distances to each cluster using the same approach as described in \citet{armstrong24}. We use a bootstrapping approach, calculating distance from median parallax from 100,000 randomly selected (with replacement) samples of cluster members and taking the 50th percentile of the posterior distributions as the cluster distance and the 16th and 84th percentiles as the distance uncertainties. We apply a parallax zero point correction \citep{lindegren21} to all cluster members, as well as a magnitude dependent correction to parallax uncertainties \citep{elbadry21}.

Distances estimated using the probabilistic approach of \citet{bailerjones21} are not optimised for stars in clusters, as the priors involved implicitly assume that sources are field stars. Calculating cluster distance by combining these estimates for multiple cluster members compounds this effect and thus estimates of distance for clusters using \citet{bailerjones21} estimates tend to be biased. 

\begin{figure} 
    {\includegraphics[width=245pt]{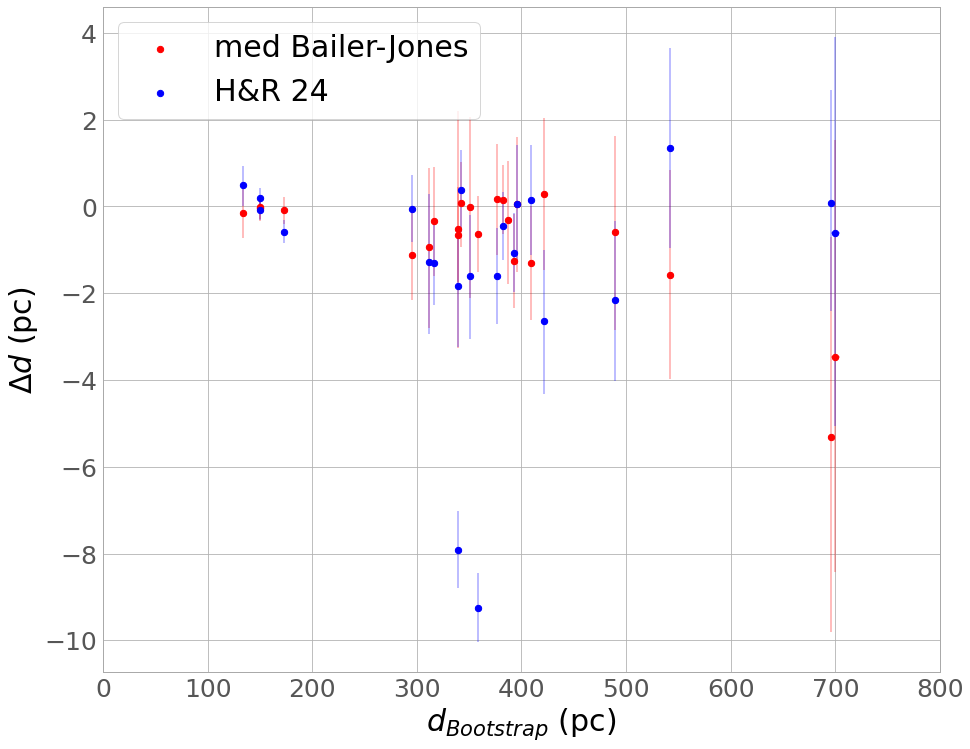} }%
    \setlength{\belowcaptionskip}{-10pt}
    \setlength{\textfloatsep}{0pt}
    \caption{Difference between our bootstrap cluster distance estimates $d_{Bootstrap}$ and median \protect\citet{bailerjones21} distances (red) and distances from \protect\citet{hunt24} (blue) $\Delta d$ versus cluster distance $d_{Bootstrap}$ (see Section \protect\ref{Distances}).} 
    \label{DistanceComparison}%
\end{figure}

The cluster catalogs of \citet{cantat-gaudin20} and \citet{hunt24} also provide estimates of distance, which are obtained using a Bayesian approach including a prior which assumes that all cluster members are at the same distance, and it is noted in \citet{cantatgaudin19c} that this is not optimal for nearby clusters where individual parallax uncertainties are smaller than the depth of the cluster. 
In Figure~\ref{DistanceComparison} we plot our distance estimates for clusters against the difference in distance between our estimate and the median \citet{bailerjones21} distance (red), or between our estimate and the cluster distance given in \citet{hunt24} (blue). The general trend is that distances from \citet{bailerjones21} or \citet{hunt24}, despite their different model priors, are generally in reasonable agreement with ours, given their uncertainties. This is likely because these clusters are relatively nearby and the parallaxes of their member stars have relatively small uncertainties, which thus constrain the distances more strongly than the model priors. The only significant exceptions are Pozzo 1 and NGC 2451B with $\Delta d>7$pc (Fig~\ref{DistanceComparison}). The differences in these distance estimates is likely due to the significant difference in cluster membership samples between \citet{cantat-gaudin20} and \citet{hunt24} for these clusters, with the \citet{hunt24} membership preferentially including more cluster members with greater parallaxes. We proceed with the following analysis using distances estimated via our bootstrapping approach. 

\section{Spatial structure}
\label{substructure}

\subsection{Cluster center and core radius}
\label{core}
The first step for both the structural and kinematic analyses is to define the central coordinates of each cluster. We follow the approach described in \citet{dario14} and AT24, which consists of determining the average position for all cluster members contained within iteratively smaller apertures, each centered on the previous aperture's center. 
In each iteration we reduce the aperture radius by 20\%, limiting the minimum aperture size to a 3 arcmin radius. 
We estimate the uncertainties of the cluster central coordinates with a bootstrapping approach, where we randomly resample (with replacement) cluster members equal to the total number of members in each iteration, and at the end we take the 16th and 84th percentiles of the posterior distribution of central coordinates as the uncertainties on the cluster central coordinates. These are reported for each cluster in Table~\ref{kinematic_table}.

We construct the radial density profile of each cluster, following the approach described in AT24. Cluster members are binned by increasing radial distance from the cluster center, $R$, in bins of 20 members each. For each bin, the areal density ($N_*$) is calculated.
We then define the core radius 
of a cluster, $R_c$, as the radius at which the areal density is half the peak density. The core radius of each cluster is reported in Table~\ref{kinematic_table}, along with the mean tangential velocities $\bar v_{l},\bar v_{b}$ of cluster members within a cluster's core.

\subsection{Q-parameter}
\label{Q-parameter}

\begin{figure} 
    {\includegraphics[width=245pt]{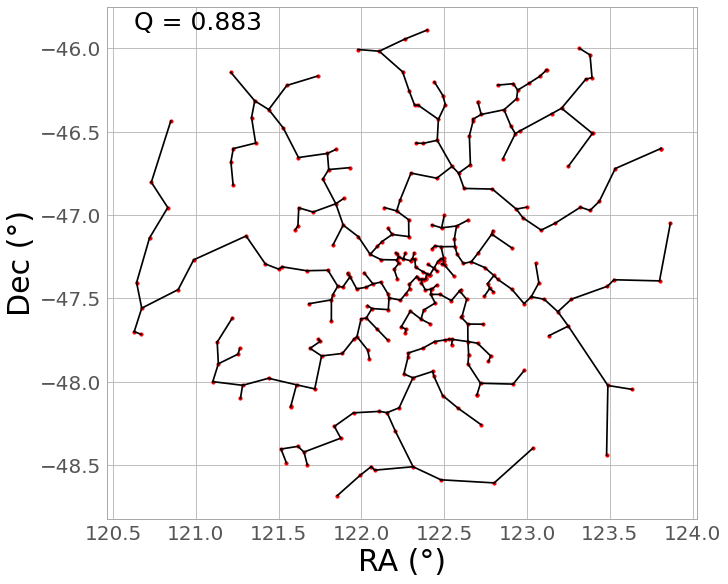} }%
    \setlength{\belowcaptionskip}{-10pt}
    \setlength{\textfloatsep}{0pt}
    \caption{Minimum spanning tree for members of Pozzo 1 from \protect\citet{cantat-gaudin20}.} 
    \label{Qp_Pozzo1}%
\end{figure}

We assess the level of substructure in these clusters using the $Q$-Parameter \citep{cartwright04}, following the same approach as AT24. This involves constructing a minimum spanning tree between the sky coordinates of cluster members and calculating the mean edge length $\Bar{m}$ and the mean edge length of the complete graph $\Bar{s}$. Note that $\Bar{m}$ is normalised to $\sqrt{NA}/({N-1})$, where $N$ is the number of cluster members and $A$ is the area of the smallest circle with radius $R$ which encompasses them. Note also that $\Bar{s}$ is normalised to $R$. The $Q$-Parameter is then the ratio between the normalised $\Bar{m}$ and $\Bar{s}$. Lower values of $Q$, i.e., $\lesssim0.7$, show a greater difference between the minimum spanning tree and the complete graph of a cluster, which indicates more substructure, whereas values of $Q$ closer to 1 indicate that a cluster has a smooth and centrally concentrated structure.

Generally, it is expected that young clusters and associations will exhibit greater degrees of substructure, since they are dynamically less evolved and still retain much of their structure associated with formation from their natal molecular clouds. On the other hand, older clusters, which are more dynamically evolved, will have erased more of their initial structure and will tend toward smoother distributions. These will also appear to be more centrally concentrated if the system retains a significant gravitationally bound core.


Fig.~\ref{Qp_Pozzo1} shows the minimum spanning tree for members of the Pozzo 1 ($\gamma$ Vel) cluster as an example, which yields a $Q$-Parameter value of 0.883. This value indicates that Pozzo 1 is mostly centrally concentrated, but still retains a small amount of substructure. Out of all of our cluster sample, Melotte 20 and Collinder 135 have the joint highest $Q$-Parameter values of 0.939, while IC 2602 has the lowest at 0.776. The vast majority, 19 out of 23 clusters, have $Q$-Parameter values between 0.77 - 0.9, which are neither entirely smooth or clumpy. 

Overall, the $Q$-Parameter seems to be largely ineffective at making distinctions between the structures of the clusters in our sample by itself, so we perform further structural analysis with alternative approaches.

\subsection{Ellipticity and angular dispersion parameter}
\label{ADP}

We fit an ellipse to the Galactic sky coordinates of members of each cluster using least squares minimisation of the edge of the ellipse to the data, given the central coordinates determined in Sect.~\ref{core} and using the same approach AT24 applied to $\lambda$~Ori. We report the orientation of the semi-major axis $\theta_a$ and the ellipticity, $e$, of the best fitting ellipse for each cluster in Table~\ref{kinematic_table}.

We analyse the structure of clusters in our sample using the Angular Dispersion Parameter (ADP; $\delta_{\rm ADP,e,N_{sect}}(r)$) \citep{dario14}, following the approach of AT24. Cluster members are divided into concentric elliptical annuli (of ellipticity $e$) using the best-fitting ellipse as described above, which are then further divided into $N_{sect}$ sectors. The standard deviation of counts per sector in a given annulus gives $\delta_{\rm ADP,e,N_{sect}}$, and thus $\delta_{\rm ADP,e,N_{sect}}(r)$ describes the ADP as a function of radial distance from the cluster center. Low values of $\delta_{\rm ADP,e,N_{sect}}<1$ indicate a smooth distribution, while high values $\delta_{\rm ADP,e,N_{sect}}>2.5$ indicate substructured and dynamically young distributions. However, for direct comparison of $\delta_{\rm ADP,e,N_{sect}}(r)$ between clusters, the same number of cluster members per annulus and the same number of sectors per annulus should be adopted.

We present the ADP values for concentric annuli of 50 cluster members each for each cluster, using either 4 or 8 sectors per annuli, in Table~\ref{ADP_table}. Clusters without sufficient cluster members to fill successive annuli have the corresponding columns left blank. 

Notably, for 17 out of 23 clusters the ADP value of innermost annulus containing the 50 cluster members closest to the cluster center ($\delta_{\rm ADP,e,N_{sect}}(50)$) is the lowest compared to other annuli. For another 3 clusters, Alessi 20, IC 348 and NGC 2547, the $\delta_{\rm ADP,e,N_{sect}}(50)$ value is consistent with the lowest value well within the typical uncertainty of $\sim0.1$. This indicates that for the majority of young clusters the dense, cluster core is smooth and dynamically evolved, whilst the sparser outskirts are less evolved and have retained a greater level of substructure. We expect this is because the higher stellar density of the cluster closer to center naturally creates a higher frequency of interactions between members, leading to rapid dynamical evolution compared to the outskirts of the cluster. A similar behavior was seen in the $\lambda$ Ori cluster by AT24 and in a sample of 22 clusters from the MYStIX survey \citep{feigelson13} in the analysis of \citet{jaehnig15}.

Some exceptions to the above general trend are Pozzo 1 and Platais 8, where $\delta_{\rm ADP,e,N_{sect}}(50)$ is greater than $\delta_{\rm ADP,e,N_{sect}}(100)$ by 0.77 and 0.37 respectively. The $\delta_{\rm ADP,e,N_{sect}}(r)$ values in Pozzo 1 show large fluctuations, and the maximum and minimum values for this cluster are at $\delta_{\rm ADP,e,N_{sect}}(250)$ and $\delta_{\rm ADP,e,N_{sect}}(100)$ respectively, while the core $\delta_{\rm ADP,e,N_{sect}}(50)$ value remains close to the mean for this cluster. For Platais 8, the minimum value is $\delta_{\rm ADP,e,N_{sect}}(100)$, but the $\delta_{\rm ADP,e,N_{sect}}(150)$ value at the outermost annulus is 2.94, indicating a high level of substructure in the outskirts of this cluster.

NGC 2451B, IC 2395, Collinder 359 and Gulliver 9 have $\delta_{\rm ADP,e,4}(r)>3$ in their outermost annuli, which indicate very high levels of substructure in their outskirts.

We also calculate the mean $\delta_{\rm ADP,e,4}(r)$ values per cluster, which are reported in Table~\ref{kinematic_table}. NGC 1977, IC 2395 and Collinder 359 all have mean $\delta_{\rm ADP,e,4}(r) > 2.5$, indicating that these clusters are highly substructured, while IC 2391, Alessi 20, ASCC 16, ASCC 19 and NGC 2547 have $\delta_{\rm ADP,e,4}(r) < 1.0$, indicating that these clusters are smoothly distributed overall. 

\begin{table*}
\caption{Angular Dispersion Parameter ($\delta_{\rm ADP,e,N_{sect}}(r)$) values for each cluster.}
\footnotesize
\begin{center}
{\renewcommand{\arraystretch}{1.2}
\begin{tabular}{|p{1.6cm}|p{0.4cm}|p{0.4cm}|p{0.4cm}|p{0.4cm}|p{0.4cm}|p{0.4cm}|p{0.4cm}|p{0.4cm}|p{0.4cm}|p{0.4cm}|p{0.4cm}|p{0.4cm}|p{0.4cm}|p{0.4cm}|p{0.4cm}|p{0.4cm}|p{0.4cm}|p{0.4cm}|p{0.4cm}|p{0.4cm}|}
\hline
Cluster & \multicolumn{10}{c}{$\delta_{\rm ADP,e,4}(r)$} & \multicolumn{10}{c}{$\delta_{\rm ADP,e,8}(r)$}   \\
  & 50 & 100 & 150 & 200 & 250 & 300 & 350 & 400 & 450 & 500 & 50 & 100 & 150 & 200 & 250 & 300 & 350 & 400 & 450 & 500 \\
\hline
$\lambda$ Ori & 1.00 & 2.20 & 2.36 & 2.13 & 1.34 & 2.22 & 1.71 & 1.24 & 1.52 & 1.32 & 0.83 & 1.59 & 1.71 & 1.56 & 1.03 & 1.79 & 1.57 & 1.05 & 1.29 & 1.10  \\
Platais 8 & 1.45 & 1.08 & 2.94 &  &  &  &  &  &  &  & 1.19 & 1.05 & 2.22 &  &  &  &  &  &  &   \\
$\sigma$ Ori & 0.51 & 2.98 &  &  &  &  &  &  &  &  & 0.47 & 2.20 &  &  &  &  &  &  &  &   \\
NGC 2451B & 0.85 & 1.05 & 0.97 & 1.20 & 4.57 &  &  &  &  &  & 0.77 & 0.85 & 0.79 & 1.02 & 3.33 &  &  &  &  &   \\
Melotte 20 & 0.49 & 0.89 & 1.52 & 1.21 & 1.07 & 1.20 & 0.60 & 0.82 & 0.98 & 1.48 & 0.55 & 0.79 & 1.13 & 0.93 & 0.80 & 0.91 & 0.52 & 0.78 & 0.82 & 1.10  \\
IC 2391 & 0.66 & 1.36 & 0.69 &  &  &  &  &  &  &  & 0.53 & 1.03 & 0.55 &  &  &  &  &  &  &   \\
Gulliver 6 & 0.76 & 1.69 & 1.42 & 2.17 & 2.57 &  &  &  &  &  & 0.60 & 1.36 & 1.14 & 1.58 & 1.88 &  &  &  &  &   \\
Pozzo 1 & 1.26 & 0.49 & 1.58 & 0.76 & 1.72 & 1.51 &  &  &  &  & 1.03 & 0.44 & 1.14 & 0.64 & 1.40 & 1.25 &  &  &  &   \\
IC 2602 & 0.43 & 1.34 & 1.05 & 2.09 & 1.89 &  &  &  &  &  & 0.44 & 1.00 & 0.79 & 1.55 & 1.46 &  &  &  &  &   \\
NGC 2232 & 0.51 & 2.63 & 1.28 &  &  &  &  &  &  &  & 0.51 & 1.89 & 0.95 &  &  &  &  &  &  &   \\
NGC 1977 & 0.98 & 4.44 &  &  &  &  &  &  &  &  & 0.78 & 3.18 &  &  &  &  &  &  &  &   \\
NGC 2264 & 0.39 & 2.96 &  &  &  &  &  &  &  &  & 0.37 & 2.13 &  &  &  &  &  &  &  &   \\
IC 2395 & 0.87 & 1.07 & 2.32 & 3.96 & 5.35 &  &  &  &  &  & 0.70 & 0.86 & 1.74 & 2.80 & 3.85 &  &  &  &  &   \\
Alessi 20 & 0.98 & 0.86 &  &  &  &  &  &  &  &  & 0.95 & 0.63 &  &  &  &  &  &  &  &   \\
Coll 359 & 0.28 & 2.04 & 3.11 & 3.20 & 3.93 &  &  &  &  &  & 0.39 & 1.46 & 2.29 & 2.26 & 2.84 &  &  &  &  &   \\
Gulliver 9 & 1.34 & 1.74 & 2.80 & 3.22 &  &  &  &  &  &  & 1.02 & 1.33 & 1.98 & 2.37 &  &  &  &  &  &   \\
ASCC 16 & 0.54 & 1.14 & 0.83 &  &  &  &  &  &  &  & 0.51 & 0.83 & 0.71 &  &  &  &  &  &  &   \\
ASCC 21 & 1.11 &  &  &  &  &  &  &  &  &  & 0.83 &  &  &  &  &  &  &  &  &   \\
IC 348 & 1.32 & 1.27 &  &  &  &  &  &  &  &  & 0.95 & 0.93 &  &  &  &  &  &  &  &   \\
ASCC 19 & 0.41 & 0.60 &  &  &  &  &  &  &  &  & 0.44 & 0.61 &  &  &  &  &  &  &  &   \\
NGC 1980 & 0.82 & 1.95 &  &  &  &  &  &  &  &  & 0.80 & 1.44  &  &  &  &  &  &  &  &   \\
NGC 2547 & 0.63 & 1.17 & 0.60 &  &  &  &  &  &  &  & 0.61 & 0.92 & 0.73 &  &  &  &  &  &  &   \\
Coll 135 & 0.94 & 1.32 & 1.48 & 1.82 & 2.00 &  &  &  &  &  & 0.67 & 1.12 & 1.11 & 1.43 & 1.42 &  &  &  &  &   \\
\hline
\end{tabular}}
\end{center}
\setlength{\belowcaptionskip}{-10pt}
\setlength{\textfloatsep}{0pt}
\tablefoot{\footnotesize Values of $\delta_{\rm ADP,e,N_{sect}}(r)$ are given for concentric elliptic annuli of 50 cluster members each of either 4 or 8 sectors, calculated following the approach described in Sect. ~\ref{ADP}. }
\label{ADP_table}
\end{table*}

\begin{figure*} 
    {\includegraphics[width=\textwidth]{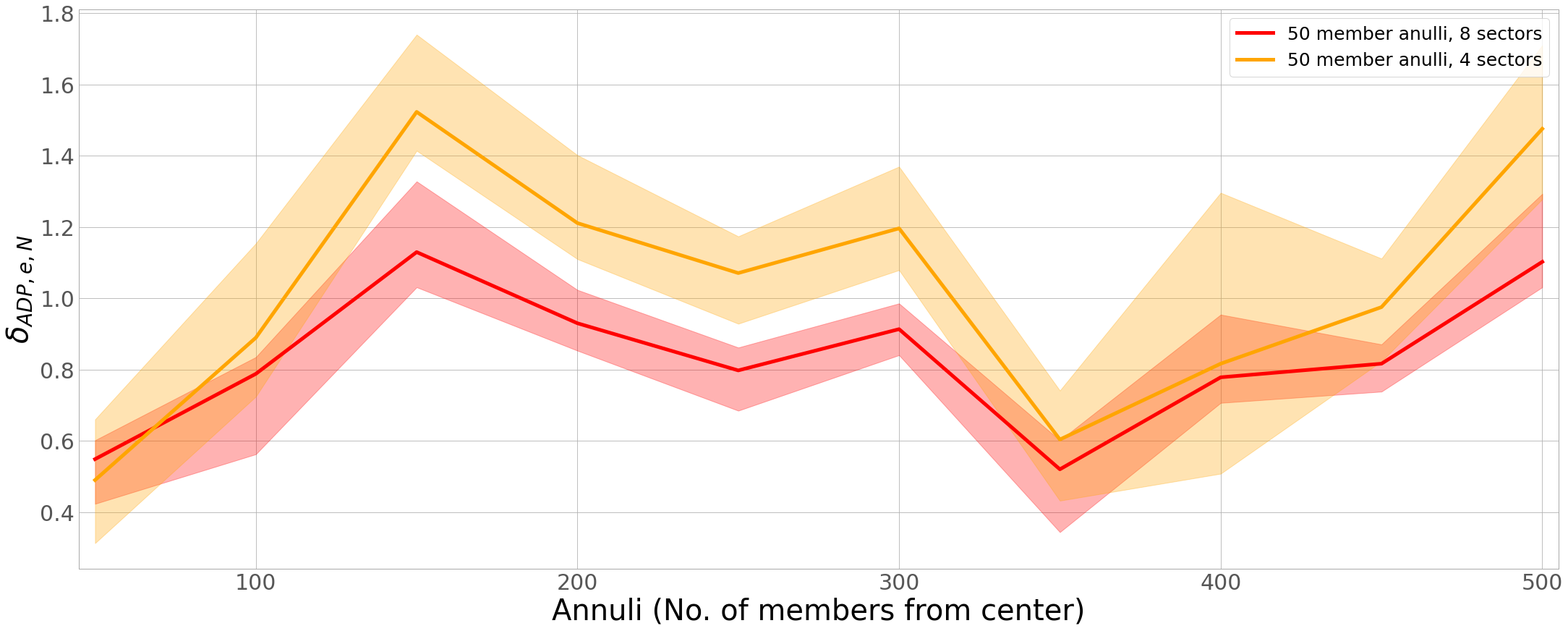} }%
    \setlength{\belowcaptionskip}{-10pt}
    \setlength{\textfloatsep}{0pt}
    \caption{Angular Dispersion Parameter ($\delta_{\rm ADP,e,N_{sect}}(r)$) for Melotte 20 with eight sectors (red)  or four sectors (yellow) per concentric annuli containing 50 members. The $\delta_{\rm ADP,e,N_{sect}}(r)$ values are calculated for orientations of sectors rotated 1$^\circ$ at a time, and we plot the 50th (solid lines), 16th, and 84th percentile values (shaded regions) for each annulus.} 
    \label{ADP_fig}%
\end{figure*}

\begin{figure*} 
    {\includegraphics[width=172pt]{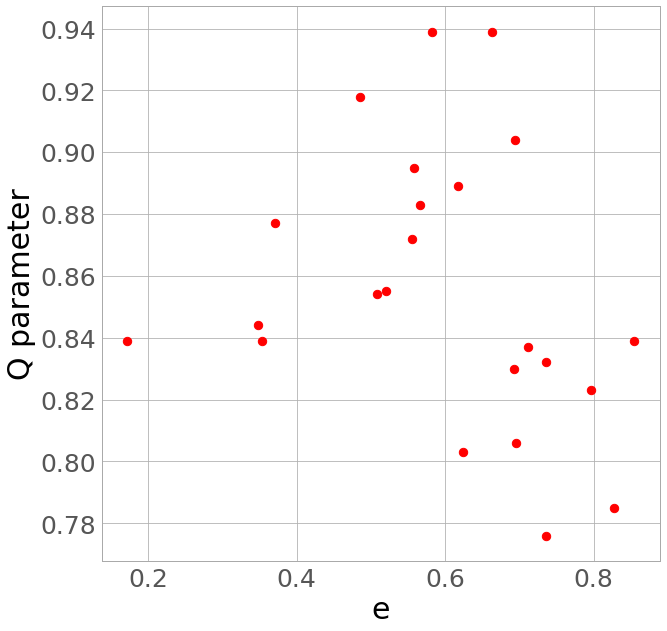} }%
    {\includegraphics[width=172pt]{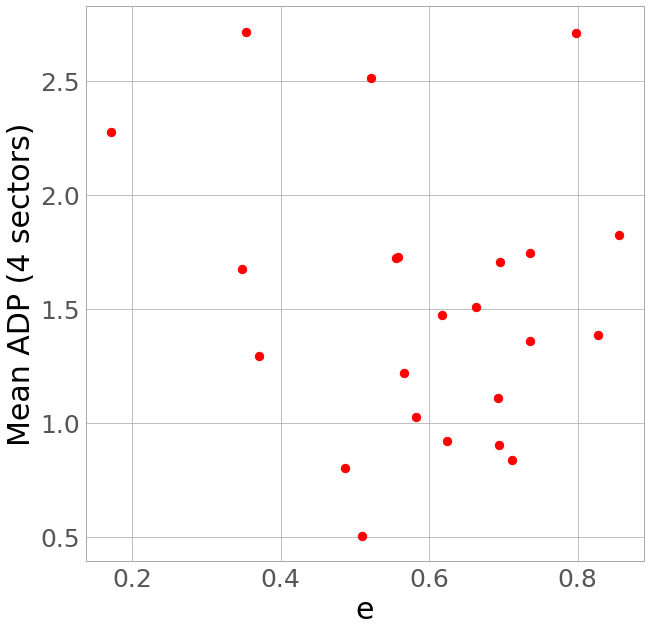}}
    {\includegraphics[width=172pt]{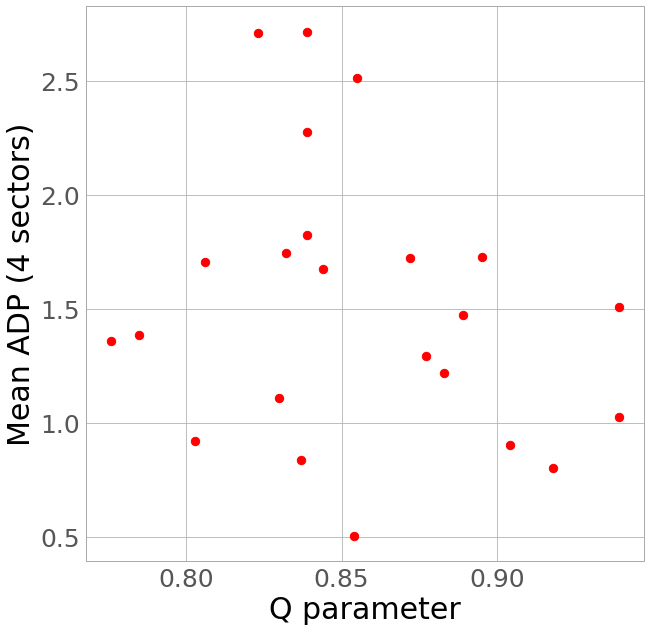}}
    \setlength{\belowcaptionskip}{-10pt}
    \setlength{\textfloatsep}{0pt}
    \caption{{\it (a) Left:} Q parameter versus ellipticity $e$. {\it (b) Middle:} Mean ADP with 4 sectors versus ellipticity $e$. {\it (c) Right:} Mean ADP with 4 sectors versus Q parameter.}
    \label{StructureComparison}%
\end{figure*}

In Fig.~\ref{StructureComparison} we compare the ellipticity, $e$, the $Q$-parameter and the mean ADP using 4 sectors ($\delta_{\rm ADP,e,4}(r)$) values for all clusters in the sample. There is no significant correlation between the three measures, which is expected in the case of ADP vs $e$ as they are determined based on the same best-fit ellipse. However, the lack of correlation between the $Q$-parameter and ADP in particular indicates the ineffectiveness of the $Q$-parameter at making distinctions between the levels of structure in cluster in this sample.

\section{Kinematics}

By investigating the kinematics of young clusters we can learn about the initial conditions of their formation and what mechanisms determine their subsequent dynamical evolution. In particular, recent studies have found that many, if not the majority of, young clusters are unbound and in the process of expanding and dispersing into the Galactic field \citep[e.g.,][]{kuhn19,wright19,armstrong20,guilherme23,wright24,armstrong24}. However, the rate of expansion in these clusters is often anisotropic, which, along with the significant spatial substructure detected, suggests that the process is more complex than the simple residual gas expulsion model. The timescale of expansion is also useful as a kinematic age estimate.

In this section we apply the kinematic analyses presented in AT24 to our larger sample of nearby young clusters to identify expansion trends and derive kinematic age estimates. 

\begin{figure} 
    {\includegraphics[width=245pt]{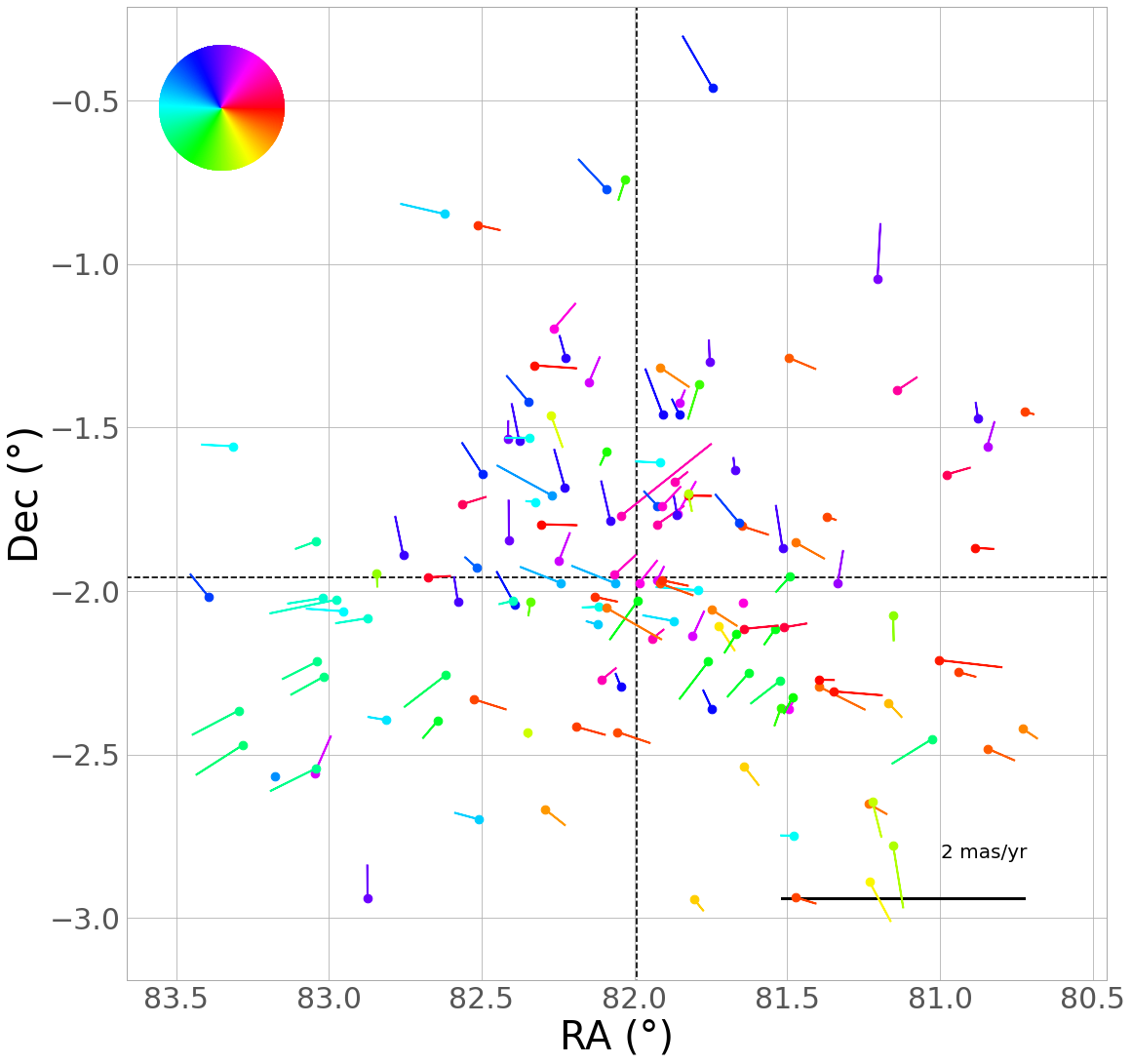} }%
    \setlength{\belowcaptionskip}{-10pt}
    \setlength{\textfloatsep}{0pt}
    \caption{Spatial distribution of the members of ASCC 19 cluster from \protect\citet{cantat-gaudin20}. The vectors indicate the proper motion relative to the cluster. Points are colour-coded based on the position angle of the proper motion vector (see the colour wheel in the top left as a key). The magnitude scale (mas yr$^{-1}$) of proper motion vectors is indicated by the scale bar in the bottom right.} 
    \label{ASCC19_vectors}%
\end{figure}

\subsection{Tangential velocities and virtual expansion correction}
\label{tangential velocities}
Figure~\ref{ASCC19_vectors} shows the sky positions of the members of ASCC 19 with vectors indicating their proper motions relative to the cluster mean, colour-coded based on the position angle of the vector. From a plot like this we can already see signs of expansion, as many cluster members have proper motion vectors consistent with moving away from the cluster center. However, to accurately measure the kinematics of the cluster, we need to account for projection effects.

As in AT24, the first step of our kinematic analysis is to transform observed proper motions into tangential velocities, to avoid projection effects caused by the curvature of the sky coordinate system, incorporating a correction for the `virtual expansion' effect using a cluster median RV following the equations of \citet{brown97}. This transformation is done with a Bayesian approach with unconstrained priors, sampling the posterior distribution with the MCMC sampler \citep{emcee}, as described in AT24. After discarding the first half of our iterations as burn-in, we take the 16th and 84th percentiles as 1$\sigma$ uncertainties. We report the mean tangential velocities in Galactic coordinates ($\bar v_{l}$, $\bar v_{b}$) for the core members of each cluster in Table~\ref{kinematic_table}.

\subsection{Expansion velocities}
\label{expansion velocities}

A simple indicator for cluster expansion is the median cluster expansion velocity, $v_{\rm out}$, i.e., the median components of velocities that are directed radially from the cluster center. As described in AT24, we measure the cluster expansion velocity by taking the median expansion velocity component of all cluster members, with uncertainties (16th and 84th percentile) estimated via Monte Carlo sampling. 
These cluster expansion velocities and uncertainties for each cluster are presented in Table ~\ref{kinematic_table}.

Our largest expansion velocity $\bar v_{\rm out}$ is $0.976^{+0.028}_{-0.028}$~km~s$^{-1}$ for the $\sigma$ Ori cluster. The smallest is $-0.029^{+0.042}_{-0.043}$~km~s$^{-1}$ for IC 348, which is consistent with no expansion or even slight contraction. These are all well within the range of expansion velocities calculated by \citet{kuhn19} for their sample of young clusters, which range from $2.07\pm1.10$ km~s$^{-1}$ for G$353.1+0.6$ to $-2.06\pm1.00$ km~s$^{-1}$ for M17, and thus none of our expansion velocities are beyond the range expected based on previous literature. There is notable improvement in the uncertainties, from a typical precision of $\sim 0.3$ km~s$^{-1}$ in \citet{kuhn19} to a typical precision of $\sim 0.03$ km~s$^{-1}$ in our analysis. Much of this is due to the improved precision of Gaia DR3 astrometry over DR2, but also partly because our sample contains less distant clusters on average than \citet{kuhn19}.

\begin{figure}
\includegraphics[width=\columnwidth]{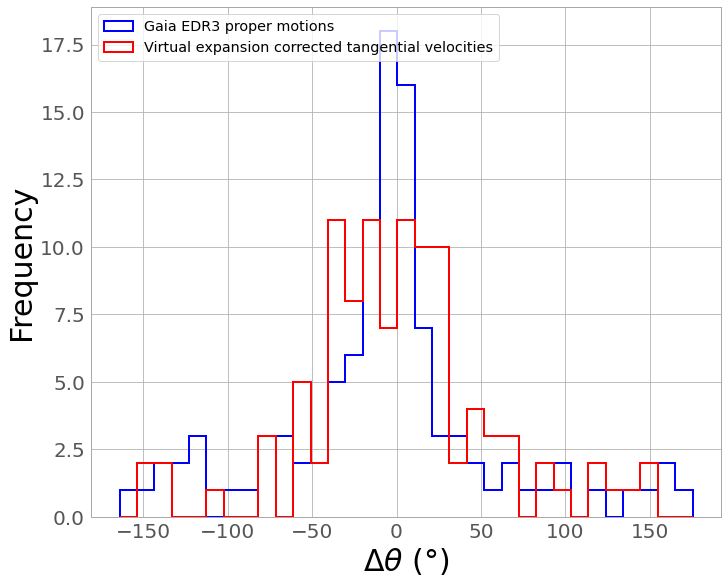}
\caption{Histogram of Velocity Position Angle, i.e., difference in angle, $\Delta \theta$, between proper motion vector and cluster member position vector relative to the cluster center for members of $\sigma$ Ori (blue histogram). The equivalent histogram using virtual expansion corrected tangential 
velocity vectors is shown in red.}
\label{Vangle}
\end{figure}

\subsection{Velocity position angle}
\label{velocity position angle}

We also investigate the distribution of directions in which cluster members are moving relative to the cluster center, by calculating the difference in angle, $\Delta\theta$, between a cluster member's plane-of-sky velocity vector and its position relative to the cluster center, i.e., the Velocity Position Angle. If a star is moving directly away from the cluster center it has $\Delta\theta = 0^\circ$. Thus the distribution of $\Delta\theta$ for stars expanding from an initially compact cluster should show a peak close to $0^\circ$.

In Fig.~\ref{Vangle} we show histograms of $\Delta\theta$ for members of $\sigma$~Ori using uncorrected Gaia EDR3 proper motions (blue) and virtual expansion corrected tangential velocities (red). The typical uncertainty on this angle is $<15^{\circ}$. The peak around 0$^{\circ}$ shows that the majority of cluster members are moving directly outward from the center of the cluster, though in this case the peak of the distribution is smaller for tangential velocities than uncorrected proper motions. We find that 66$\%$ of cluster members have $|\Delta\theta| < 45^\circ$ and 80$\%$ $|\Delta\theta| < 90^\circ$, showing that the majority of cluster members are moving outward from the center, i.e., strong evidence that this cluster is unbound and expanding from a more compact initial configuration. 

In Figs.~\ref{VangleA1}, \ref{VangleA2}, \ref{VangleA3}, and \ref{VangleA4} we show $\Delta\theta$ histograms for each cluster in our sample. Not all clusters have as clearly defined central peaks as either $\sigma$ Ori or $\lambda$ Ori, which is also reflected in the fractions of cluster members with $|\Delta\theta| <45^\circ$ or $<90^\circ$ given for all clusters in Table~\ref{kinematic_table}. However, most clusters still have a sufficient peak to be considered evidence for expansion. We also note the differences in distributions of $\Delta\theta$ in these figures calculated using uncorrected proper motions (blue) and 'virtual expansion' corrected tangential velocities (red). In many clusters the significance of central peaks in these histograms changes between the corrected and uncorrected velocities, illustrating the importance of the 'virtual expansion' correction when identifying plane-of-sky kinematic signatures in nearby clusters.

\subsection{Radial expansion rate}
\label{1d expansion}

By comparing the expansion velocity, $v_{\rm out}$, of cluster members to their radial distance from the cluster center, $R$, we can establish radial expansion gradients, e.g., measured in km~s$^{-1}\:$pc$^{-1}$. These gradients can then be inverted to give expansion timescales (see Sect.~\ref{expansion timescale}).

\begin{figure*}
\includegraphics[width=\textwidth]{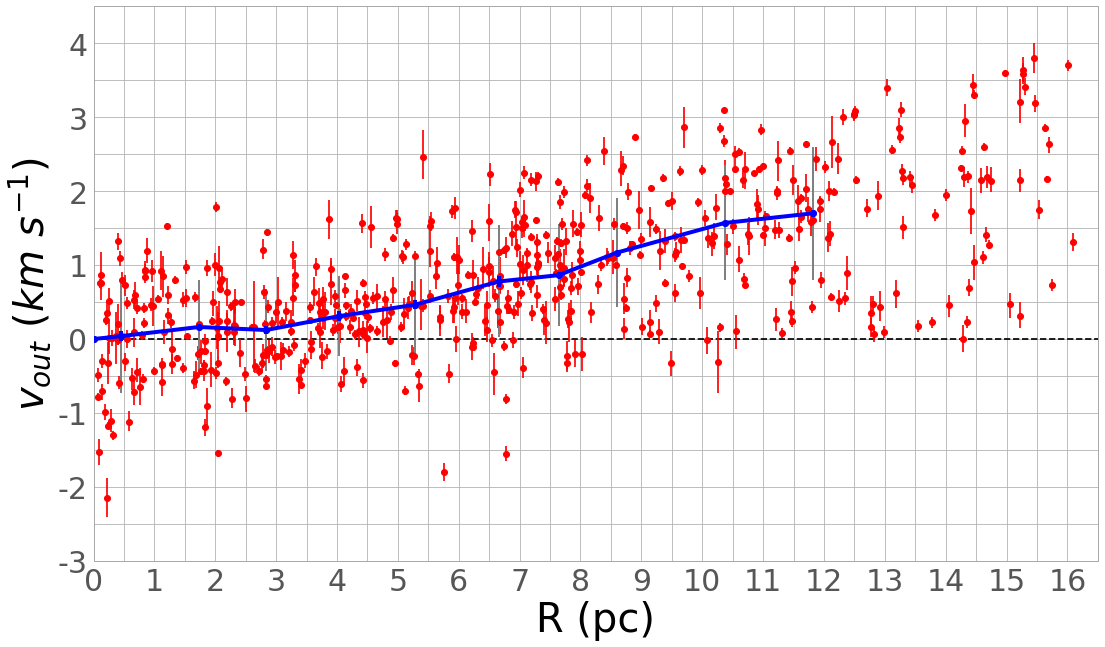}
\caption{Expansion velocity, $v_{\rm out}$, versus radial distance from the cluster center, $R$, for cluster members of $\lambda$ Ori (red), with median expansion velocity $\bar{v}_{\rm out}$ (km~s$^{-1}$) per concentric elliptical annulus containing 50 cluster members each (blue; Table~\ref{vout_table}) and the expansion velocity dispersion $\sigma_{v_{\rm out}}$ per concentric elliptical annulus containing 50 cluster members each (grey). }
\label{VoutProfile}
\end{figure*}

We sort cluster members into bins of 50 members each, using the same concentric elliptical annuli as for our ADP analysis (Sect.~\ref{ADP}). We calculate the median $v_{\rm out}$ of cluster members in each bin for 10,000 iterations, each time randomly sampling uncertainties in $v_{\rm out}$ for individual cluster members per bin, and then taking the 50th percentile value of the posterior distribution as the median $v_{\rm out}$ and the 16th and 84th percentiles as the negative and positive standard errors, respectively. These median $v_{\rm out}$ values and their standard errors are reported in Table~\ref{vout_table}. In Fig.~\ref{VoutProfile} we plot $v_{\rm out}$ against $R$ for cluster members of $\lambda$ Ori (red), similar to Figure 10 of AT24, with the median $v_{\rm out}$ per 50 member bin plotted (blue) at the median radial distance $R$ for members of each bin, and with the standard deviation of expansion velocities $\sigma_{v_{\rm out}}$ per bin given by gray errorbars.

Notably, from Table~\ref{vout_table} it is apparent that the median $v_{\rm out}$ per bin increases with distance from the cluster center $R$ for most clusters. Exceptions to this are Alessi 20 and ASCC 21, which do not have enough cluster members from \citet{cantat-gaudin20} to populate more than one bin each, i.e., they have $<100$ YSOs, and NGC 2547 where there is a slight decrease in median $v_{\rm out}$ towards the edge of the cluster. Also, IC 348 and Pozzo 1 have significantly negative median $v_{\rm out}$ in their innermost bins, indicating contraction of their cores.

Though the median $v_{\rm out}$ values generally increase with distance, this is not always monotonic. In NGC 2451B, for example, the median $v_{\rm out}$ decreases from $0.17 \pm 0.09$ km~s$^{-1}$ to $0.04 \pm 0.02$ km~s$^{-1}$ between the 3rd and 4th concentric annuli. However, in no case is such a decrease of $>3\sigma$ significance.

Following the method described in AT24, we also estimate best-fitting linear rates of expansion via Bayesian inference, using MCMC to sample the posterior distribution function. The model parameters include the linear gradient, $y$-axis intercept and fractional amount by which uncertainties are underestimated. In this analysis we assume that errors follow a Gaussian distribution and are independent. We also use linear least squares to estimate the maximum likelihood. We perform 2000 iterations with 200 walkers and discard the first half as burn-in, reporting the 50th, 16th and 84th percentiles of the posterior distribution as linear best-fit parameters and their respective uncertainties, which we report in Table~\ref{kinematic_table}.

\subsection{Direction of maximum expansion and expansion asymmetry}
\label{direction of maximum expansion}

As AT24 established in $\lambda$ Ori, along with other recent studies \citep{wright24,Armstrong25}, expansion in young clusters and star-forming regions is often anisotropic. For OB associations in particular this anisotropy has been argued to provide evidence against the classical model of expansion from a monolithic compact cluster following residual gas expulsion \citep{wright18,Armstrong22,Armstrong25}. Since the rate of expansion is greater or smaller in different directions, it indicates that the initial configurations of these systems are not monolithic, but substructured. Thus, the level of anisotropy, as well as the direction in which expansion is at its maximum, can provide information of the star-formation history and early evolution of these populations.

To find the direction of maximum expansion, following AT24 for the $\lambda$ Ori cluster, we calculate the 1D rate of expansion along a given orientation axis ($l$ vs $v_{l}$) (using the MCMC method, described above in Sect.~\ref{1d expansion}), then rotate the axis by 
2$^{\circ}$ and repeat until we have calculated rates of expansion for axes rotated by up to 180$^{\circ}$.

\begin{figure}
    {\includegraphics[width=245pt]{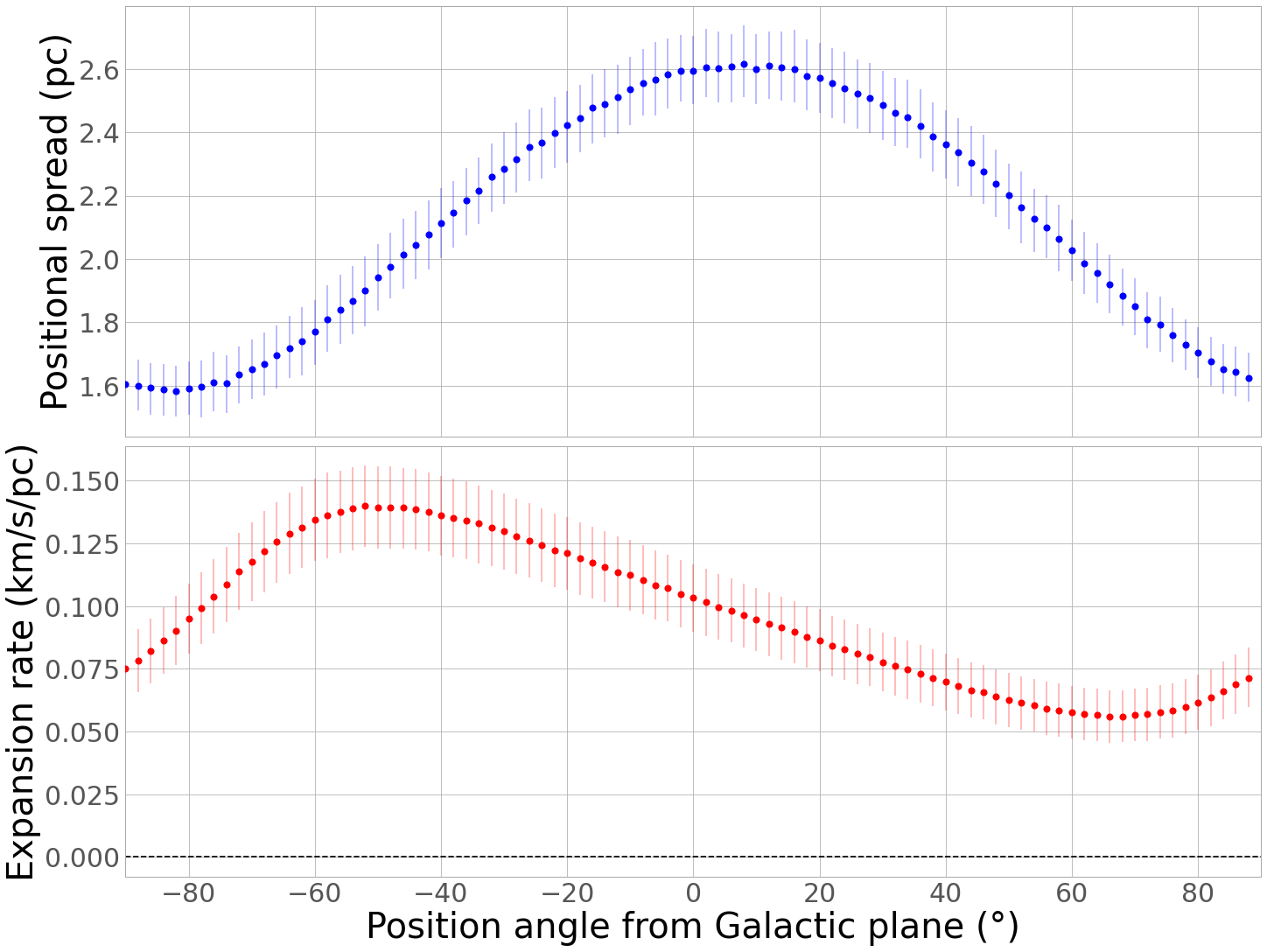} }%
    \setlength{\belowcaptionskip}{-10pt}
    \setlength{\textfloatsep}{0pt}
    \caption{Positional spread (blue) and rate of expansion (red) with uncertainties versus position angle from the Galactic plane for members of IC 2395. The direction of maximum expansion is shown to be at -46$^{\circ}$ below the Galactic plane with increasing longitude, while minimum expansion is in the direction 70$^{\circ}$ above the Galactic plane. The rates of maximum and minimum expansion are different at the $>$4$\sigma$ confidence level, making the plane-of-sky expansion of IC 2395 significantly anisotropic. }%
    \label{ExpansionAsymmetry}%
\end{figure}

\begin{figure} 
    {\includegraphics[width=245pt]{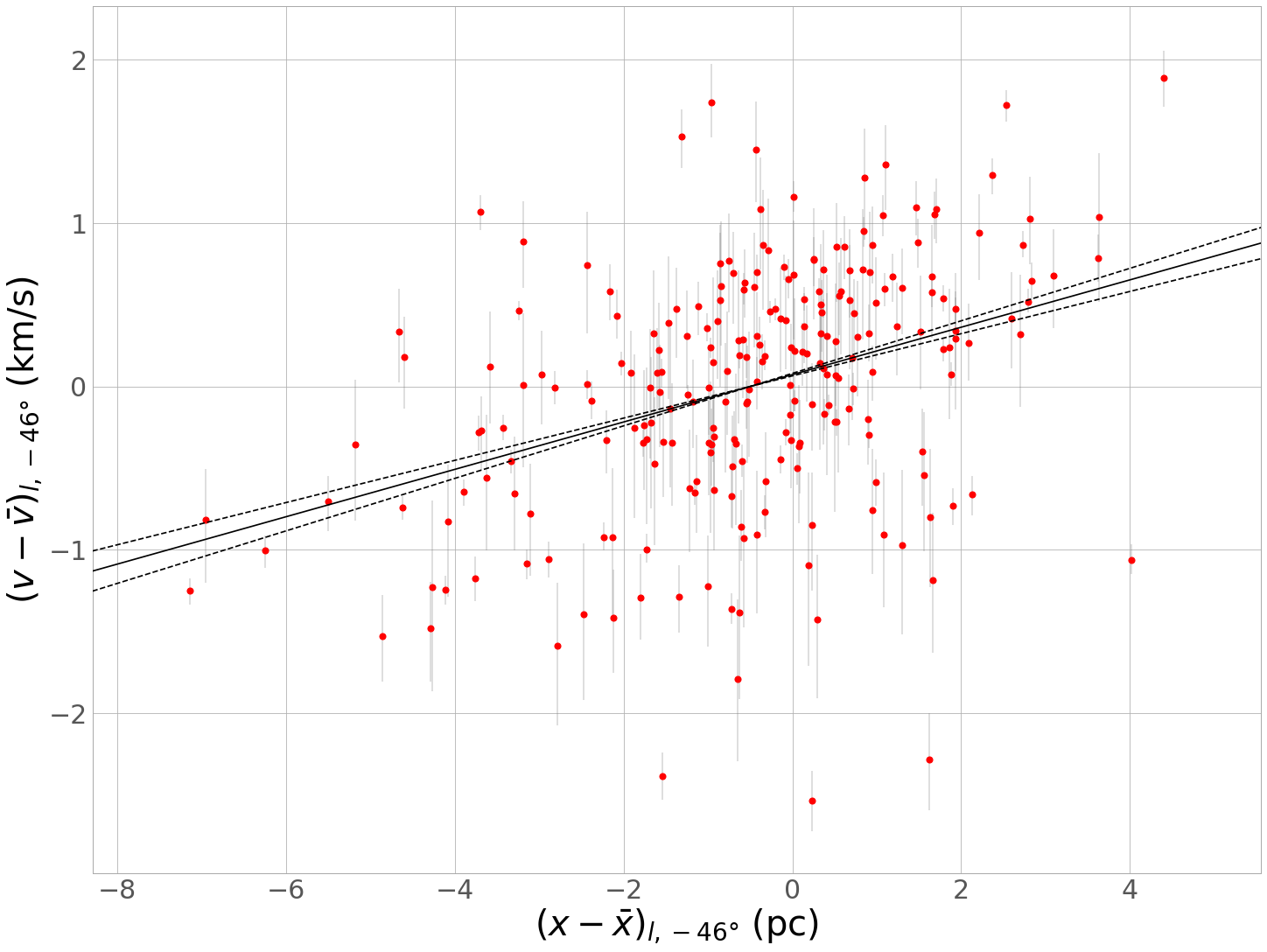} }%
    \setlength{\belowcaptionskip}{-10pt}
    \setlength{\textfloatsep}{0pt}
    \caption{
    Position vs velocity in the direction of maximum expansion for members of IC 2395 with the best fit linear gradient and 1$\sigma$ errors of $0.146^{+0.016}_{-0.016}$ km~s$^{-1}$ pc$^{-1}$, which indicates a significant expansion trend with a corresponding expansion timescale of $6.849^{+0.843}_{-0.676}$ Myr.} 
    \label{MaxExpansion}%
\end{figure}

As an example, we plot the expansion rate against axis orientation for IC~2395 in Fig.~\ref{ExpansionAsymmetry}. For this cluster we find the maximum rate of expansion to be $0.146^{+0.016}_{-0.016}$ km~s$^{-1}$pc$^{-1}$ along an axis that is at -46$^\circ$ clockwise from the Galactic plane, which is plotted in Fig.~\ref{MaxExpansion}. We note this is a 9$\sigma$ detection of expansion for this cluster. We report the maximum and minimum expansion rates and their plane-of-sky orientations for each cluster in Table ~\ref{kinematic_table}.

For seven clusters in our sample, the minimum rate of expansion is significantly negative, which indicates an overall cluster contraction in this direction, but in all cases the maximum rate of expansion is significantly positive. This means that clusters can be simultaneously expanding in one direction whilst contracting in another. 


Clusters vary in the significance of expansion asymmetry. The most significant asymmetries ($>10 \sigma$) are found in the most nearby clusters, Melotte 20, IC 2602, Platais 8 \& IC 2391, despite having relatively small maximum rates of expansion. This is due to their relatively large number of cluster members and large spread of cluster members on the sky, which make the measured expansion rates for these cluster more precise, thus improving the significance of their asymmetry. These clusters are also among the oldest in our sample, all being $\tau_{iso}>25$ Myr. While not being as significant as in the $\tau_{iso}>25$ Myr clusters mentioned before, younger clusters ($\tau_{iso}<10$ Myr) such as $\lambda$ Ori, Gulliver 6, Pozzo 1 \& ASCC 21 still exhibit expansion asymmetry at the $>5 \sigma$ significance level. These metrics of maximum and minimum expansion rates and degree of asymmetry are constraints on cluster formation models.

\subsection{Expansion timescales}
\label{expansion timescale}

From the linear gradients fit to $R$ and $v_{out}$ and their uncertainties we derive the corresponding expansion timescales $\tau_{1D,linear}$, which we we report in Table~\ref{kinematic_table}.

Only 10 clusters in our sample have positive 1D linear expansion rates of $>2\sigma$ significance. Of these, $\lambda$ Ori, ASCC 19, Pozzo 1, Collinder 135 \& Platais 8 have radial expansion timescales $\tau_{1D}$ in good agreement with their isochronal ages from \citet{hunt24} $\tau_{iso,HR24}$ ($5.64^{+1.11}_{-0.8}~\textrm{Myr}$, $7.75^{+4.43}_{-2.13}~\textrm{Myr}$, $9.74^{+4.87}_{-2.49}~\textrm{Myr}$, $25.58^{+15.35}_{-6.98}~\textrm{Myr}$ \& $27.65^{+6.45}_{-4.4}~\textrm{Myr}$, respectively). However, $\sigma$ Ori, IC 2395, ASCC 21, ASCC 16 \& Gulliver 9 have expansion ages much greater than their isochronal ages $\tau_{1D}>\tau_{iso, HR24}$ ($5.85^{+1.79}_{-1.15}~\textrm{Myr}$, $60.18^{+67.7}_{-20.83}~\textrm{Myr}$, $29.23^{+19.49}_{-9.17}~\textrm{Myr}$, $29.23^{+13.4}_{-7.46}~\textrm{Myr}$ \& $17.05^{+3.83}_{-2.64}~\textrm{Myr}$, respectively). 

This shows that fitting 1D radial expansion models to young clusters, without allowing for anisotropy, does not always provide robust evidence for expansion that may be evident through other measures, such as expansion velocities $\bar v_{out}$ (Sect.~\ref{expansion velocities}) or radial expansion which accounts for anisotropy (see Sect.~\ref{direction of maximum expansion}). Also, even if a 1D radial expansion rate is significantly positive, the associated expansion timescale $\tau_{1D,linear}$ will not necessarily be in good agreement with other age estimates. For clusters where expansion rates are anisotropic, there will likely be considerable scatter in $v_{out}$ for cluster members as a function of $R$, which makes their associated timescales less reliable, as AT24 concluded for $\lambda$ Ori (see also the scatter of $v_{out}$ in Fig.~\ref{VoutProfile}).   

As well as the radial expansion gradients, we also consider estimating cluster expansion ages by calculating a radial expansion age per star from $R/v_{out}$ and taking the median value as the cluster expansion age. We do this in a Monte Carlo approach with 10,000 iterations, each time sampling the uncertainties on $R$ and $v_{out}$ for each star. From the posterior distribution of median $R/v_{out}$ ages, we take the 50th percentile as the cluster age, and the 16th and 84th percentile values as the lower and upper uncertainties respectively. We do this both for all members per cluster and also only for cluster members with $|\Delta\theta|<45^\circ$. The resulting ages $\tau_{1D,median}$, $\tau_{1D,median,|\Delta\theta|<45^\circ}$ and their respective uncertainties are reported per cluster in Table~\ref{kinematic_table}.

\begin{figure} 
    {\includegraphics[width=245pt]{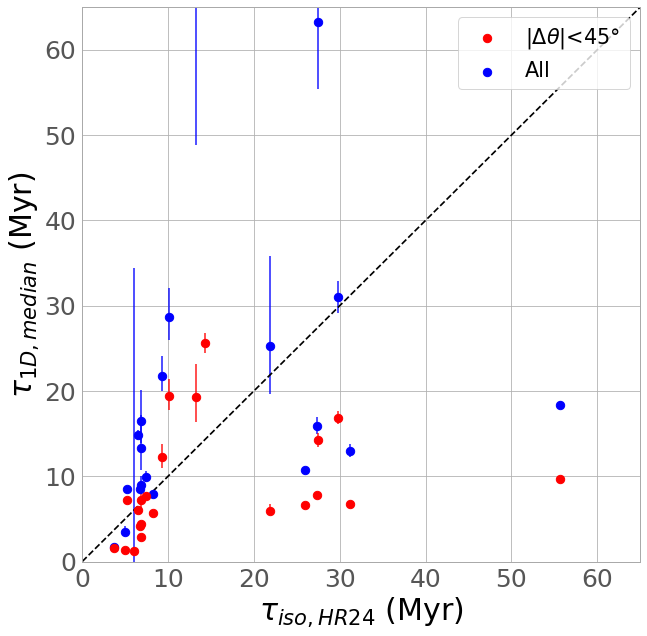} }%
    \setlength{\belowcaptionskip}{-10pt}
    \setlength{\textfloatsep}{0pt}
    \caption{Cluster ages from \protect\citet{hunt24} versus $\tau_{1D,median}$ ages for all cluster members (blue) and for only cluster members with $|\Delta\theta|<45^\circ$ (red). } 
    \label{Age_Rovervout}%
\end{figure}

Median ages using all cluster members are significantly positive for all clusters, except for IC~348, which has a median $v_{out}$ consistent with zero or slight contraction (Sect.~\ref{expansion velocities}). However, the ages themselves do not necessarily correlate well with isochronal age estimates from the cluster catalog, as shown in Fig.~\ref{Age_Rovervout} (blue points). The median ages estimated using only cluster members with $|\Delta\theta|<45^\circ$ (red) show better agreement with isochronal ages, especially for the younger clusters $\tau_{iso,HR24}<10$ Myr. However, there are still a number of clusters with isochronal ages $\tau_{iso,HR24}>20$ Myr where $\tau_{1D,median,|\Delta\theta|<45^\circ}$ are much smaller. This may be evidence that these clusters are expanding, but began doing so later on in their evolution, i.e., perhaps because of a prolonged period of gradual cluster formation in a gas-rich, bound state \citep[e.g.,][]{2019MNRAS.483.4999F,2023MNRAS.523.2083F}.

In order to derive more reliable expansion timescales than those calculated without accounting for anisotropy (Sect.~\ref{1d expansion}), we derive expansion timescales for each cluster from the rate of expansion in the direction of maximum expansion $\tau_{max}$ (Sect.~\ref{direction of maximum expansion}), which are given in Table ~\ref{kinematic_table}.

For many clusters these expansion timescales are in good agreement with their isochronal ages \citep{cantat-gaudin20,dias21,hunt24},
i.e., $\tau_{\rm max}\approx\tau_{\rm iso}$, 
especially for clusters with $\tau_{\rm iso} < 10\:$Myr. The agreement between the expansion timescale and isochronal age for $\lambda$~Ori was discussed in AT24, where the expansion timescale $\sim7$ Myr is close to various isochronal age estimates of 4 - 6 Myr \citep{zari19,cao22}. However, some clusters, particularly those with $\tau_{\rm iso} > 25$ Myr, tend to have expansion timescales much smaller than their isochronal ages. Again, this may indicate that they have started expanding later in their evolution and not immediately after gas expulsion. We discuss this possibility further in Section~\ref{discussion}.

In particular, we consider the 10 clusters for which we calculated 1D radial expansion timescales, $\tau_{1D}$ (Sect.~\ref{1d expansion}), as they had positive 1D linear expansion rates of $>2\sigma$ significance. We note that for the 5 clusters with $\tau_{1D}$ in good agreement with $\tau_{iso,HR24}$, timescales derived from the maximum expansion rates $\tau_{max}$ are also in good agreement with $\tau_{iso, HR24}$, except for Platais 8 where  $\tau_{max}<\tau_{iso,HR24}$ and $\lambda$ Ori where $\tau_{max}>\tau_{iso,HR24}$. On the other hand, for the 5 clusters with $\tau_{1D}>\tau_{iso, HR24}$, timescales derived from the maximum expansion rates are significantly smaller and in better agreement with isochronal age, except for Gulliver 9 where both $\tau_{1D}, \tau_{max}>\tau_{iso, HR24}$, and $\sigma$ Ori where $\tau_{max}<\tau_{iso, HR24}$. The expansion timescales of Gulliver 9 are in much better agreement with the isochronal age estimated in \citet{cantat-gaudin20} rather than in \citet{hunt24} (17.8 Myr rather 10.1 Myr), but it is the only cluster out of these 10 where this is the case.

\begin{figure} 
    {\includegraphics[width=245pt]{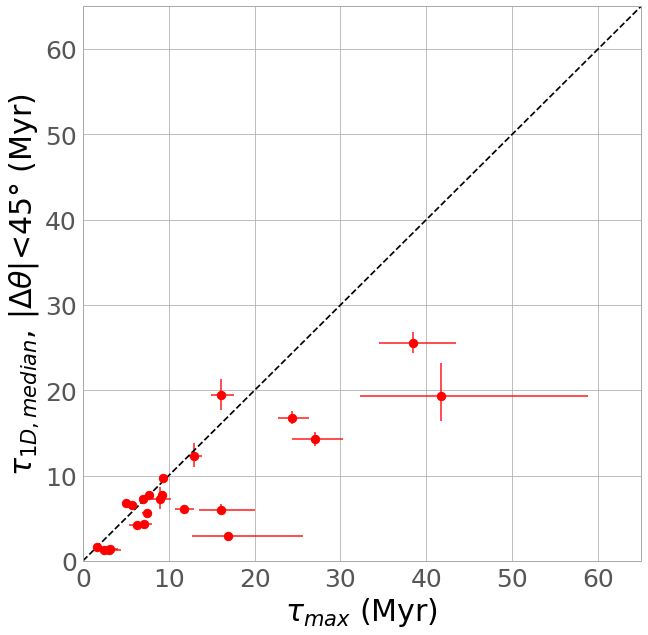} }%
    \setlength{\belowcaptionskip}{-10pt}
    \setlength{\textfloatsep}{0pt}
    \caption{Maximum expansion timescale $\tau_{max}$ versus $\tau_{1D,median}$ ages for cluster members with $|\Delta\theta|<45^\circ$. } 
    \label{MaxExpansion_Rovervout}%
\end{figure}

The expansion timescales $\tau_{max}$ are in better agreement with $\tau_{1D,median,|\Delta\theta|<45^\circ}$, as shown in Fig.~\ref{MaxExpansion_Rovervout}. The agreement is best for clusters with maximum expansion timescales $<10$ Myr, but for clusters with maximum expansion timescales $>10$ Myr $\tau_{1D,median,|\Delta\theta|<45^\circ}$ tend to be relatively underestimated.




\begin{figure*}
	\includegraphics[width=\textwidth]{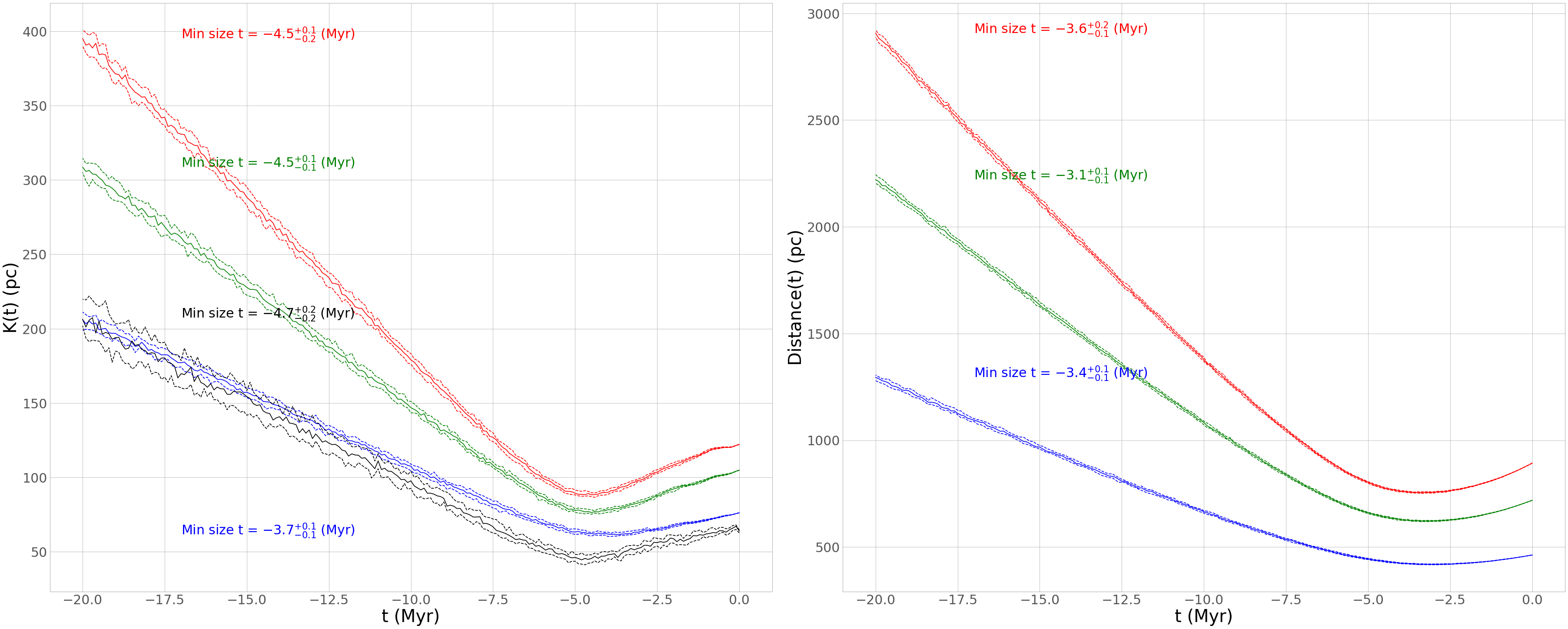}
	\setlength{\belowcaptionskip}{-10pt}
	\setlength{\textfloatsep}{0pt}
	\caption{2D cluster size traceback for IC 2602. \textit{Left:} Minimum spanning-tree total length as a function of trace-back time with no filter for outliers (red), $3\sigma$ velocity outliers removed (green), 2$\sigma$ velocity outliers removed (blue) and 32\% longest branches removed (black) with their respective uncertainties. \textit{Right:} Sum of distances for each star to the association subgroup center as a function of trace-back time with no filter for outliers (red), 3$\sigma$ velocity outliers removed (green), 2$\sigma$ velocity outliers removed (blue). }
	\label{MinAreaPlot}
\end{figure*}

\begin{figure*}
	\includegraphics[width=\textwidth]{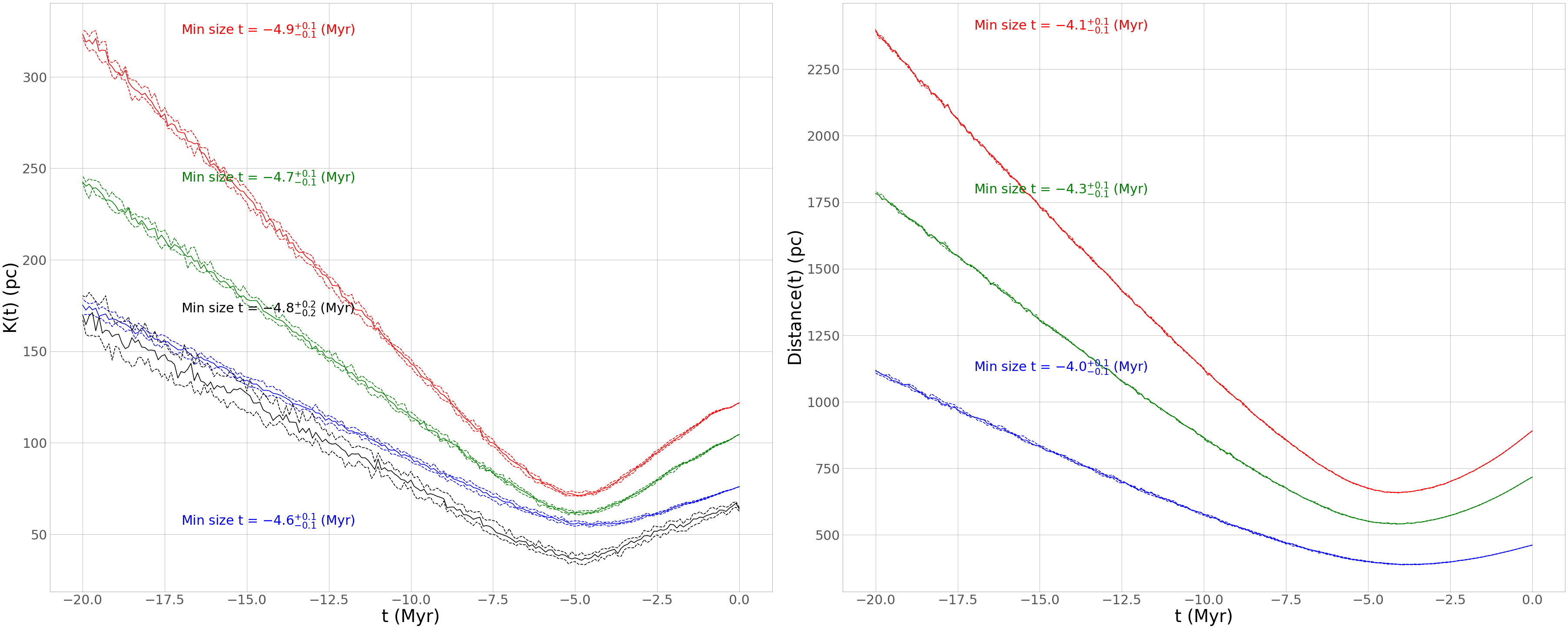}
	\setlength{\belowcaptionskip}{-10pt}
	\setlength{\textfloatsep}{0pt}
	\caption{2D cluster size traceback for IC 2602 with the correction for estimated size inflation due to error propagation. \textit{Left:} Minimum spanning-tree total length as a function of trace-back time with no filter for outliers (red), 3$\sigma$ velocity outliers removed (green), 2$\sigma$ velocity outliers removed (blue) and 32\% longest branches removed (black) with their respective uncertainties. \textit{Right:} Sum of distances for each star to the association subgroup center as a function of trace-back time with no filter for outliers (red), 3$\sigma$ velocity outliers removed (green), 2$\sigma$ velocity outliers removed (blue). }
	\label{MinAreaPlot_2}
\end{figure*}

An alternative method of estimating kinematic age is via 'traceback'. By estimating the relative positions of cluster members in the past, we can estimate when a cluster will have been in its most compact configuration, yielding a kinematic age for the cluster. Following the approach described in \citet{armstrong24}, we calculate 2D Cartesian past positions $X(t)$, $Y(t)$, for cluster members with a linear approximation in 0.1 Myr steps, according to their relative tangential velocities. At each step we estimate the cluster size using MST total length \citep{squicciarini21} and sum of distances \citep{quintana22} between members. We do this with a Monte Carlo approach, randomly sampling from observational uncertainties for 1000 iterations and taking 1$\sigma$ uncertainties on the past cluster size metrics from the 84th and 16th percentiles of the posterior distribution. We also apply several filters on cluster members for calculating past cluster size, removing 3$\sigma$ or 2$\sigma$ velocity outliers and the 32\% longest branches for the MST metric, which are plotted in green, blue and black in Fig.~\ref{MinAreaPlot} using IC 2602 as an example, with red for the full sample of cluster members.

However, as is discussed in Section 5.7.1 of \citet{Armstrong25}, these estimates of size will be inflated by scatter due to observational uncertainties in parallax and proper motions which are propagated through the traceback calculation. Following the approach described in \citet{Armstrong25}, we estimate the inflation of the size metrics due to uncertainties by performing numerical simulations, where we create a synthetic population of $n$ stars equal to the number of cluster members for a given cluster, and place them at the origin of a 2D Cartesian space. We then add perturbations in 2D positions and velocities randomly sampled from the observed position and velocity uncertainties to the synthetic population of stars and calculate the size metrics. We perform 10 000 iterations of this simulation at each time step and take the median of the posterior distributions of $Dsum(\tau)$ and $MSTlength(\tau)$ as the predicted inflation of the size estimates, and the 16th and 84th percentiles values as the $1\sigma$ uncertainties. We then subtract these inflation factors from the size metrics calculated at each step of the traceback analysis. This is also done for each filtered subset of cluster members. We show the corrected traceback size of IC 2602 in Fig.~\ref{MinAreaPlot_2} as an example.

\begin{figure} 
    {\includegraphics[width=245pt]{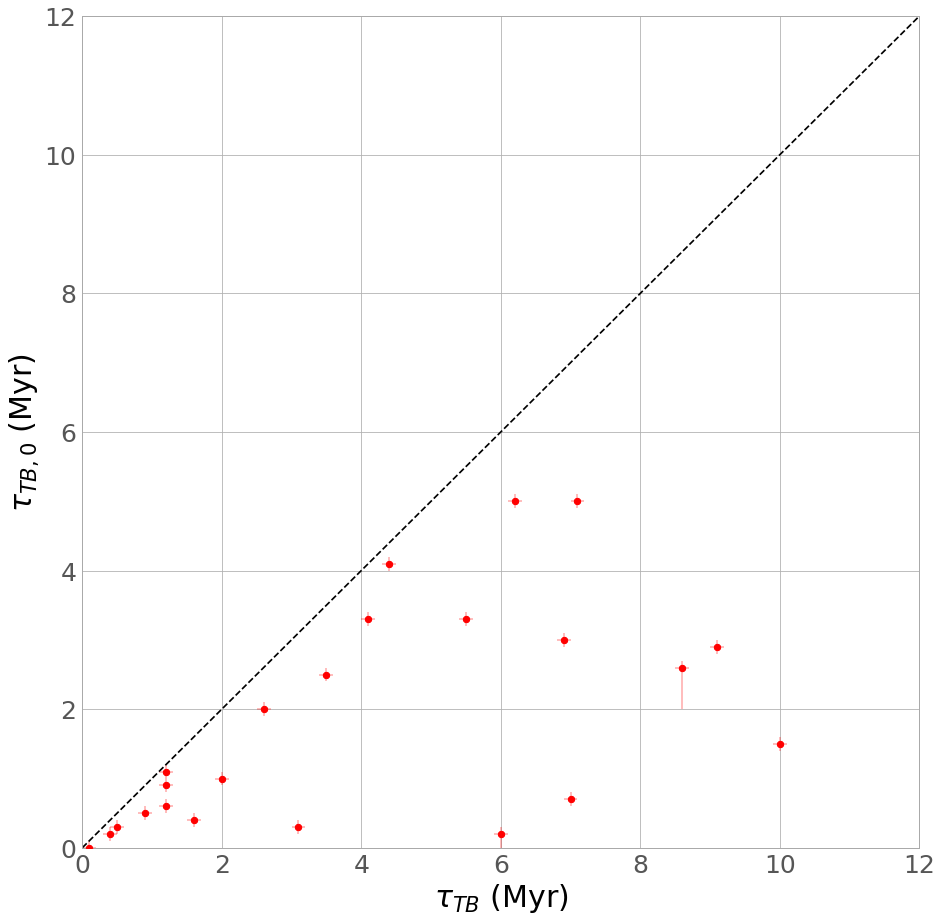} }%
    \setlength{\belowcaptionskip}{-10pt}
    \setlength{\textfloatsep}{0pt}
    \caption{Traceback timescales with the correction for size inflation due to astrometric uncertainties $\tau_{TB}$ against uncorrected traceback timescales $\tau_{TB,0}$.} 
    \label{TBtimeComp}%
\end{figure}

The resulting corrected traceback timescales $\tau_{TB}$ are older than the uncorrected timescales $\tau_{TB,0}$ by varying amounts, as illustrated in Fig.~\ref{TBtimeComp}. The oldest traceback timescale is now $10 \pm 0.1$ Myr for Pozzo 1 ($\gamma$ Vel). We report both the corrected and uncorrected traceback timescales in Table~\ref{kinematic_table} for the no-outlier case (red) using the sum of distances metric \citet{quintana22}.

\begin{figure*} 
    {\includegraphics[width=172pt]{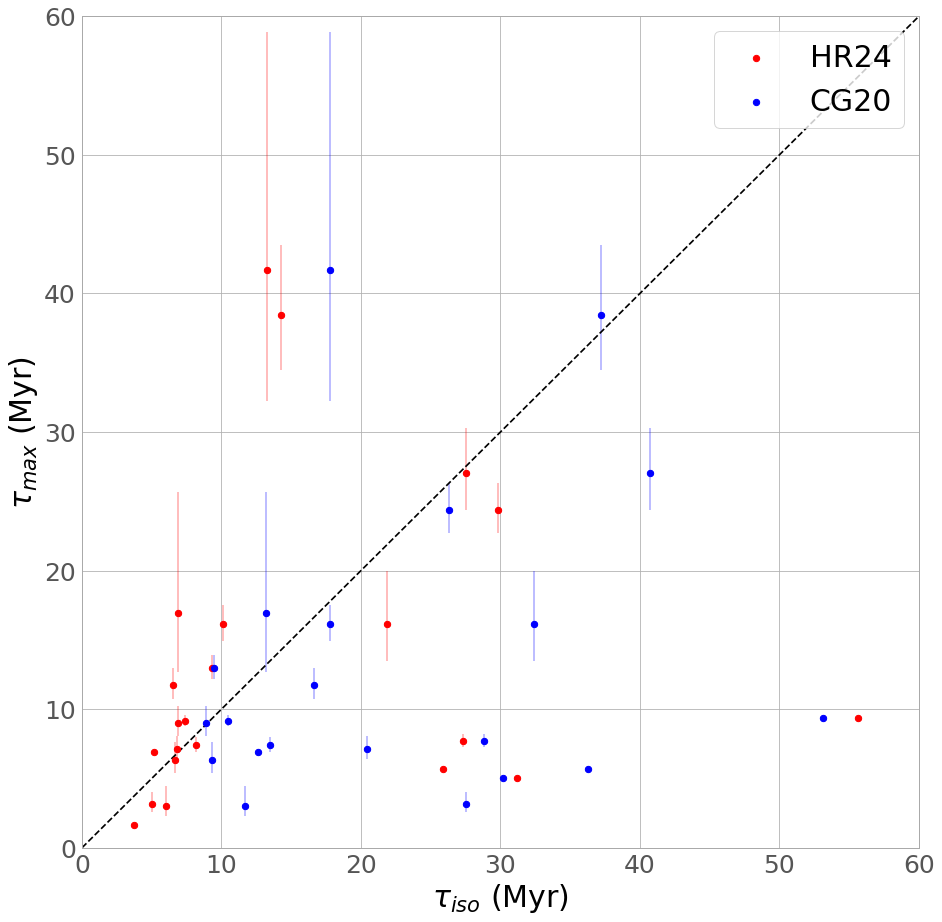} }%
    {\includegraphics[width=172pt]{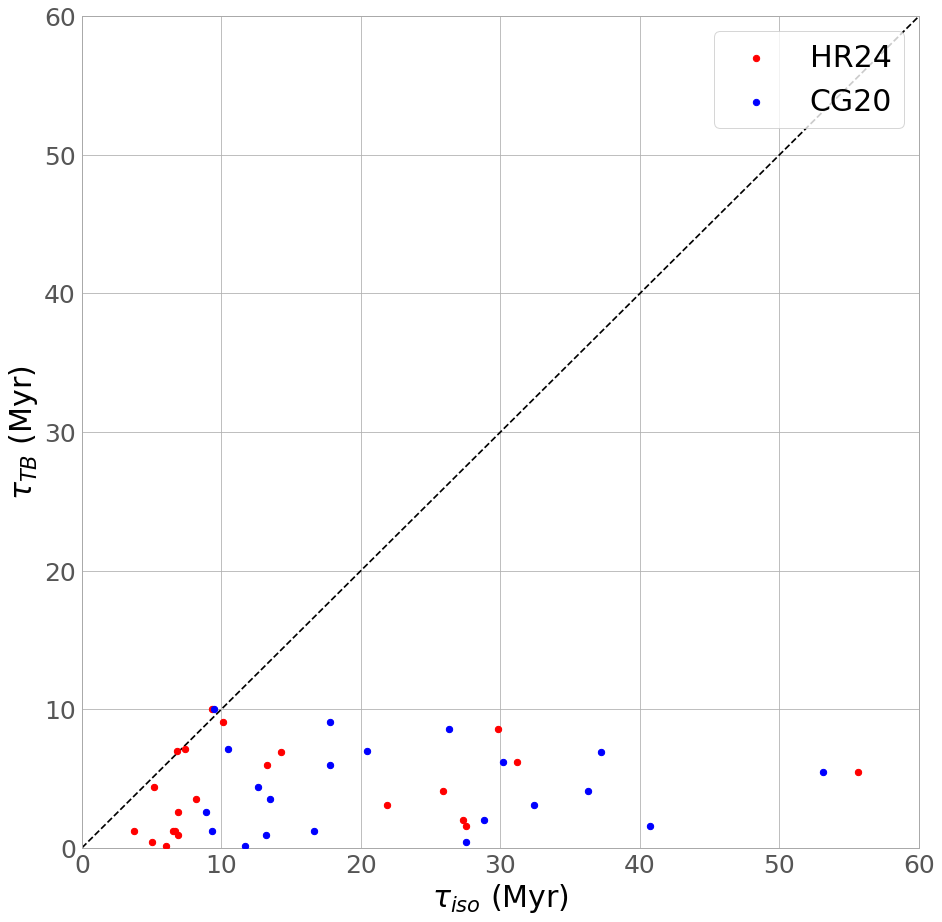}}
    {\includegraphics[width=172pt]{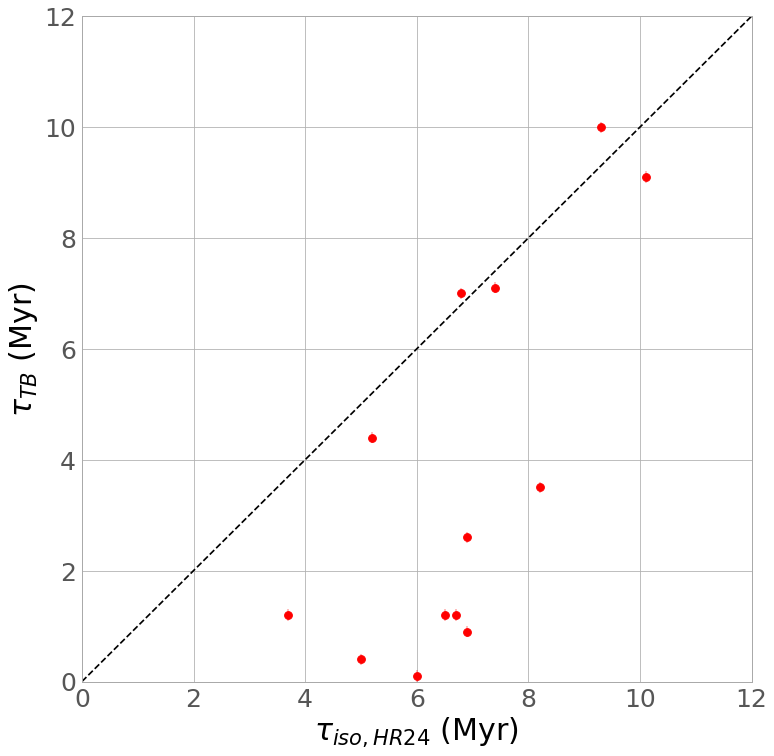}}
    \setlength{\belowcaptionskip}{-10pt}
    \setlength{\textfloatsep}{0pt}
    \caption{{\it (a) Left:} Expansion age based on the maximum rate of expansion per cluster, $\tau_{max}$, versus isochronal age, $\tau_{\rm iso}$, with estimates shown from \protect\citet{hunt24} (red symbols) and \protect\citet{cantat-gaudin20} (blue symbols). {\it (b) Middle:} Linear traceback age, $\tau_{\rm TB}$, versus isochronal age, $\tau_{\rm iso}$. {\it (c) Right:} As (b), but zooming in to 0 to 12 Myr ages.}
    \label{AgeExpansionTime}%
\end{figure*}

In Fig.~\ref{AgeExpansionTime} we plot cluster isochronal ages, $\tau_{\rm iso}$ \citep{cantat-gaudin20,hunt24} against the expansion timescale derived from the cluster's maximum rate of expansion $\tau_{\rm max}$ (\textit{left}), and against the traceback timescale $\tau_{TB}$ (\text{middle \& right}), with respective uncertainties. Generally, isochronal ages from \citet{hunt24} tend to be younger for these clusters than ages from \citet{cantat-gaudin20}. We also see that traceback timescales are always younger than expansion timescales. 

Many clusters have expansion timescales consistent with their isochronal ages, i.e., $\tau_{\rm max}\approx\tau_{\rm iso}$ (Fig.~\ref{AgeExpansionTime}a) even up to relatively old ages; NGC 2451B has an isochronal age of 27.5 Myr from \citet{hunt24} and an expansion timescale of $27.027^{+3.276}_{-2.637}$ Myr and Collinder 359 has an isochronal age of 37.2 Myr from \citet{cantat-gaudin20} and an expansion timescale of $38.461^{+5.017}_{-3.979}$ Myr. A close agreement between kinematic age (such as an expansion timescale) and isochronal age is consistent with the cluster becoming gravitationally unbound shortly after formation. 

However, there are also many clusters in our sample with expansion timescales much shorter than their isochronal ages $\tau_{max}<\tau_{iso}$. Platais 8, for example, has isochronal ages of 30.2 and 31.2 Myr respectively from \citet{cantat-gaudin20} and \citet{hunt24}, but its maximum expansion rate of $0.199^{+0.006}_{-0.006}$ km~s$^{-1}$ pc$^{-1}$ implies a timescale of $5.025^{+0.156}_{-0.147}$ Myr. The traceback timescale of Platais 8 is $\tau_{TB}=6.2\pm0.1$ Myr, and the median 1D age is $\tau_{1D,median,|\Delta\theta|<45^\circ}=6.78^{+0.36}_{-0.30}$ Myr, both also in reasonable agreement with the expansion timescale. The signatures of expansion for Platais 8 are very significant, but the timescale indicates that this expansion only began $\sim25$ Myr into its evolution.

For clusters with measured $\tau_{1D}$ as well as $\tau_{max}$, $\tau_{max}$ is in most cases in better agreement with isochronal ages $\tau_{iso}$, except for Platais 8, and in most cases $\tau_{1D}$ and $\tau_{max}$ are in better agreement with isochronal ages from \citet{hunt24} rather than from \citet{cantat-gaudin20}, except for Gulliver 9.

In Fig.~\ref{AgeExpansionTime} (\textit{left}) we show isochronal ages from \citet{hunt24} $\tau_{iso,HR24}$ against traceback timescales $\tau_{TB}$ for clusters with ages up to 12 Myr. Five of these clusters have traceback timescales in good agreement with their isochronal ages, while 8 others have traceback timescales significantly smaller than isochronal ages $\tau_{TB}<\tau_{iso,HR24}$. This offset between isochronal age and kinematic age has been seen in other regions, particularly among subgroups of Upper Scorpius \citep{miret-roig24}, where it has been claimed as evidence for an ``embedded phase'' in early cluster evolution. In this scenario, a cluster forms in a molecular cloud and the surrounding gas accounts for the majority of the cluster's binding mass, keeping the cluster gravitationally bound before feedback from massive stars begin to dispel the gas and the cluster begins to expand from the loss of binding mass. The offsets between traceback timescale and isochronal age we see for these 8 clusters in our sample would then indicate ``embedded phases'' typically between $\sim3-6$ Myr. 

However, the linear traceback ages, $\tau_{TB}$, that we calculate do not always agree with the expansion timescales, $\tau_{max}$, for these clusters. Some difference may be expected between these two kinematic age methods simply due to the spatial and kinematic structure of the cluster members. In Fig.~\ref{KinematicAgesDiagram} we show a simple illustrative example where a 2D arrangement of four stars with different relative velocities per row result in different kinematic age estimates via linear traceback (\textit{left column}) or the gradient of maximum expansion (\textit{right column}). Ultimately, the difference is due to the implicit assumptions in each method of what the initial configuration of stars is likely to be. 

On the other hand, the oldest traceback age is $\tau_{TB}=10 \pm 0.1$ Myr for Pozzo 1 ($\gamma$ Vel). Many clusters in our sample have older isochronal ages and expansion timescales than this, but the traceback timescales remain $\tau_{TB}<10$ Myr, as seen in Fig.~\ref{AgeExpansionTime} (\textit{lower panel}). A possible reason is that these traceback calculations do not account for Galactic orbits, and are thus linear approximations, and so for older clusters this approximation underestimates expansion timescales. In order to perform precise dynamical traceback calculations for every cluster, we would need precise spectroscopic RVs for the majority of their cluster members, which in many cases are not currently available. However, for a few clusters in our sample with quality RVs for $>40$ cluster members ($\lambda$ Ori, Pozzo 1, Melotte 20 \& NGC 2547) we perform 3D traceback calculations following the approach described in \citet{Armstrong22,Armstrong25} (for further details, see Appendix~\ref{3Dtraceback}). Overall, we find no evidence that traceback with orbital equations increases the kinematic age estimates for these clusters. We also perform numerical simulations of the traceback calculations using mock clusters with velocities and respective uncertainties sampled from our cluster membership lists, but with positions set to give the mock clusters a 'true' expansion age of 20 Myr. We perform these simulations using velocities and respective uncertainties sampled from Melotte 20, $\lambda$ Ori and NGC 2264 (for further details, see Appendix~\ref{Traceback test}). In all cases, these simulations produce $\tau_{TB}\sim20$ Myr, showing that the apparent 10 Myr upper limit in our results is not an artifact of the methodology, and we therefore argue that our 2D linear traceback calculations are accurate given the available observational data. 

For nearby clusters in close proximity to GMCs, the gravitational potential of the GMC may have a greater influence than the Galactic potential. For example, the Galactocentric acceleration of the Solar system is $\sim 2.3 \times 10^{-10} ~\textrm{m\: s}^{-2}$ \citep{xu12,klioner21}, whereas the acceleration due to a GMC of mass $10^6\:M_\odot$ at $10\:$pc is $\sim 1 \times 10^{-9} ~\textrm{m\: s}^{-2}$. Simulation work by \citet{za19} indicate that the gravitational potential of dense gas dispersed by feedback in the early evolution of a star cluster can pull the stellar population outwards and accelerate expansion, and in particular, that asymmetric gas expulsion would lead to asymmetric cluster expansion. Further observational studies of dense gas structures in the vicinities of young clusters is needed to better constrain such effects and inform the interpretation of expansion signatures in young clusters and their respective timescales.

Thus, overall, while isochronal ages and kinematic ages are often found to be in good agreement for young clusters, many systems also show large discrepancies. Some clusters have kinematic ages slightly lower than isochronal ages, which could be attributed to an 'embedded phase' \citep[e.g.,][]{miret-roig24}, while others have kinematic ages much lower (with a delay $>10$ Myr). One possibility is that clusters with ages $\sim20-60\:$Myr that are detected in the analysis of Gaia data by \cite{cantat-gaudin20} are ones that have remained gravitational bound for a more prolonged period and are only recently seen to be expanding.
Other recent investigations into the timescales of cluster emergence have suggested dependence on other cluster properties, namely total mass \citep{pedrini26}, which could go some way to explaining the variation seen in young nearby clusters.  

It should also be noted that cluster age estimates from stellar evolution models $\tau_{iso}$ are still uncertain. Isochronal age estimates are dependent on choice of stellar evolution models (e.g., magnetic or non-magnetic), choice of photometric colour \citep[see][]{ratzenbock23,cheshire25}, cluster membership and isochrone-fitting approach \citep[considering the differences in ages for this sample from][]{cantat-gaudin20,hunt24} as well as observational factors including extinction, variability and unresolved binarity. Kinematic ages also vary between approaches, either 1D $\tau_{1D}$ or anisotropic expansion timescales $\tau_{max}$, linear or orbital traceback $\tau_{TB}$. Kinematic age estimates also neglect N-body dynamics in the cluster, and approximate the past motion of cluster members as linear. For a cluster that has a sparse configuration at the onset of expansion this may be a good approximation, but less so for a cluster that was significantly more compact in the past. 

\section{Discussion}
\label{discussion}

Here we discuss the results of our spatial and kinematic analyses and compare them to results from other relevant literature.

\subsection{Evidence for expansion}
\label{ExpansionDiscussion}

Out of 23 clusters, 18 have expansion velocities $\bar v_{\rm out}$ which are positive at the 5$\sigma$ significance level or greater. Of the remaining five clusters, ASCC 21 and NGC 1980 have expansion velocities $\bar v_{\rm out}$ which are positive at the 4$\sigma$ significance level and NGC 2547 at the 2$\sigma$ significance level. Only NGC 2232 and IC 348 have expansion velocities $\bar v_{\rm out}$ consistent with zero. None of our clusters have expansion velocities $\bar v_{\rm out}$ that are significantly negative indicating contraction.

Out of 23 clusters, 19 have at least 50\% of their member stars moving away from the cluster center on the plane-of-sky. The remaining four clusters are NGC 2451B, NGC 2232, IC 348 \& NGC 2547, which have 40\%, 42\%, 39\% \& 47\% of their members consistent with moving away from the cluster center, respectively. In addition, $\lambda$ Ori, $\sigma$ Ori, ASCC 16 \& ASCC 19 each have at least 70\% of their members moving away from their centers, with at least 50\% of their members having tangential velocities directed $<45\deg$ relative to their position angle around the cluster center, indicating strong radial expansion. In Figs.~\ref{VangleA1}, \ref{VangleA2}, \ref{VangleA3}, and \ref{VangleA4} we show $\Delta\theta$ histograms for each cluster in our sample. In many cases a peak at $\Delta\theta\sim0^{\circ}$ is clearly visible, indicating cluster members moving radially away from the cluster center.

Out of 23 clusters, 19 have significantly positive maximum expansion rates at the 5$\sigma$ level. Of the remaining four clusters, NGC 2264 has a positive maximum expansion rate of 4$\sigma$ significance, NGC 2232 and IC 348 have positive maximum expansion rates of 3$\sigma$ significance, and NGC 1980 has a positive maximum expansion rate of 2.5$\sigma$ significance.

Notably, $\sigma$ Ori has the strongest signatures of expansion in the sample, having the largest proportion of members consistent with moving away from the center (80\%), the greatest expansion velocity ($v_{\rm out} = 0.989 \pm 0.029$ km~s$^{-1}$) and the greatest maximum expansion rate ($0.604 \pm 0.03$ km~s$^{-1}$pc$^{-1}$). It is also the youngest cluster in the sample according to the age estimates of the \citet{hunt24} catalog $\tau_{iso,HR24}$, at 3.7 Myr.

Out of 23 clusters, 3 have significantly negative minimum expansion rates at the 5$\sigma$ level; Platais 8, NGC 2451B \& Melotte 20, IC 2602 has a negative minimum expansion rate of 4$\sigma$ significance, NGC 2547 and IC 348 have negative minimum expansion rates of 2$\sigma$ significance, and NGC 1977 has a negative minimum expansion rate of 1.5$\sigma$ significance. However, in all these clusters, with the exception of NGC 2451B, the magnitude of their positive maximum expansion rates are greater than their negative minimum expansion rates. Notably, the clusters with the most significantly negative minimum expansion rates, Platais 8, NGC 2451B, Melotte 20 \& IC 2602, are among the oldest clusters in the sample, with \citet{hunt24} age estimates of 31.2, 27.5, 55.6 \& 25.9 Myr, respectively.



Out of 23 clusters, 7 have significantly positive minimum expansion rates at the 5$\sigma$ level. Additionally, according to our kinematic traceback analysis (Sect.~\ref{expansion timescale}), all clusters except for NGC 2264 and IC 348 were significantly more compact in the past than their present-day configuration. All of these results give clear evidence that the majority of nearby young clusters are in the process of expanding.

Despite this evidence, the 1D linear expansion trends (Sect.~\ref{1d expansion}) fitted to radial distance from the cluster center, $R$, against expansion velocity, $v_{\rm out}$, often do not show evidence for significant expansion. Only Platais 8, $\sigma$ Ori, ASCC 16, Gulliver 9 and $\lambda$ Ori have 1D expansion trends which are positive at the 3$\sigma$ level. This is likely due to the scatter of $v_{out}$ values for cluster members at a given distance $R$, which in turn is due to anisotropy in cluster expansion (Sect.~\ref{1d expansion}). Thus, we will focus on expansion rates determined from the direction of maximum expansion in the following discussion.

\subsection{Expansion anisotropy}
\label{Expansion_Anisotropy_discussion}

Out of 23 clusters, 13 have significantly anisotropic expansion rates at the 5$\sigma$ level and 6 have significantly anisotropic expansion rates at the 3$\sigma$ level. Of the remaining four clusters, NGC 2232, Alessi 20 \& ASCC 19 have anisotropic expansion rates of 2$\sigma$ significance, while NGC 1980 is the only cluster with no significant evidence for expansion anisotropy.

This indicates that, not only are the majority of young clusters expanding, but they are expanding anisotropically. Thus expansion rates and expansion timescales derived from them vary depending on the direction in which they are measured across a cluster. There is even the possibility that methods of detecting expansion without accounting for anisotropy risk missing these signatures altogether. For example, \citet{guilherme23} analysed 1,237 clusters with membership from \citet{cantatgaudin19c} and found clear evidence for expansion in 14 of them, but failed to detect expansion in clusters such as IC 2602, for which we find clear evidence. We note in Sect.~\ref{1d expansion} that only 10 of our cluster sample have significant positive 1D expansion trends, while all of our clusters show positive expansion trends in the direction of maximum expansion of at least $2\sigma$ significance.

\subsection{Age trends}
\label{Age_trends}

In Table~\ref{correlations_table} we present Kendalls' $\tau$ coefficients and their respective $p$-values (in brackets) for pair-wise combinations of cluster parameters. Results indicating strong evidence for correlations are highlighted in red, and moderate evidence for correlations are highlighted in yellow. For Kendalls' $\tau$ coefficient, values $0.00 -0.06$ are considered to indicate negligible correlation, $0.06 -0.26$ weak correlation, $0.26 -0.49$ moderate correlation, $0.49 -0.71$ strong correlation and $>0.71$ very strong correlation, while negative coefficients indicate inverse correlations of varying strength.

In particular, we find evidence for strong correlations between cluster age $\tau_{iso,HR24}$ and each of cluster core radius $R_c$ and expansion asymmetry, while we find evidence for strong negative correlations between cluster age and each of cluster distance $D$ 
and the absolute value of orientation angle of maximum expansion.

\begin{figure} 
    {\includegraphics[width=245pt]{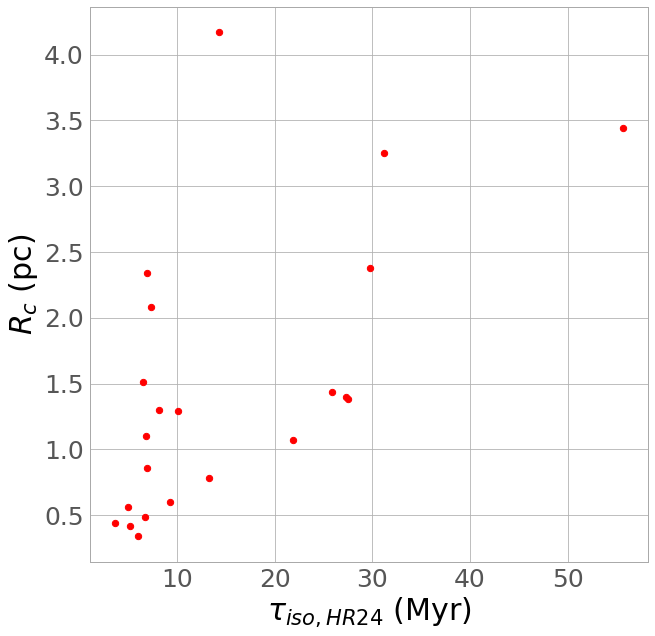} }%
    \setlength{\belowcaptionskip}{-10pt}
    \setlength{\textfloatsep}{0pt}
    \caption{Cluster ages from \protect\citet{hunt24} ($\tau_{iso,HR24}$) against core radii $R_c$ for each cluster.} 
    \label{AgeCoreRadii}%
\end{figure}

In Fig.~\ref{AgeCoreRadii} we plot cluster ages from \citet{hunt24} $\tau_{iso,HR24}$ against the core radius $R_c$ for each cluster measured following the approach described in Sect.~\ref{core}. We perform rank correlation tests and find a Kendall correlation coefficient of 0.52 with a $p$-value 0.001 and a Spearman rank coefficient of 0.69 with a $p$-value of 0.0003. These indicate a strong correlation between a cluster's age and its core radius $R_c$. A similar correlation has been identified in previous studies \citep{pfalzner09,getman18,wright24} between age and cluster (effective) radius, which is usually interpreted as a direct consequence of clusters expanding over time. With this strong age dependency also for core radius, we infer that even the densest concentration of cluster members at the core gradually relaxes and becomes unbound, or less tightly bound, over time, and that in many cases clusters will completely dissolve into the field.

We perform rank correlation tests on cluster ages from \citet{hunt24} $\tau_{iso,HR24}$ against expansion timescales $\tau_{max}$ (see Fig.~\ref{AgeExpansionTime} \textit{upper}) and we find a Kendall correlation coefficient of 0.38 with a $p$-value 0.014 and a Spearman rank coefficient of 0.486 with a $p$-value of 0.0217. These indicate a mild correlation between a cluster's age and its expansion timescale across the sample. As the expansion timescale is the inverse of the maximum expansion rate, this correlation is given as -0.38 (0.014) in Table.~\ref{correlations_table} between Age$_{HR24}$ and $MaxExpRate$. This is a stronger correlation than that found by \citet{wright24} who reported a weak correlation between expansion timescale and age with a p-value of 0.25 from their cluster sample. The main reason for the stronger correlation is likely the difference in cluster sample, as the sample studied by \citet{wright24} was restricted to spectroscopic members of clusters observed in the Gaia-ESO survey \citep{randich13}. Thus their sample contained fewer clusters with fewer members on average than our sample, and often at greater distances (7 of their 18 clusters are at distances $>900$ pc, further away than any cluster in our sample), which all lead to our expansion rates and timescales being more precise on average.

\begin{figure} 
    {\includegraphics[width=245pt]{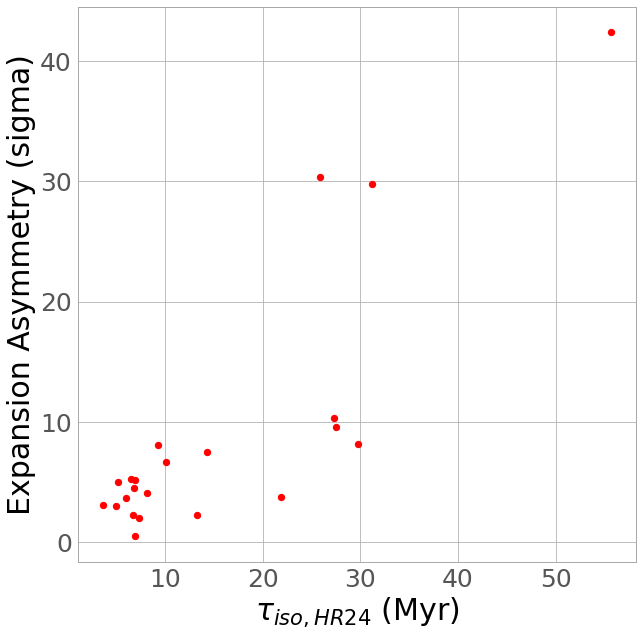} }%
    \setlength{\belowcaptionskip}{-10pt}
    \setlength{\textfloatsep}{0pt}
    \caption{Cluster ages from \protect\citet{hunt24} ($\tau_{iso,HR24}$) against expansion asymmetry for each cluster.} 
    \label{AgeExpAsymmetry}%
\end{figure}

In Fig.~\ref{AgeExpAsymmetry} we plot cluster ages from \citet{hunt24} $\tau_{iso,HR24}$ against the expansion asymmetry for each cluster measured following the approach described in Sect.~\ref{direction of maximum expansion}. We perform rank correlation tests and find a Kendall correlation coefficient of 0.51 with a $p$-value 0.001 and a Spearman rank coefficient of 0.69 with a $p$-value of 0.0003. These indicate a strong correlation between a cluster's age and the asymmetry of its expansion signatures, which indicates that the dynamical evolution of these clusters may be impacted by external forces, such as Galactic potential, causing them to become more kinematically asymmetric over time. This may be a similar mechanism to that which produces tidal tails over $\sim100$ Myr timescales, as have been found in many nearby clusters \citep{risbud25}.

\subsection{Sample-wide trends}
\label{trends}

In Fig.~\ref{frac45_expv} we plot the fraction of cluster members with velocities oriented at $\Delta\theta<|45^\circ|$ relative to the cluster center against the expansion velocity for each cluster measured following the approach described in Sect.~\ref{expansion velocities}. We perform rank correlation tests and find a Kendall correlation coefficient of 0.61 with a $p$-value 0.0001 and a Spearman rank coefficient of 0.755 with a $p$-value of 0.00003. These indicate a very strong correlation between a the proportion of cluster members moving away from the cluster center and the cluster expansion velocity, which further supports the use of both of these properties as measures of cluster expansion.

In Table~\ref{correlations_table} there are is a mild correlation between mean ADP and cluster distance $D$, but otherwise there is no evidence for correlation between mean ADP and any other cluster property. In particular, this indicates that the average level of structure across clusters in this sample is not primarily dependent on age $\tau_{iso}$ or the rate of cluster expansion. Rather, structure can be preserved in older ($>20$ Myr) clusters and younger clusters ($<10$ Myr) may become smoothly distributed on short timescales in different cases. This may depend more strongly on the cluster's internal kinematic properties apart from expansion, such as the velocity dispersion.


In Fig.~\ref{exp_angles} we plot a histogram of orientation angle of maximum expansion rate for clusters in our sample. These orientation angles appear consistent with being uniformly distributed, i.e., there is no obvious preferential direction in which young clusters typically expand. This may indicate that the direction of maximum expansion for young clusters is dependent on their initial structure and velocities inherited from their parent molecular cloud, or local tidal forces from nearby molecular clouds, rather than Galactic tidal forces in the way that tidal tails are produced in older open clusters \citep{risbud25}. There is also evidence for this in Fig.~\ref{age_exp_angles} where we plot the orientation angle of maximum expansion against cluster age from \citet{hunt24}. Younger ($\tau_{iso}< 20$ Myr) clusters have a wide range of orientation angles, but the orientations seemingly tend toward $0^\circ$ (parallel to the Galactic plane) for older clusters. Indeed, in Table.~\ref{correlations_table} we report a strong negative correlation (Kendall; -0.48, $p$-value; 0.002) between cluster age and the absolute value of orientation angle of maximum expansion. This may indicate that, over time, Galactic tidal forces begin to dominate the continued expansion of clusters and the orientation of that expansion tends towards being parallel with the Galactic plane.

\subsection{Comparison to \citet{wright24}}
\label{Wright24}

\citet{wright24} studied the kinematics of 18 young clusters and associations observed as part of the Gaia-ESO survey \citep{randich13}, using cluster membership as determined by spectroscopic youth criteria such as equivalent-widths of Li. They measured 
expansion trends and found that the majority of their clusters showed significant evidence for expansion, and that the majority of these were expanding anisotropically. 

Only 3 of our clusters overlap with the sample analysed by \citet{wright24}; NGC 2264 (S Mon cluster), Pozzo 1 (Gamma Vel) and $\lambda$ Ori (including B30 and B35 clusters) as analysed in AT24. Other clusters they examined either are not present in the \citet{cantat-gaudin20} catalog (e.g., Rho Ophiuchus), are more distant than the $1$kpc limit we impose (e.g., NGC 2244) or have too few RVs in \citet{tsantaki22} (e.g., Collinder 197). 

For NGC 2264, our sample of cluster members (149) here is smaller than that of \citet{wright24} (244). However, the main expansion trends remain in excellent agreement; $\bar v_{out} = 0.333^{+0.06}_{-0.06} ~\textrm{km~s}^{-1}$ and $0.40^{+0.01}_{-0.06} ~\textrm{km~s}^{-1}$, and the maximum expansion rates $0.316^{+0.068}_{-0.068} ~\textrm{km~s}^{-1}\textrm{pc}^{-1}$ and $0.35^{+0.04}_{-0.04} ~\textrm{km~s}^{-1}\textrm{pc}^{-1}$, respectively. Thus the expansion timescale of $\sim3$ Myr is robust between membership samples. The directions of maximum expansion given in \citet{wright24} are reported as orientation angle relative to the RA ($\alpha$) axis, whereas we report orientation angle relative to the axis of galactic longitude $l$. After transforming these orientation angles into the same reference frame, we find that our result for the orientation angle of maximum expansion in NGC 2264 is offset from that of \citet{wright24} by $\sim45^\circ$, though this is a difference of $<2\sigma$ significance given the orientation uncertainties.


For Pozzo 1, our sample of cluster members (312) is significantly larger than that of \citet{wright24} (97), likely due to the small FOV in which the Gaia-ESO survey observed this cluster, but also the distinction between the cluster and the overlapping Vela OB2 association made in \citet{wright24}, following the criteria described in \citet{armstrong20}, which the \citet{cantat-gaudin20} catalog does not make. As a result, there is a notable difference (of $3\sigma$ significance) in the cluster expansion velocities; $\bar v_{out} = 0.16^{+0.026}_{-0.026} ~\textrm{km~s}^{-1}$ compared to $-0.06^{+0.06}_{-0.01} ~\textrm{km~s}^{-1}$. On the other hand, the maximum expansion rates are in good agreement; $0.077^{+0.005}_{-0.005} ~\textrm{km~s}^{-1}\textrm{pc}^{-1}$ compared to $0.05^{+0.04}_{-0.04} ~\textrm{km~s}^{-1}\textrm{pc}^{-1}$. This may indicate that the maximum expansion rate is more robust to differences in cluster membership samples than the expansion velocity. This maximum expansion rate is also in good agreement with the rate of $0.062^{+0.012}_{-0.013} ~\textrm{km~s}^{-1}\textrm{pc}^{-1}$ found in the Cartesian X direction by \citet{Armstrong22}. However, our result for the orientation angle of maximum expansion in Pozzo 1 is offset from that of \citet{wright24} by $\sim37^\circ$, which is a difference of $>2\sigma$ significance.

For $\lambda$ Ori, our sample of cluster members (563) is also significantly larger than that of \citet{wright24}, even considering that we include Barnard 30 \& 35 with $\lambda$ Ori \citep{armstrong24} while \citet{wright24} separate them (144, 57 \& 57 members respectively for a total of 258). Again there is a notable difference (this time of $7\sigma$ significance) in the cluster expansion velocities; $\bar v_{out} = 0.711^{+0.021}_{-0.021} ~\textrm{km~s}^{-1}$ compared to $0.24^{+0.06}_{-0.01} ~\textrm{km~s}^{-1}$. But similarly to Pozzo 1, the maximum expansion rates are in better agreement; $0.144^{+0.003}_{-0.003} ~\textrm{km~s}^{-1}\textrm{pc}^{-1}$ compared to $0.2^{+0.03}_{-0.04} ~\textrm{km~s}^{-1}\textrm{pc}^{-1}$, again hinting that the maximum expansion rate is the more robust expansion indicator. Our result for the orientation angle of maximum expansion in $\lambda$ Ori is offset from that of \citet{wright24} by $\sim13^\circ$, which are thus in good agreement.

\subsection{Comparison to \citet{kuhn19}}
\label{KUhn19}

\citet{kuhn19} analysed the kinematics of 28 young clusters and associations observed as part of the MYStIX survey \citep{feigelson13}, which identified members on the basis of IR and X-ray criteria and cross-matched these to Gaia DR2 sources. They found evidence for expansion in a majority ($75\%$) of their sample.

Of our sample of clusters only NGC 2264 (S Mon) and IC 348 overlap with the sample of \citet{kuhn19}. Again, this is because clusters they examined either are not present in the \citet{cantat-gaudin20} catalog (e.g., Orion Nebula cluster), are more distant than the $1$kpc limit we impose (e.g., NGC 6530) or have too few RVs in \citet{tsantaki22} (e.g., NGC 1333).

For NGC 2264, our sample of cluster members (149) here is smaller than that of \citet{kuhn19} (242). However, the expansion velocities remain in excellent agreement; $\bar v_{out} = 0.333^{+0.06}_{-0.06} ~\textrm{km~s}^{-1}$ and $0.39^{+0.15}_{-0.15} ~\textrm{km~s}^{-1}$.

Also, for IC 348, our sample of cluster members (132) here is smaller than that of \citet{kuhn19} (180). However, the expansion velocities remain consistent within their relatively large uncertainties; $\bar v_{out} = -0.035^{+0.043}_{-0.043} ~\textrm{km~s}^{-1}$ and $0.16^{+0.18}_{-0.18} ~\textrm{km~s}^{-1}$.

As these comparisons between our results and those of \citet{wright24} and \citet{kuhn19} suggest, analyses using greatly differing cluster membership samples with different selection criteria and biases in many cases still produce good agreement for metrics of cluster expansion. We acknowledge the caveats that possible contamination and incompleteness of cluster membership may affect the significance of detected expansion signatures, but new Gaia-based cluster membership catalogues (such as \citealt{hunt23}) which generally expand the membership lists compared to the more conservative selection of \citet{cantat-gaudin20} will preferentially add new cluster members which are distributed sparsely and further from the cluster center and will be kinematically distinct from the cluster average. The inclusion of such members will likely increase the detectability of expansion signatures for future studies. 



\subsection{Cluster formation and dissolution}\label{Cluster_formation}

There is a growing body of evidence that young stellar clusters are not typically dense and monolithic, but rather form with significant spatial \citep[e.g.,][]{kuhn14,arnold24} and kinematic structure \citep[as traced by anisotropic velocity dispersions; e.g.,][]{wright24}. The spatial structure begins to be erased as the clusters evolve dynamically, which happens more quickly in the dense cores of clusters where the frequency of dynamical interactions is higher \citep{jaehnig15}. However, in many cases clusters can preserve their spatial structure outside their dense cores up to $>10$ Myr timescales (e.g., NGC 2451B, Collinder 359 and Gulliver 9; Table~\ref{kinematic_table}). This is likely aided by cluster expansion causing cluster members to move apart from each other and reducing the frequency of dynamical interactions.

Recent kinematic studies making use of precise astrometry from Gaia are now consistently showing that the majority of young clusters are expanding \citep{kuhn19,wright19,armstrong20,guilherme23,dellacroce24,wright24,cheshire25} and that in many cases this cluster expansion is anisotropic, i.e., clusters expand at greater rates in a preferred direction, as we have shown in this study. Clusters that exhibit rapid expansion ($\sigma$ Ori) are likely to dissolve quickly ($<10$ Myr), while clusters with slower expansion rates may survive as distinct populations for longer ($20 - 50$ Myr). 
For young clusters, the orientation of the maximum expansion rate is likely determined by the turbulent kinematics of their natal molecular cloud, but there is some evidence that this orientation moves closer to parallel with the Galactic plane over time (Fig.~\ref{age_exp_angles}).

\begin{figure} 
    {\includegraphics[width=245pt]{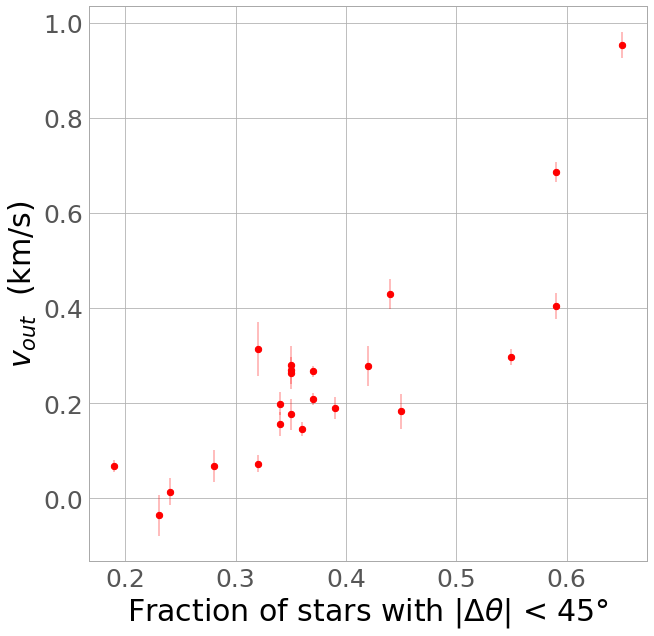} }%
    \setlength{\belowcaptionskip}{-10pt}
    \setlength{\textfloatsep}{0pt}
    \caption{Fraction of cluster members with $\Delta\theta<|45^\circ|$ against expansion velocity $\bar v_{out}$ for each cluster.} 
    \label{frac45_expv}%
\end{figure}


\begin{figure} 
    {\includegraphics[width=245pt]{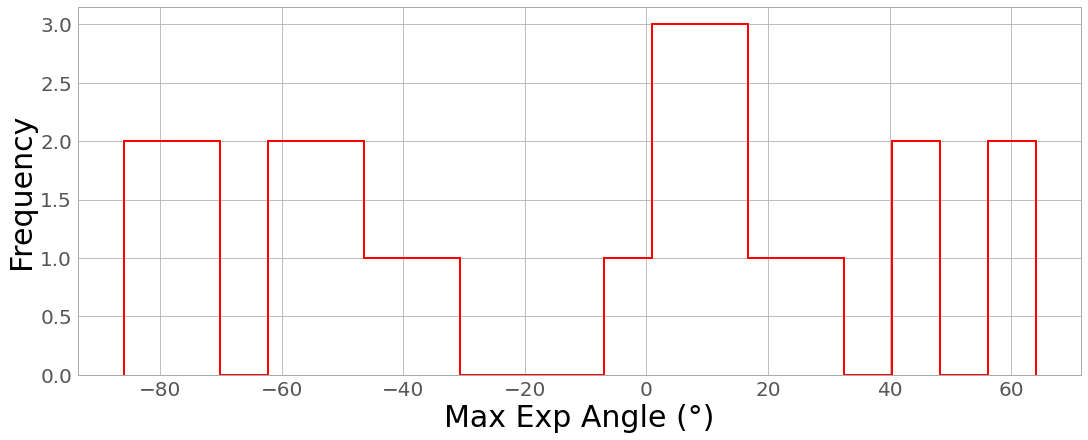} }%
    \setlength{\belowcaptionskip}{-10pt}
    \setlength{\textfloatsep}{0pt}
    \caption{Histogram of orientation angle of maximum expansion rate.} 
    \label{exp_angles}%
\end{figure}

\begin{figure} 
    {\includegraphics[width=245pt]{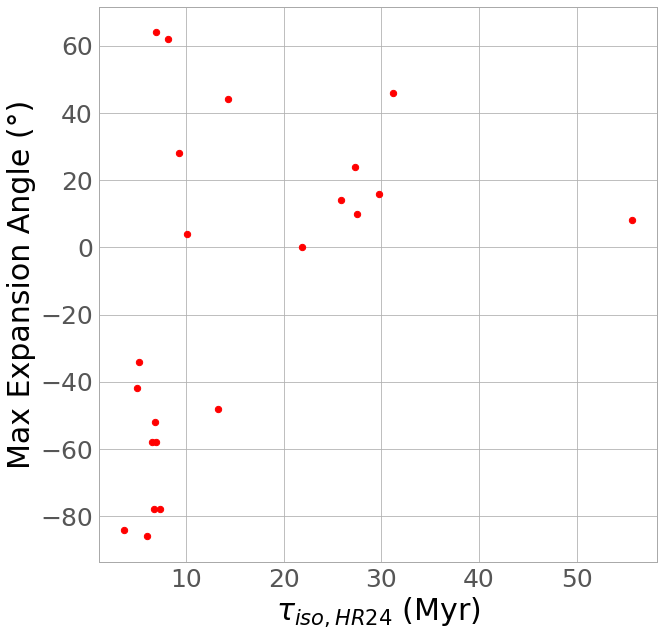} }%
    \setlength{\belowcaptionskip}{-10pt}
    \setlength{\textfloatsep}{0pt}
    \caption{Cluster ages from \protect\citet{hunt24} ($\tau_{iso,HR24}$) against the orientation angle of maximum expansion rate.} 
    \label{age_exp_angles}%
\end{figure}

\begin{table*}
\caption{Cluster property correlations}
\footnotesize
\begin{center}
{\renewcommand{\arraystretch}{1.5}
\begin{tabular}{|p{2.4cm}|p{1.0cm}|p{1.0cm}|p{1.0cm}|p{1.0cm}|p{1.0cm}|p{1.0cm}|p{1.0cm}|p{1.0cm}|p{1.0cm}|}
\hline
 & $R_c$ & Age$_{HR24}$ & $D$ & ADP & $\bar{v}_{out}$ & Max Exp Rate & Exp Asym & N members & 
 |Max Exp Ang| \\
\hline
$R_c$           & \cellcolor{black} &  &  &  &  &  &  &  &  \\
Age$_{HR24}$    & \cellcolor{red}0.52 (0.001) & \cellcolor{black} &  &  &  &  &  &  & \\
D               & -0.25 (0.091) & \cellcolor{red}-0.44 (0.004) & \cellcolor{black} &  &  &  &  &  &  \\
ADP             & -0.02 (0.916) & -0.07 (0.652) & \cellcolor{yellow}0.29 (0.057) & \cellcolor{black} &  &  &  &  & \\
$\bar{v}_{out}$ & 0.02 (0.895) & -0.20 (0.185) & 0.18 (0.224) & 0.16 (0.291) & \cellcolor{black}  &  &  &  &\\
Max Exp Rate    & \cellcolor{yellow}-0.29 (0.054) & \cellcolor{yellow}-0.38 (0.014) & 0.01 (0.958) & 0.02 (0.874) & \cellcolor{red}0.43 (0.005) & \cellcolor{black}  &  & & \\
Exp Asym         & \cellcolor{red}0.40 (0.008) & \cellcolor{red}0.51 (0.001) & \cellcolor{yellow}-0.31 (0.039) & 0.14 (0.342) & -0.03 (0.853) & -0.07 (0.653) & \cellcolor{black}  & & \\
N members  & 0.24 (0.107) & \cellcolor{yellow}0.31 (0.045) & -0.06 (0.711) & 0.06 (0.711) & 0.01 (0.979) & \cellcolor{yellow}-0.27 (0.077) & \cellcolor{red}0.39 (0.009) & \cellcolor{black} & \\
|Max Exp Ang|  & -0.15 (0.315) & \cellcolor{red}-0.48 (0.002) & 0.04 (0.771) & -0.12 (0.443) & 0.08 (0.579) & 0.21 (0.169) & \cellcolor{red}-0.43 (0.004) & \cellcolor{yellow}-0.36 (0.017) & 
\cellcolor{black} \\
\hline
\end{tabular}}
\end{center}
\setlength{\belowcaptionskip}{-10pt}
\setlength{\textfloatsep}{0pt}
\tablefoot{\footnotesize Kendall's $\tau$ coefficients with p-values in brackets for combinations of cluster parameters; core radius ($R_c$), isochronal age from \protect\citep{hunt24} (Age$_{HR24}$), cluster distance ($D$), mean ADP with 4 sectors, expansion velocity ($\bar v_{out}$), maximum expansion rate, expansion anisotropy, number of cluster members and orientation of the direction of maximum expansion.  Results indicating strong evidence for correlations are highlighted in red, and moderate evidence for correlations are highlighted in yellow. Negative values of $\tau$ indicate inverse correlations. }
\label{correlations_table}
\end{table*}

\section{Summary}
\label{summary}

\begin{itemize}
        \item We calculate cluster median RVs using the RV catalog of \citet{tsantaki22} and find that, in general, they are $\sim5$ km~s$^{-1}$ greater than cluster RVs from \citet{hunt24} using only RVs from Gaia. We attribute this to the unreliability of Gaia RV measurements for young stars.
        \item We estimate cluster distances using the bootstrapping approach of \citet{armstrong24} on the Gaia DR3 parallaxes of individual cluster members. We find that, in general, our cluster distances are larger than those of \citet{hunt24} by an amount that increases non-linearly with distance. We attribute this to the unsuitability of prior assumptions in the approach of \citet{hunt24} for nearby clusters where individual parallax uncertainties are smaller than the depth of the cluster.
        \item We calculate cluster core radii $R_c$ using the approach described in \citet{tarricq22}. We find a large range of core radii for clusters in our sample, from 0.34 pc to 4.35 pc, which in general increase with cluster age $\tau_{iso,HR24}$.
        \item We quantify cluster substructure using $Q$-Parameter and Angular Dispersion Parameter (ADP) methods. We find that most young clusters in our sample have smoother distributions towards their centers, while their outskirts retain significantly more clumpy substructure. This is in agreement with the findings of \citet{jaehnig15} who analysed clusters observed in the MYStIX survey, and concluded that the dense centers of clusters quickly become smooth due to the high rate of dynamical interactions between members erasing initial structure over short timescales, compared to the sparse outskirts where interactions between cluster members are less frequent.
        \item NGC 1977, IC 2395 and Collinder 359 all have mean $\delta_{\rm ADP,e,4}(r) > 2.5$, indicating that, overall, these clusters are dynamically less evolved and still retain much of their initial structure.
        \item  IC 2391, Alessi 20, ASCC 16, ASCC 19 and NGC 2547 have $\delta_{\rm ADP,e,4}(r) < 1.0$, which indicates that, overall, these clusters are smoothly distributed and are sufficiently dynamically evolved to have erased much of their initial structure. 
        \item We investigate structure (ADP) as a function of radial distance from the center of each cluster $R$. We find that, for the majority of clusters, the level of substructure generally increases with radial distance $R$ and the central sectors of the clusters are often the most smoothly distributed. This indicates that the dense cores of clusters experience a rapid dynamical evolution, erasing their initial structure on short timescales (a few Myr) whereas the sparse cluster outskirts or 'halos' remain dynamically unevolved for longer timescales (tens of Myr).
        \item The majority of nearby young clusters that we analyse show strong evidence for expansion, with the exceptions of NGC 2232, IC 348 \& NGC 1980, based on their expansion velocities $\bar v_{out}$ and expansion rates.
        \item The majority of clusters have expansion trends that are significantly anisotropic, 13 of at least the 5$\sigma$ level and another 6 of at least the 3$\sigma$ level.
        \item We calculate expansion timescales from each cluster's maximum rate of expansion $\tau_{max}$ and find many of them are in good agreement with isochronal ages $\tau_{iso}$ given in either the \citet{cantat-gaudin20} or \citet{hunt24} catalogs, mostly for clusters younger than $10$ Myr, but also in some cases $>20$ Myr, such as for NGC 2451B and Collinder 359.
        \item Many other clusters in our sample show significant evidence for expansion but have expansion timescales much lower than their isochronal ages $\tau_{max}<\tau_{iso}$. We consider that expansion in these clusters may not have begun immediately following residual gas expulsion in their early evolution, but either became unbound gradually as they evolved dynamically, or experienced catastrophic events such as encounters with molecular clouds which caused them to become unbound later in their evolution.
        \item NGC 2232 has an expansion timescale much greater than its isochronal age. A possible explanation for this is that NGC 2232 formed in a sparse configuration, much like an OB association. Since the expansion timescale assumes that expansion begins from a compact configuration, a cluster expanding from an initially sparse configuration would cause an expansion timescale to overestimate the age cluster age. However, NGC 2232 also shows some of the weakest evidence for expansion in our sample with regards to expansion velocities $\bar v_{out}$, $\Delta\theta$ and linear expansion rates (Sect.~\ref{ExpansionDiscussion}), so this timescale may not be a reliable age indicator for this cluster.
        \item Most clusters are consistent with having been more compact in the past according to our linear traceback calculations, and we calculate traceback ages $\tau_{TB}$ from when in the past clusters would have been at their most compact. However, these linear traceback calculations seem to reach an upper limit of $\sim10$ Myr for clusters in our sample despite many having ages derived from stellar evolution models $\tau_{iso}>10$ Myr.
        \item We find strong correlations between measures for expansion, such as the fraction of cluster members moving away from the center and the cluster expansion velocity $\bar v_{out}$ (e.g., Fig.~\ref{frac45_expv}), providing evidence that each of these measures are valid methods of identifying cluster expansion. 
        \item The orientations in which clusters have their maximum rate of expansion are consistent with being uniformly distributed (Fig.~\ref{exp_angles}), indicating that the anisotropy may be dependent on the turbulent motion inherited from the natal molecular cloud at formation. We also note a strong negative correlation with the orientation of maximum expansion and the age of the cluster (Fig.~\ref{age_exp_angles}), indicating that older clusters tend to expand preferentially parallel to the Galactic plane. This may be due to the influence of the Galactic potential which can eventually create tidal tails in clusters over $\sim100$ Myr timescales \citep{risbud25}.
\end{itemize}

In future work, we will make a detailed study of other kinematic properties for this cluster sample, including velocity dispersions and angular rotation signatures, and compare those to the results presented in this work.

The upcoming Gaia DR4 will provide improved precision on 5-parameter astrometry which will facilitate more precise measurements of cluster kinematic properties, such as expansion, and allow us to expand the cluster sample to greater distances. Gaia DR4 will also include a wealth of valuable auxiliary data, including non-single star solutions which will assist in identifying binary systems in clusters, which otherwise can confound kinematic signatures and bias age estimation.

Better understanding of the Gaia selection function and cluster selection function will enable modeling of cluster incompleteness, as well as more robust membership determination in future cluster catalogs, so that kinematically distinct cluster members are better retained in membership lists.

Further spectroscopic observations of the full populations of these clusters would provide spectroscopic youth confirmation, allowing membership determination independent of kinematics, and greater RV coverage, allowing full 3D kinematic analysis. Upcoming spectroscopic surveys from facilities such as WEAVE \citep{dalton18} and 4MOST \citep{dejong19} will observe thousands of members of young clusters and associations.

\begin{acknowledgements}
      J.J.A. acknowledges support from a Chalmers Initiative on Cosmic Origins (CICO) postdoctoral fellowship. J.C.T. acknowledges support from ERC Advanced Grant MSTAR (788829) and SNSA-R grant 2025-00239. This work has made use of data from the ESA space mission Gaia (http://www.cosmos.esa.int/gaia), processed by the Gaia Data Processing and Analysis Consortium (DPAC, http://www.cosmos.esa.int/web/gaia/dpac/consortium). Funding for DPAC has been provided by national institutions, in particular the institutions participating in the Gaia Multilateral Agreement. This research made use of the Simbad and Vizier catalog access tools (provided by CDS, Strasbourg, France), Astropy \citep{astr13} and TOPCAT \citep{tayl05}.
\end{acknowledgements}

\bibliographystyle{aa} 
\bibliography{references2}

\appendix

\begin{landscape}

\begin{table}
\caption{Kinematic properties for nearby young clusters}
\begin{centering}
{\renewcommand{\arraystretch}{1.5}

\tiny
\begin{supertabular}{|p{4.6cm}|p{1.5cm}|p{1.5cm}|p{1.5cm}|p{1.5cm}|p{1.5cm}|p{1.5cm}|p{1.5cm}|p{1.5cm}|p{1.5cm}|p{1.5cm}|}
\hline
Cluster Name & $\lambda$ Ori (AT24) & Platais 8 & $\sigma$ Ori & NGC 2451B & Melotte 20 & IC 2391 & Gulliver 6 & Pozzo 1 & IC 2602 & NGC 2232 \\
\hline
catalog Age $\tau_{iso}$ (Myr) & $12.6^{1}$, $5.0^{2}$, $5.2^{3}$ & $30.2^{1}$, $31.2^{3}$ & $3.7^{3}$ & $40.7^{1}$, $44.7^{2}$, $27.5^{3}$ & $53.1^{1}$, $55.6^{3}$ & $28.8^{1}$, $45.7^{2}$, $27.3^{3}$ & $16.6^{1}$, $6.5^{3}$ & $9.5^{1}$, $9.3^{3}$ & $36.3^{1}$, $25.9^{3}$ & $17.8^{1}$, $53.7^{2}$, $13.3^{3}$ \\
No. Members with RUWE $<$ 1.4 & 563 & 173 & 105 & 252 & 662 & 194 & 295 & 312 & 280 & 177 \\
Center coordinate l ($^\circ$) & $195.085^{+0.012}_{-0.023}$ & $277.367^{+0.722}_{-0.214}$ & $206.823^{+0.006}_{-0.010}$ & $252.407^{+0.062}_{-0.123}$ & $147.203^{+0.322}_{-0.100}$ & $270.331^{+0.157}_{-0.079}$ & $205.231^{+0.061}_{-0.053}$ & $262.832^{+0.008}_{-0.025}$ & $289.765^{+0.186}_{-0.223}$ & $214.531^{+0.022}_{-0.052}$ \\ 
Center coordinate b ($^\circ$) & $-11.978^{+0.006}_{-0.009}$ & $-7.835^{+0.211}_{-0.156}$ & $-17.338^{+0.010}_{-0.013}$ & $-6.817^{+0.048}_{-0.059}$ & $-6.493^{+0.19}_{-0.267}$ & $-6.881^{+0.26}_{-0.042}$ & $-18.265^{+0.032}_{-0.082}$ & $-7.691^{+0.047}_{-0.023}$ & $-4.888^{+0.195}_{-0.154}$ & $-7.369^{+0.026}_{-0.028}$ \\ 
Core radius $R_c$ (pc) & 0.42 & 3.25 & 0.44 & 1.38 & 3.44 & 1.4 & 1.51 & 0.6 & 1.44 & 0.78  \\
Cluster core velocity $\bar v_{l}$ (km~s$^{-1}$) & $3.904^{+0.076}_{-0.074}$ & $-10.516^{+0.022}_{-0.023}$ & $1.725^{+0.040}_{-0.040}$ & $-13.565^{+0.050}_{-0.053}$ & $21.864^{+0.025}_{-0.025}$ & $-19.713^{+0.033}_{-0.034}$ & $0.311^{+0.039}_{-0.038}$ & $-17.134^{+0.105}_{-0.102}$ & $-8.468^{+0.019}_{-0.016}$ & $-0.744^{+0.025}_{-0.032}$ \\ 
Cluster core velocity $\bar v_{b}$ (km~s$^{-1}$) & $-0.646^{+0.043}_{-0.047}$ & $2.489^{+0.011}_{-0.010}$ & $1.801^{+0.059}_{-0.063}$ & $-7.359^{+0.05}_{-0.043}$ & $-10.302^{+0.017}_{-0.017}$ & $1.7^{+0.008}_{-0.008}$ & $0.009^{+0.026}_{-0.025}$ & $2.13^{+0.041}_{-0.044}$ & $4.187^{+0.01}_{-0.011}$ & $-7.646^{+0.048}_{-0.046}$ \\ 
Q-Parameter & 0.806 & 0.839 & 0.832 & 0.895 & 0.939 & 0.904 & 0.872 & 0.883 & 0.776 & 0.889 \\ 
2D Ellipse $\theta_a$ ($^\circ$) & 24.3 & -11.4 & -69.9 & -1.6 & -38.3 & -23.7 & -20.9 & -51.7 & 7.3 & 8.2 \\ 
2D Ellipse e & 0.695 & 0.855 & 0.736 & 0.558 & 0.582 & 0.694 & 0.555 & 0.566 & 0.736 & 0.617 \\ 
Mean ADP 4 & 1.704 & 1.823 & 1.745 & 1.727 & 1.025 & 0.903 & 1.721 & 1.220 & 1.361 & 1.475 \\ 
No. RVs & 189 & 26 & 76 & 66 & 234 & 51 & 62 & 109 & 67 & 19  \\ 
Median RV (km~s$^{-1}$) & $27.47^{+0.08}_{-0.08}$ & $22.04^{+0.06}_{-0.07}$ & $30.83^{+0.07}_{-0.07}$ & $15.21^{+0.10}_{-0.10}$ & $0.41^{+0.09}_{-0.08}$ & $14.53^{+0.18}_{-0.17}$ & $30.27^{+0.14}_{-0.14}$ & $17.20^{+0.06}_{-0.06}$ & $17.36^{+0.15}_{-0.16}$ & $25.96^{+1.15}_{-1.1}$ \\ 
Median LSR RV (km~s$^{-1}$) & $12.37^{+0.08}_{-0.08}$ & $10.44^{+0.06}_{-0.07}$ & $13.94^{+0.07}_{-0.07}$ & $-0.57^{+0.10}_{-0.10}$ & $-3.09^{+0.09}_{-0.08}$ & $1.57^{+0.18}_{-0.17}$ & $13.51^{+0.14}_{-0.14}$ & $2.82^{+0.06}_{-0.06}$ & $9.00^{+0.15}_{-0.16}$ & $9.08^{+1.15}_{-1.1}$ \\ 
Median Distance (pc) & $393.96^{+0.61}_{-0.64}$ & $134.22^{+0.34}_{-0.34}$ & $396.087^{+0.820}_{-0.848}$ & $358.929^{+0.427}_{-0.424}$ & $173.092^{+0.161}_{-0.158}$ & $150.398^{+0.103}_{-0.102}$ & $410.294^{+0.505}_{-0.512}$ & $339.623^{+0.654}_{-0.649}$ & $150.513^{+0.154}_{-0.153}$ & $317.067^{+0.863}_{-0.829}$  \\ 
Distance $BS$ (pc) & $392.71^{+0.88}_{-0.87}$ & $134.08^{+0.44}_{-0.47}$ & $396.14^{+1.31}_{-1.32}$ & $358.29^{+0.78}_{-0.77}$ & $173.00^{+0.26}_{-0.26}$ & $150.38^{+0.24}_{-0.25}$ & $409.00^{+1.2}_{-1.21}$ & $338.97^{+0.88}_{-0.86}$ & $150.48^{+0.26}_{-0.25}$ & $316.73^{+0.89}_{-0.93}$  \\ 
Fraction of members with $|\Delta\theta|<45^\circ$ & 0.59 & 0.44 & 0.65 & 0.19 & 0.35 & 0.36 & 0.37 & 0.34 & 0.37 & 0.24 \\ 
Fraction of members with $|\Delta\theta|<90^\circ$ & 0.74 & 0.55 & 0.80 & 0.40 & 0.54 & 0.59 & 0.62 & 0.58 & 0.56 & 0.42 \\ 
$\tau_{1D,median}$ (Myr) & $8.52^{+0.21}_{-0.18}$ & $12.91^{+0.89}_{-0.70}$ & $1.65^{+0.09}_{-0.08}$ & $63.21^{+11.16}_{-7.83}$ & $18.31^{+0.47}_{-0.44}$ & $15.92^{+0.99}_{-0.93}$ & $14.86^{+0.55}_{-0.50}$ & $21.80^{+2.29}_{-1.85}$ & $10.71^{+0.32}_{-0.30}$ & $99.84^{+195.47}_{-50.96}$ \\ 
$\tau_{1D,median}$ with $|\Delta\theta|<45^\circ$ (Myr) & $7.20^{+0.15}_{-0.14}$ & $6.78^{+0.36}_{-0.30}$ & $1.57^{+0.03}_{-0.03}$ & $14.24^{+0.86}_{-0.76}$ & $9.68^{+0.36}_{-0.35}$ & $7.76^{+0.34}_{-0.31}$ & $6.06^{+0.12}_{-0.11}$ & $12.26^{+1.55}_{-1.24}$ & $6.58^{+0.17}_{-0.16}$ & $19.28^{+3.91}_{-2.87}$ \\ 
Expansion Velocity ($v_{\rm out}$ km~s$^{-1}$) & $0.686^{+0.02}_{-0.02}$ & $0.43^{+0.032}_{-0.032}$ & $0.953^{+0.028}_{-0.028}$ & $0.068^{+0.012}_{-0.012}$ & $0.271^{+0.016}_{-0.016}$ & $0.146^{+0.015}_{-0.015}$ & $0.209^{+0.013}_{-0.013}$ & $0.156^{+0.025}_{-0.025}$ & $0.267^{+0.012}_{-0.012}$ & $0.014^{+0.028}_{-0.028}$  \\ 
1D Expansion Rate (km~s$^{-1}$pc$^{-1}$) & $0.181^{+0.030}_{-0.030}$ & $0.037^{+0.007}_{-0.007}$ & $0.175^{+0.043}_{-0.041}$ & $0.003^{+0.003}_{-0.003}$ & $0.018^{+0.004}_{-0.018}$ & $0.005^{+0.007}_{-0.005}$ & $0.000^{+0.003}_{-0.003}$ & $0.105^{+0.036}_{-0.035}$ & $-0.001^{+0.001}_{-0.002}$ & $-0.001^{+0.001}_{-0.002}$ \\ 
$\tau_{1D,linear}$ (Myr) & $5.64^{+1.11}_{-0.80}$ & $27.65^{+6.45}_{-4.40}$ & $5.85^{+1.79}_{-1.15}$ & - & - & - & - & $9.74^{+4.87}_{-2.49}$ & - & - \\ 
Max Expansion Rate (km~s$^{-1}$pc$^{-1}$) & $0.144^{+0.003}_{-0.003}$ & $0.199^{+0.006}_{-0.006}$ & $0.604^{+0.030}_{-0.030}$ & $0.037^{+0.004}_{-0.004}$ & $0.107^{+0.002}_{-0.002}$ & $0.13^{+0.008}_{-0.008}$ & $0.085^{+0.008}_{-0.008}$ & $0.077^{+0.005}_{-0.005}$ & $0.175^{+0.003}_{-0.003}$ & $0.024^{+0.007}_{-0.007}$  \\ 
Max Expansion Angle ($^\circ$) & $-34$ & $46$ & $-84$ & $10$ & $8$ & $24$ & $-58$ & $28$ & $14$ & $-48$  \\ 
$\tau_{max}$ (Myr) & $6.944^{+0.148}_{-0.142}$ & $5.025^{+0.156}_{-0.147}$ & $1.626^{+0.086}_{-0.078}$ & $27.027^{+3.276}_{-2.637}$ & $9.346^{+0.178}_{-0.171}$ & $7.692^{+0.504}_{-0.446}$ & $11.765^{+1.222}_{-1.012}$ & $12.987^{+0.902}_{-0.792}$ & $5.714^{+0.100}_{-0.096}$ & $41.667^{+17.157}_{-9.409}$ \\ 
Min Expansion Rate (km~s$^{-1}$pc$^{-1}$) & $0.119^{+0.004}_{-0.004}$ & $-0.034^{+0.005}_{-0.005}$ & $0.431^{+0.047}_{-0.047}$ & $-0.049^{+0.008}_{-0.008}$ & $-0.046^{+0.003}_{-0.003}$ & $-0.010^{+0.011}_{-0.011}$ & $0.001^{+0.014}_{-0.014}$ & $0.025^{+0.004}_{-0.004}$ & $-0.029^{+0.006}_{-0.007}$ & $0.001^{+0.007}_{-0.007}$ \\ 
Min Expansion Angle ($^\circ$) & $80$ & $-88$ & $12$ & $-88$ & $-88$ & $-70$ & $20$ & $-74$ & $86$ & $58$  \\ 
$\tau_{min}$ (Myr) & $8.403^{+0.292}_{-0.273}$ & - & $2.32^{+0.284}_{-0.228}$ & - & - & - & - & $40.000^{+7.619}_{-5.517}$ & - & - \\ 
Expansion Asymmetry ($\sigma$) & 5.0 & 29.8 & 3.1 & 9.6 & 42.4 & 10.3 & 5.3 & 8.1 & 30.4 & 2.3  \\ 
$\tau_{TB,0}$ (Myr) & $4.1^{+0.1}_{-0.1}$ & $5.0^{+0.1}_{-0.1}$ & $1.1^{+0.1}_{-0.1}$ & $0.4^{+0.1}_{-0.1}$ & $3.3^{+0.1}_{-0.1}$ & $1.0^{+0.1}_{-0.1}$ & $0.9^{+0.1}_{-0.1}$ & $1.5^{+0.1}_{-0.1}$ & $3.3^{+0.1}_{-0.1}$ & $0.2^{+0.1}_{-0.2}$ \\
$\tau_{TB}$ (Myr) & $4.4^{+0.1}_{-0.1}$ & $6.2^{+0.1}_{-0.1}$ & $1.2^{+0.1}_{-0.1}$ & $1.6^{+0.1}_{-0.1}$ & $5.5^{+0.1}_{-0.1}$ & $2.0^{+0.1}_{-0.1}$ & $1.2^{+0.1}_{-0.1}$ & $10.0^{+0.1}_{-0.1}$ & $4.1^{+0.1}_{-0.1}$ & $6.0^{+0.1}_{-0.1}$ \\
\hline

\end{supertabular}}
\tablefoot{Kinematic properties for nearby young clusters with cluster members identified by \protect\citet{cantat-gaudin20}. See the text for a discussion of how these quantities were derived. Ages are from $^{1}$ \protect\citet{cantat-gaudin20}, $^{2}$ \protect\citet{dias21}, $^{3}$ \protect\citet{hunt24}.  }
\label{kinematic_table}
\end{centering}
\end{table}

\end{landscape}

\begin{landscape}
\begin{table}
\begin{centering}
{\renewcommand{\arraystretch}{1.5}
\tiny
\begin{supertabular}{|p{4.6cm}|p{1.4cm}|p{1.4cm}|p{1.4cm}|p{1.4cm}|p{1.4cm}|p{1.4cm}|p{1.4cm}|p{1.4cm}|p{1.4cm}|p{1.4cm}|}
\hline
Cluster Name & NGC 1977 & NGC 2264 & IC 2395 & Alessi 20 & Collinder 359 & Gulliver 9 & ASCC 16 & ASCC 21 & IC 348 \\
\hline
catalog Age $\tau_{iso}$ (Myr) & $97.7^{1}$, $12.0^{2}$ & $27.5^{1}$, $8.9^{2}$, $5.0^{3}$ & $20.4^{1}$, $6.3^{2}$, $6.8^{3}$ & $9.3^{1}$, $3.0^{2}$, $6.7^{3}$ & $37.2^{1}$, $14.3^{3}$ & $17.8^{1}$, $10.1^{3}$ & $13.5^{1}$, $8.2^{3}$ & $8.9^{1}$, $12.9^{2}$, $6.9^{3}$ & $11.7^{1}$, $43.7^{2}$, $6.0^{3}$ \\
No. Members with RUWE $<$ 1.4 & 102 & 149 & 265 & 111 & 252 & 231 & 160 & 82 & 132 \\
Center coordinate l ($^\circ$) & $208.499^{+0.019}_{-0.03}$ & $202.958^{+0.010}_{-0.016}$ & $266.664^{+0.022}_{-0.025}$ & $117.643^{+0.008}_{-0.005}$ & $29.932^{+0.085}_{-0.094}$ & $265.129^{+0.048}_{-0.043}$ & $201.065^{+0.039}_{-0.069}$ & $199.869^{+0.040}_{-0.090}$ & $160.504^{+0.015}_{-0.044}$ \\ 
Center coordinate b ($^\circ$) & $-19.080^{+0.024}_{-0.014}$ & $2.2^{+0.008}_{-0.037}$ & $-3.588^{+0.034}_{-0.022}$ & $-3.713^{+0.014}_{-0.018}$ & $12.623^{+0.084}_{-0.106}$ & $-5.282^{+0.015}_{-0.030}$ & $-18.368^{+0.084}_{-0.121}$ & $-16.650^{+0.043}_{-0.066}$ & $-17.814^{+0.012}_{-0.007}$ \\ 
Core radius $R_c$ (pc) & 0.66 & 0.56 & 1.1 & 0.49 & 4.17 & 1.29 & 1.3 & 2.34 & 0.34 \\
Cluster core velocity $\bar v_{l}$ (km~s$^{-1}$) & $2.693^{+0.034}_{-0.034}$ & $8.51^{+0.106}_{-0.103}$ & $-14.553^{+0.088}_{-0.086}$ & $7.666^{+0.055}_{-0.055}$ & $-19.602^{+0.125}_{-0.125}$ & $-18.435^{+0.136}_{-0.112}$ & $1.095^{+0.027}_{-0.027}$ & $2.018^{+0.023}_{-0.025}$ & $10.593^{+0.089}_{-0.09}$ \\ 
Cluster core velocity $\bar v_{b}$ (km~s$^{-1}$) & $1.401^{+0.032}_{-0.031}$ & $-10.307^{+0.128}_{-0.123}$ & $-0.879^{+0.027}_{-0.029}$ & $-5.965^{+0.053}_{-0.054}$ & $-11.928^{+0.092}_{-0.092}$ & $1.887^{+0.034}_{-0.028}$ & $2.072^{+0.041}_{-0.045}$ & $1.452^{+0.025}_{-0.028}$ & $-3.659^{+0.05}_{-0.06}$ \\ 
Q-Parameter  & 0.823 & 0.844 & 0.839 & 0.803 & 0.855 & 0.839 & 0.837 & 0.830 & 0.877 \\ 
2D Ellipse $\theta_a$ ($^\circ$) & -16.8 & -59.3 & 1.4 & 35.9 & -48.2 & -34.2 & -38.9 & 9.8 & -4.1 \\ 
2D Ellipse e  & 0.797 & 0.348 & 0.353 & 0.624 & 0.521 & 0.172 & 0.712 & 0.693 & 0.371 \\ 
Mean ADP 4 & 2.707 & 1.675 & 2.715 & 0.922 & 2.512 & 2.274 & 0.837 & 1.112 & 1.296 \\ 
No. RVs & 75 & 114 & 7 & 18 & 29 & 16 & 52 & 15 & 118 \\ 
Median RV (km~s$^{-1}$) & $30.75^{+0.08}_{-0.08}$ & $21.68^{+0.09}_{-0.09}$ & $20.76^{+3.69}_{-3.19}$ & $-5.7^{+0.85}_{-0.86}$ & $-4.69^{+1.35}_{-1.04}$ & $23.55^{+2.23}_{-1.57}$ & $20.91^{+0.11}_{-0.11}$ & $19.84^{+0.25}_{-0.2}$ & $15.51^{+0.08}_{-0.07}$ \\ 
Median LSR RV (km~s$^{-1}$) & $13.64^{+0.08}_{-0.08}$ & $6.98^{+0.09}_{-0.09}$ & $7.46^{+3.69}_{-3.19}$ & $-0.49^{+0.85}_{-0.86}$ & $9.8^{+2.23}_{-1.57}$ & $4.62^{+0.11}_{-0.11}$ & $3.77^{+0.25}_{-0.2}$ & $7.22^{+0.08}_{-0.07}$ &  \\ 
Median Distance (pc) & $387.295^{+0.461}_{-0.463}$ & $703.228^{+2.426}_{-2.423}$ & $701.018^{+3.772}_{-3.825}$ & $421.392^{+0.669}_{-0.673}$ & $543.757^{+0.917}_{-0.931}$ & $490.200^{+1.333}_{-1.353}$ & $342.405^{+0.377}_{-0.370}$ & $339.455^{+2.398}_{-2.332}$ & $312.605^{+0.921}_{-0.937}$ \\ 
Distance $BS$ (pc) & $386.98^{+1.3}_{-1.4}$ & $699.78^{+4.35}_{-4.3}$ & $695.77^{+2.53}_{-2.39}$ & $421.67^{+1.6}_{-1.64}$ & $542.19^{+2.22}_{-2.23}$ & $489.61^{+1.77}_{-1.81}$ & $342.48^{+0.88}_{-0.93}$ & $338.91^{+1.32}_{-1.37}$ & $311.67^{+1.56}_{-1.63}$ \\ 
Fraction of members with $|\Delta\theta|<45^\circ$ & 0.35 & 0.32 & 0.35 & 0.34 & 0.35 & 0.42 & 0.55 & 0.45 & 0.23 \\ 
Fraction of members with $|\Delta\theta|<90^\circ$ & 0.55 & 0.56 & 0.54 & 0.6 & 0.59 & 0.58 & 0.75 & 0.66 & 0.39 \\ 
$\tau_{1D,median}$ (Myr) & $3.84^{+0.39}_{-0.33}$ & $3.45^{+0.76}_{-0.51}$ & $8.96^{+1.02}_{-0.79}$ & $8.54^{+0.59}_{-0.50}$ & $105.48^{+14.89}_{-12.55}$ & $28.70^{+3.35}_{-2.76}$ & $7.95^{+0.28}_{-0.25}$ & $13.36^{+3.80}_{-2.58}$ & $-14.51^{+48.93}_{-28.14}$  \\ 
$\tau_{1D,median}$ with $|\Delta\theta|<45^\circ$ (Myr) & $1.30^{+0.08}_{-0.06}$ & $1.35^{+0.13}_{-0.12}$ & $4.34^{+0.34}_{-0.29}$ & $4.21^{+0.21}_{-0.17}$ & $25.56^{+1.22}_{-1.17}$ & $19.42^{+1.92}_{-1.71}$ & $5.66^{+0.18}_{-0.15}$ & $7.22^{+1.42}_{-1.11}$ & $1.26^{+0.13}_{-0.11}$  \\
Expansion Velocity (km~s$^{-1}$) & $0.264^{+0.034}_{-0.034}$ & $0.314^{+0.057}_{-0.057}$ & $0.281^{+0.04}_{-0.041}$ & $0.199^{+0.024}_{-0.024}$ & $0.177^{+0.033}_{-0.033}$ & $0.278^{+0.042}_{-0.042}$ & $0.298^{+0.017}_{-0.017}$ & $0.183^{+0.037}_{-0.038}$ & $-0.035^{+0.043}_{-0.043}$ \\ 
1D Expansion Rate (km~s$^{-1}$pc$^{-1}$) & $-0.308^{+0.145}_{-0.156}$ & $-0.014^{+0.037}_{-0.040}$ & $0.017^{+0.009}_{-0.009}$ & $0.006^{+0.011}_{-0.006}$ & $0.002^{+0.002}_{-0.001}$ & $0.060^{+0.011}_{-0.011}$ & $0.035^{+0.012}_{-0.011}$ & $0.035^{+0.016}_{-0.014}$ & $-0.012^{+0.017}_{-0.028}$  \\ 
$\tau_{1D,linear}$ (Myr) & - & - & $60.18^{+67.70}_{-20.83}$ & - & - & $17.05^{+3.83}_{-2.64}$ & $29.23^{+13.40}_{-7.46}$ & $29.23^{+19.49}_{-9.17}$ & - \\ 
Max Expansion Rate (km~s$^{-1}$pc$^{-1}$) & $0.401^{+0.050}_{-0.050}$ & $0.316^{+0.068}_{-0.068}$ & $0.14^{+0.016}_{-0.016}$ & $0.158^{+0.027}_{-0.027}$ & $0.026^{+0.003}_{-0.003}$ & $0.062^{+0.005}_{-0.005}$ & $0.135^{+0.010}_{-0.010}$ & $0.111^{+0.013}_{-0.013}$ & $0.334^{+0.108}_{-0.111}$ \\ 
Max Expansion Angle ($^\circ$) & $6$ & $-42$ & $-52$ & $-78$ & $44$ & $4$ & $62$ & $64$ & $-86$ \\ 
$\tau_{max}$ (Myr) & $2.494^{+0.355}_{-0.276}$ & $3.165^{+0.868}_{-0.56}$ & $7.143^{+0.922}_{-0.733}$ & $6.329^{+1.304}_{-0.924}$ & $38.461^{+5.017}_{-3.979}$ & $16.129^{+1.415}_{-1.203}$ & $7.407^{+0.593}_{-0.51}$ & $9.009^{+1.195}_{-0.944}$ & $2.994^{+1.49}_{-0.732}$  \\ 
Min Expansion Rate (km~s$^{-1}$pc$^{-1}$) & $-0.117^{+0.062}_{-0.062}$ & $0.04^{+0.061}_{-0.059}$ & $0.056^{+0.01}_{-0.01}$ & $0.082^{+0.017}_{-0.017}$ & $-0.001^{+0.002}_{-0.002}$ & $0.026^{+0.002}_{-0.002}$ & $0.08^{+0.009}_{-0.009}$ & $0.016^{+0.013}_{-0.013}$ & $-0.187^{+0.086}_{-0.086}$ \\ 
Min Expansion Angle ($^\circ$) & $-76$ & $46$ & $68$ & $6$ & $-34$ & $-88$ & $-14$ & $-42$ & $10$ \\ 
$\tau_{min}$ (Myr) & - & - & $17.857^{+3.882}_{-2.706}$ & $12.195^{+3.189}_{-2.094}$ & - & $38.462^{+3.205}_{-2.747}$ & $12.5^{+1.585}_{-1.264}$ & - & - \\ 
Expansion Asymmetry ($\sigma$) & 6.5 & 3.0 & 4.5 & 2.3 & 7.5 & 6.7 & 4.1 & 5.2 & 3.7  \\ 
$\tau_{TB,0}$ (Myr) & $0.3^{+0.1}_{-0.1}$ & $0.2^{+0.1}_{-0.1}$ & $0.7^{+0.1}_{-0.1}$ & $0.6^{+0.1}_{-0.1}$ & $3.0^{+0.1}_{-0.1}$ & $2.9^{+0.1}_{-0.1}$ & $2.5^{+0.1}_{-0.1}$ & $2.0^{+0.1}_{-0.1}$ & $0.0^{+0.1}_{-0.1}$  \\
$\tau_{TB}$ (Myr) & $0.5^{+0.1}_{-0.1}$ & $0.4^{+0.1}_{-0.1}$ & $7.0^{+0.1}_{-0.1}$ & $1.2^{+0.1}_{-0.1}$ & $6.9^{+0.1}_{-0.1}$ & $9.1^{+0.1}_{-0.1}$ & $3.5^{+0.1}_{-0.1}$ & $2.6^{+0.1}_{-0.1}$ & $0.1^{+0.1}_{-0.1}$ \\
\hline

\end{supertabular}}
\tablefoot{Continuation of Table~\ref{kinematic_table}.}
\end{centering}
\end{table}
\end{landscape}

\begin{landscape}
\begin{table}
\begin{centering}
{\renewcommand{\arraystretch}{1.5}
\tiny
\begin{supertabular}{|p{4.6cm}|p{1.4cm}|p{1.4cm}|p{1.4cm}|p{1.5cm}|}
\hline
Cluster Name & ASCC 19 & NGC 2547 & Collinder 135 & NGC 1980 \\
\hline
catalog Age $\tau_{iso}$ (Myr) & $10.5^{1}$, $43.7^{2}$, $7.4^{3}$ & $32.4^{1}$, $38.0^{2}$, $21.9^{3}$ & $26.3^{1}$, $25.7^{2}$, $29.8^{3}$ & $13.2^{1}$, $4.7^{2}$, $6.9^{3}$ \\
No. Members with RUWE $<$ 1.4 & 138 & 198 & 290 & 113 \\
Center coordinate l ($^\circ$) & $204.87^{+0.099}_{-0.221}$ & $264.472^{+0.035}_{-0.04}$ & $248.826^{+0.167}_{-0.056}$ & $209.5^{+0.039}_{-0.017}$ \\
Center coordinate b ($^\circ$) & $-19.431^{+0.076}_{-0.058}$ & $-8.6^{+0.072}_{-0.023}$ & $-10.993^{+0.057}_{-0.135}$ & $-19.596^{+0.013}_{-0.02}$ \\
Core radius $R_c$ (pc) & 2.08 & 1.07 & 2.38 & 0.86 \\
Cluster core velocity $\bar v_{l}$ (km~s$^{-1}$) & $2.58^{+0.022}_{-0.023}$ & $-11.906^{+0.05}_{-0.049}$ & $-12.886^{+0.035}_{-0.035}$ & $-0.005^{+0.021}_{-0.021}$ \\
Cluster core velocity $\bar v_{b}$ (km~s$^{-1}$) & $0.56^{+0.021}_{-0.020}$ & $-4.503^{+0.018}_{-0.017}$ & $-6.496^{+0.025}_{-0.027}$ & $2.389^{+0.031}_{-0.03}$ \\
Q-Parameter & 0.854 & 0.918 & 0.939 & 0.785 \\
2D Ellipse $\theta_a$ ($^\circ$) & 38.0 & -7.8 & -11.9 & -35.7 \\
2D Ellipse e & 0.509 & 0.486 & 0.663 & 0.827 \\
Mean ADP 4 & 0.507 & 0.801 & 1.510 & 1.385 \\
No. RVs & 40 & 85 & 46 & 62 \\
Median RV (km~s$^{-1}$) & $22.52^{+0.1}_{-0.13}$ & $12.67^{+0.08}_{-0.08}$ & $16.25^{+0.72}_{-0.52}$ & $26.67^{+0.09}_{-0.09}$ \\ 
Median LSR RV (km~s$^{-1}$) & $5.76^{+0.1}_{-0.13}$ & $-1.52^{+0.08}_{-0.08}$ & $-0.27^{+0.72}_{-0.52}$ & $9.46^{+0.09}_{-0.09}$ \\ 
Median Distance (pc) & $350.484^{+1.587}_{-1.578}$ & $381.785^{+0.234}_{-0.234}$ & $296.92^{+0.74}_{-0.742}$ & $376.526^{+0.687}_{-0.702}$ \\
Distance $BS$ (pc) & $350.48^{+1.33}_{-1.39}$ & $381.94^{+0.76}_{-0.76}$ & $295.8^{+0.77}_{-0.73}$ & $376.69^{+1.07}_{-1.08}$  \\ 
Fraction of members with $|\Delta\theta|<45^\circ$ & 0.59 & 0.28 & 0.39 & 0.32 \\
Fraction of members with $|\Delta\theta|<90^\circ$ & 0.79 & 0.47 & 0.59 & 0.54 \\
$\tau_{1D,median}$ (Myr) & $9.95^{+0.69}_{-0.61}$ & $25.31^{+10.56}_{-5.62}$ & $30.98^{+1.90}_{-1.83}$ & $16.46^{+3.69}_{-2.57}$ \\
$\tau_{1D,median}$ with $|\Delta\theta|<45^\circ$ (Myr) & $7.72^{+0.48}_{-0.41}$ & $5.97^{+0.72}_{-0.54}$ & $16.79^{+0.81}_{-0.72}$ & $2.92^{+0.13}_{-0.13}$ \\
Expansion Velocity (km~s$^{-1}$) & $0.404^{+0.028}_{-0.027}$ & $0.068^{+0.034}_{-0.034}$ & $0.191^{+0.023}_{-0.023}$ & $0.073^{+0.018}_{-0.018}$ \\
1D Expansion Rate (km~s$^{-1}$pc$^{-1}$) & $0.132^{+0.05}_{-0.048}$ & $-0.006^{+0.005}_{-0.005}$ & $0.04^{+0.015}_{-0.015}$ & $-0.019^{+0.017}_{-0.014}$ \\
$\tau_{1D,linear}$ (Myr) & $7.75^{+4.43}_{-2.13}$ & - & $25.58^{+15.35}_{-6.98}$ & - \\
Max Expansion Rate (km~s$^{-1}$pc$^{-1}$) & $0.109^{+0.005}_{-0.005}$ & $0.062^{+0.012}_{-0.012}$ & $0.041^{+0.003}_{-0.003}$ & $0.059^{+0.02}_{-0.02}$ \\
Max Expansion Angle ($^\circ$) & $-78$ & $0$ & $16$ & $-58$ \\
$\tau_{max}$ (Myr) & $9.174^{+0.441}_{-0.402}$ & $16.129^{+3.871}_{-2.616}$ & $24.39^{+1.926}_{-1.663}$ & $16.949^{+8.692}_{-4.291}$ \\
Min Expansion Rate (km~s$^{-1}$pc$^{-1}$) & $0.093^{+0.006}_{-0.006}$ & $-0.03^{+0.012}_{-0.012}$ & $0.006^{+0.003}_{-0.004}$ & $0.041^{+0.029}_{-0.029}$ \\
Min Expansion Angle ($^\circ$) & $4$ & $84$ & $-88$ & $44$ \\
$\tau_{min}$ (Myr) & $10.753^{+0.742}_{-0.652}$ & - & - & $24.39^{+58.943}_{-10.105}$ \\
Expansion Asymmetry ($\sigma$) & 2.0 & 3.8 & 8.2 & 0.5 \\
$\tau_{TB,0}$ (Myr) & $5.0^{+0.1}_{-0.1}$ & $0.3^{+0.1}_{-0.1}$ & $2.6^{+0.1}_{-0.1}$ & $0.5^{+0.1}_{-0.1}$ \\
$\tau_{TB}$ (Myr) & $7.1^{+0.1}_{-0.1}$  & $3.1^{+0.1}_{-0.1}$ & $8.6^{+0.1}_{-0.1}$ & $0.9^{+0.1}_{-0.1}$ \\

\hline

\end{supertabular}}
\tablefoot{Continuation of Table~\ref{kinematic_table}.}
\end{centering}
\end{table}
\end{landscape}

\section{3D dynamical traceback}\label{3Dtraceback}

\begin{figure*}
	\includegraphics[width=\textwidth]{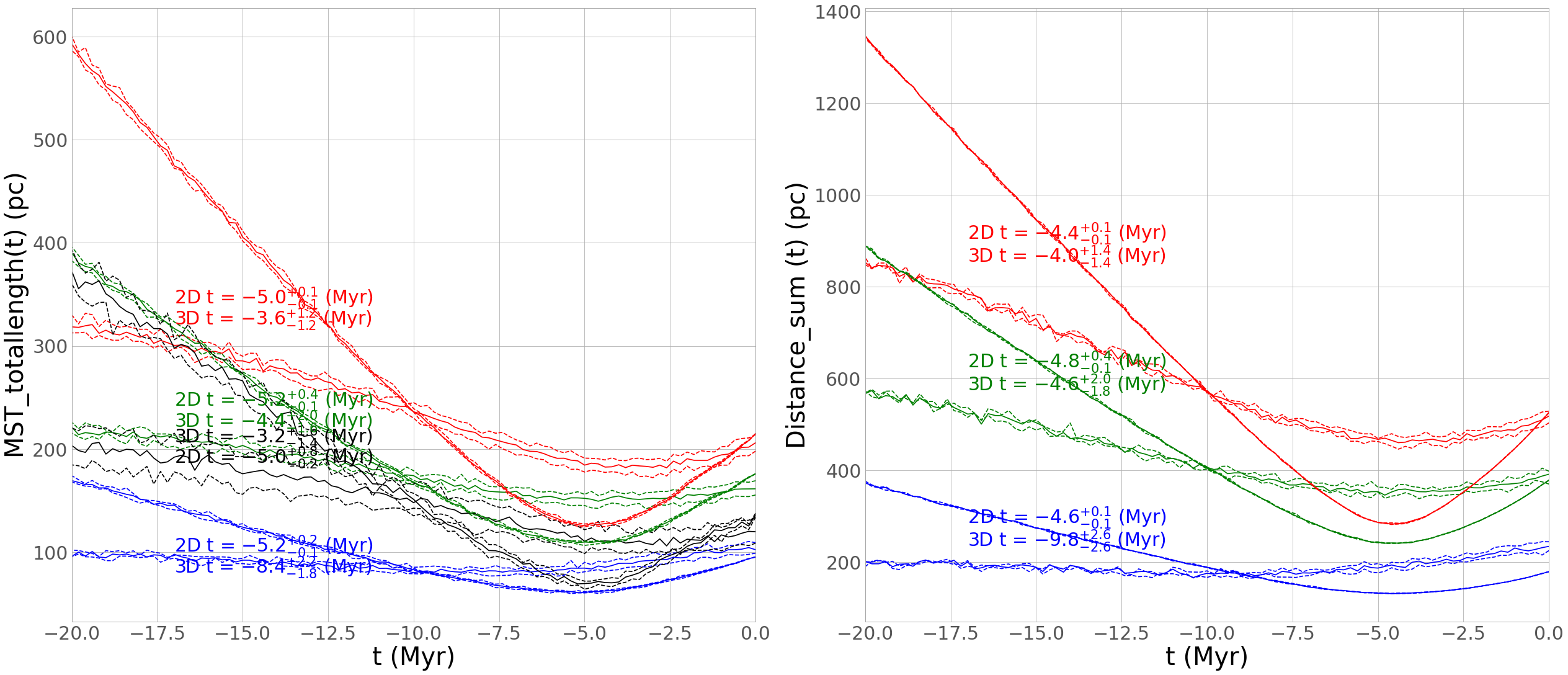}
	\setlength{\belowcaptionskip}{-10pt}
	\setlength{\textfloatsep}{0pt}
	\caption{2D and 3D cluster size traceback for $\lambda$ Ori cluster members, with precise 3D velocities, using the correction for estimated size inflation due to error propagation. \textit{Left:} Minimum spanning-tree total length as a function of trace-back time with no filter for outliers (red), 3$\sigma$ velocity outliers removed (green), 2$\sigma$ velocity outliers removed (blue) and 32\% longest branches removed (black) with their respective uncertainties. \textit{Right:} Sum of distances for each star to the association subgroup center as a function of trace-back time with no filter for outliers (red), 3$\sigma$ velocity outliers removed (green), 2$\sigma$ velocity outliers removed (blue). 3D traceback results show much larger uncertainties, due to the inclusion of parallax and RV uncertainties in the calculations, but generally the times of minimum cluster size are in good agreement.   }
	\label{MinAreaPlot_3D}
\end{figure*}

For clusters with enough members with precise 3D position and velocity information, we transform 5-parameter astrometry and RVs into 3D Heliocentric Cartesian coordinates X, Y, Z and velocities U, V, W, with their respective uncertainties, and perform 3D dynamical traceback estimation using orbital equations as described in \citet{Armstrong22,Armstrong25}, for comparison with our plane-of-sky traceback (Sect.~\ref{expansion timescale}). As errors will propagate with increasing timesteps and inflate cluster size metrics, we limit the cluster members used in this analysis to those with $\sigma_{RV}< 0.5~\textrm{km~s}^{-1}$, $flag_{rv}<2$ and $flag_{binary}~!= binary$ from \citet{tsantaki22}, as well as $\sigma_X,\sigma_Y<8~\textrm{pc}$, and incorporate the size metric correction as described in Sect.~\ref{expansion timescale}.

For $\lambda$ Ori, we have 67 cluster members that meet these criteria. We show the results of the 3D traceback calculations for these in Fig.~\ref{MinAreaPlot_3D} with the 2D traceback results overlaid. The resulting traceback times vary depending on the filtering of cluster members, but generally are consistent with the 2D traceback age of $4.4\pm0.1$ Myr within their uncertainties. 

For Pozzo 1 we have 62 cluster members that meet the 3D criteria. However, the results are consistent with the cluster being at its most compact between $0 - 3$ Myr ago, much less than the $10\pm0.1$ Myr timescale indicated by the 2D traceback. 

For Melotte 20 we have 54 cluster members that meet the 3D criteria. The results are consistent with the cluster being most compact between $2 - 6$ Myr ago, consistent with the 2D traceback age of $5.5\pm0.1$ Myr. 

For NGC 2547 we have 44 cluster members that meet the 3D criteria. The results are consistent with the cluster being most compact between $0 - 3$ Myr ago, consistent with the 2D traceback age of $3.1\pm0.1$ Myr. 

Overall, the 3D traceback with orbital equations does not clearly increase the kinematic age estimate $\tau_{TB}$ compared to the 2D linear approximations presented in Sect.~\ref{expansion timescale}. 

\begin{figure*}
	\includegraphics[width=470pt]{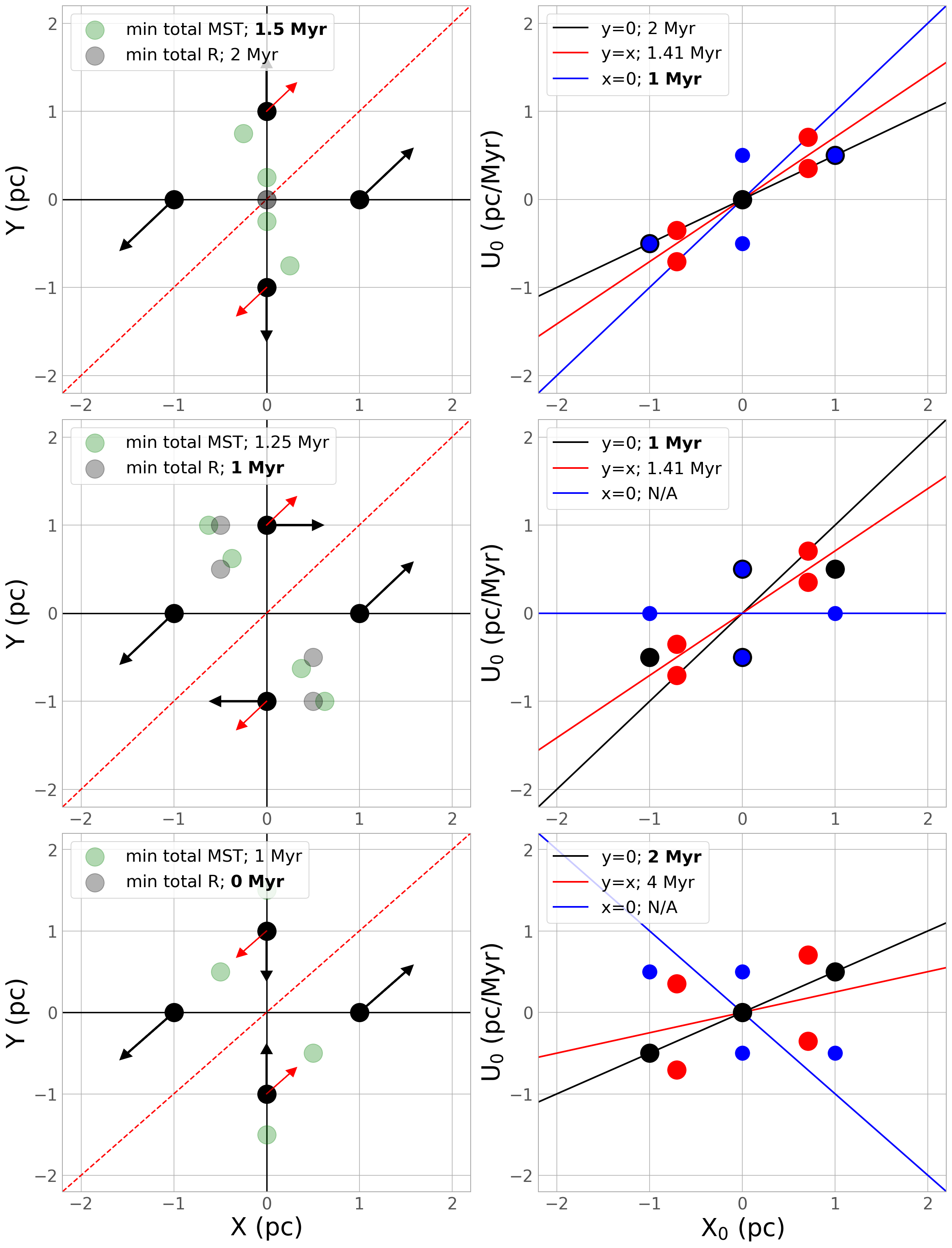}
	\setlength{\belowcaptionskip}{-10pt}
	\setlength{\textfloatsep}{0pt}
	\caption{Illustrative diagram of a simple cluster where rows show kinematic configurations of members that result in differences between expansion timescales and traceback timescales. \textit{Left:} 2D positions of cluster members (black) with vectors indicating relative velocities, and points indicating past configurations at minimum MST total length (faint green) and minimum total $R$ (faint grey). \textit{Right:} Position X$_0$ versus velocity U$_0$ parallel to axes $y=0$, $y=x$ and $x=0$, with coloured points indicating the positions and velocities of cluster members parallel to those axes (black, red \& blue, respectively). Coloured lines indicate the best-fit linear expansion gradients to the coloured points.}
	\label{KinematicAgesDiagram}
\end{figure*}

\section{Numerical simulations for 2D traceback}\label{Traceback test}
We perform numerical simulations to test the apparent $\sim10$ Myr upper limit on $\tau_{TB}$ and to test the validity of our inflation correction approach. We perform the traceback method as presented in Sect.~\ref{expansion timescale}, using tangential velocities U, V and their respective uncertainties randomly sampled from cluster members of $\lambda$ Ori, with positions X, Y set as $(U,V) \times 20/1.023$, where 1.023 is the conversion factor between kms$^{-1}$ and pc~Myr$^{-1}$, and with the ‘true’ traceback age set to 20 Myr. We obtain a resulting uncorrected $\tau_{TB}$ of 18.8 Myr and a corrected $\tau_{TB}$ of 19.2 Myr. We also perform this test using velocities and uncertainties from Melotte 20 and NGC 2264, which give corrected $\tau_{TB}$ of 19.1 Myr and 17.3 Myr, respectively. These results not only show that the apparent 10 Myr limit is not due to methodology, and rather reflects the kinematics of the clusters, but also validates the correction approach we employ to offset the age bias caused by uncertainty propagation. 
    
\section{Differences between Linear Expansion Timescales and Minimum Area Traceback}\label{KinematicAgeDifferences}
Kinematic ages for clusters derived from different approaches can differ for the same cluster membership sample. This may be due to differences in underlying assumptions implicit in these approaches and not necessarily because of observational uncertainties in the data.

We show in Fig.~\ref{KinematicAgesDiagram} an illustrative diagram, showing a simplified cluster (containing 4 stars) with simplistic spatial and kinematic distributions. In this plot, the three rows are three cases; \textit{top} where Traceback t $>$ Max Expansion t, \textit{middle} where Traceback t $=$ Max Expansion t and \textit{bottom} where Traceback t $<$ Max Expansion t. We illustrate how these cases can arise only by differences in the relative velocities of the cluster members. 

The left column shows the 2D spatial positions of the cluster members in a Cartesian coordinate system with the origin placed at the geometric center of the cluster, with the relative velocities as black vectors. Faint grey and faint green points represent past positions of stars (going backwards from their motion vectors) to the configurations that would give the minimum MST total length and minimum total R, the two metrics used to define the minimum cluster size in the traceback analysis in Sect.~\ref{expansion timescale}. In the legend is given the past time when these configurations are reached.

The right column shows position X$_0$ versus velocity U$_0$ for these same stars per row, to demonstrate what the corresponding maximum expansion gradient and timescales would be. In black is plotted the positions and velocities in the direction parallel to $y=0$, in red parallel to $y=x$ and in blue parallel to $x=0$. Then the best-fitting linear gradient (by least squares) for each is plotted in the same colour. In the legend is given the expansion timescale corresponding to each gradient, the lowest (for the maximum expansion rate) in bold.

The lower row seems to be the case that applies for many of the clusters in the sample analysed in this paper, where the Traceback time $\tau_{TB}$ $<$ Expansion Age $\tau_{\rm max}$. 

\begin{table*}
\footnotesize
\begin{center}
{\renewcommand{\arraystretch}{1.5}
\begin{tabular}{|p{1.6cm}|p{1.1cm}|p{1.1cm}|p{1.1cm}|p{1.1cm}|p{1.1cm}|p{1.1cm}|p{1.1cm}|p{1.1cm}|p{1.1cm}|p{1.1cm}|}
\hline
Cluster & \multicolumn{10}{c}{$\bar{v_{out}}$ (km~s$^{-1}$) per concentric annulus}  \\
  & 50 & 100 & 150 & 200 & 250 & 300 & 350 & 400 & 450 & 500 \\
\hline
$\lambda$ Ori & $0.02^{+0.07}_{-0.07}$ & $0.15^{+0.05}_{-0.05}$ & $0.12^{+0.05}_{-0.05}$ & $0.30^{+0.06}_{-0.06}$ & $0.44^{+0.07}_{-0.07}$ & $0.70^{+0.07}_{-0.07}$ & $0.89^{+0.04}_{-0.05}$ & $1.08^{+0.01}_{-0.01}$ & $1.60^{+0.01}_{-0.01}$ & $1.73^{+0.04}_{-0.04}$  \\
Platais 8 & $0.19^{+0.02}_{-0.02}$ & $0.51^{+0.05}_{-0.05}$ & $0.69^{+0.03}_{-0.03}$ &  &  &  &  &  &  &   \\
$\sigma$ Ori & $0.30^{+0.02}_{-0.02}$ & $1.23^{+0.03}_{-0.03}$ &  &  &  &  &  &  &  &    \\
NGC 2451B & $-0.12^{+0.08}_{-0.08}$ & $-0.04^{+0.06}_{-0.05}$ & $0.17^{+0.09}_{-0.09}$ & $0.04^{+0.02}_{-0.02}$ & $0.13^{+0.04}_{-0.04}$ &  &  &  &  &   \\
Melotte 20 & $0.02^{+0.03}_{-0.02}$ & $0.07^{+0.03}_{-0.03}$ & $0.13^{+0.02}_{-0.02}$ & $0.21^{+0.05}_{-0.05}$ & $0.13^{+0.04}_{-0.04}$ & $0.23^{+0.03}_{-0.04}$ & $0.13^{+0.10}_{-0.10}$ & $0.32^{+0.01}_{-0.01}$ & $0.36^{+0.01}_{-0.01}$ & $0.47^{+0.09}_{-0.09}$  \\
IC 2391 & $0.03^{+0.04}_{-0.04}$ & $0.10^{+0.01}_{-0.01}$ & $0.16^{+0.02}_{-0.02}$ &  &  &  &  &  &  &    \\
Gulliver 6 & $0.08^{+0.02}_{-0.02}$ & $0.26^{+0.06}_{-0.05}$ & $0.22^{+0.01}_{-0.01}$ & $0.35^{+0.06}_{-0.06}$ & $0.29^{+0.02}_{-0.02}$ &  &  &  &  &    \\
Pozzo 1 & $-0.12^{+0.05}_{-0.05}$ & $0.14^{+0.04}_{-0.04}$ & $0.14^{+0.03}_{-0.04}$ & $0.11^{+0.08}_{-0.07}$ & $0.28^{+0.04}_{-0.04}$ & $0.33^{+0.07}_{-0.07}$ &  &  &  &    \\
IC 2602 & $0.08^{+0.03}_{-0.03}$ & $0.15^{+0.03}_{-0.03}$ & $0.28^{+0.01}_{-0.01}$ & $0.54^{+0.02}_{-0.02}$ & $0.42^{+0.04}_{-0.04}$ &  &  &  &  &   \\
NGC 2232 & $-0.05^{+0.03}_{-0.03}$ & $0.06^{+0.02}_{-0.02}$ & $0.07^{+0.03}_{-0.03}$ &  &  &  &  &  &  &   \\
NGC 1977 & $0.24^{+0.03}_{-0.03}$ & $0.30^{+0.02}_{-0.02}$ &  &  &  &  &  &  &  &   \\
NGC 2264 & $0.26^{+0.15}_{-0.15}$ & $0.41^{+0.08}_{-0.08}$ &  &  &  &  &  &  &  &   \\
IC 2395 & $0.08^{+0.17}_{-0.21}$ & $0.10^{+0.18}_{-0.18}$ & $0.15^{+0.07}_{-0.07}$ & $0.34^{+0.06}_{-0.06}$ & $0.43^{+0.05}_{-0.05}$ &  &  &  &  &   \\
Alessi 20 & $0.03^{+0.04}_{-0.04}$ &  &  &  &  &  &  &  &  &    \\
Coll 359 & $-0.10^{+0.10}_{-0.10}$ & $0.12^{+0.26}_{-0.22}$ & $0.34^{+0.06}_{-0.06}$ & $0.30^{+0.05}_{-0.05}$ & $0.37^{+0.18}_{-0.17}$ &  &  &  &  &   \\
Gulliver 9 & $-0.01^{+0.06}_{-0.06}$ & $0.12^{+0.09}_{-0.08}$ & $0.24^{+0.11}_{-0.11}$ & $0.53^{+0.04}_{-0.05}$ &  &  &  &  &  &   \\
ASCC 16 & $0.17^{+0.01}_{-0.01}$ & $0.37^{+0.01}_{-0.01}$ & $0.36^{+0.02}_{-0.02}$ &  &  &  &  &  &  &   \\
ASCC 21 & $0.15^{+0.04}_{-0.04}$ &  &  &  &  &  &  &  &  &   \\
IC 348 & $-0.17^{+0.05}_{-0.05}$ & $-0.05^{+0.06}_{-0.06}$ &  &  &  &  &  &  &  &  \\
ASCC 19 & $0.18^{+0.02}_{-0.02}$ & $0.37^{+0.05}_{-0.04}$ &  &  &  &  &  &  &  &   \\
NGC 1980 & $-0.01^{+0.01}_{-0.01}$ & $0.18^{+0.02}_{-0.02}$ &  &  &  &  &  &  &  &   \\
NGC 2547 & $0.13^{+0.04}_{-0.04}$ & $0.12^{+0.02}_{-0.02}$ & $0.06^{+0.02}_{-0.02}$ &  &  &  &  &  &  &   \\
Coll 135 & $0.07^{+0.02}_{-0.02}$ & $0.16^{+0.03}_{-0.02}$ & $0.17^{+0.06}_{-0.07}$ & $0.20^{+0.02}_{-0.02}$ & $0.34^{+0.08}_{-0.07}$ &  &  &  &  &   \\
\hline
\end{tabular}}
\end{center}
\setlength{\belowcaptionskip}{-10pt}
\setlength{\textfloatsep}{0pt}
\caption{\footnotesize 1D expansion velocity $v_{out}$ (km~s$^{-1}$) values for each cluster. Values are given for concentric elliptic annuli of 50 cluster members each, as used for measuring the Angular Dispersion Parameter described in Sect. ~\ref{ADP}. }
\label{vout_table}
\end{table*}

\begin{figure*}
\subfloat{{\includegraphics[width=245pt]{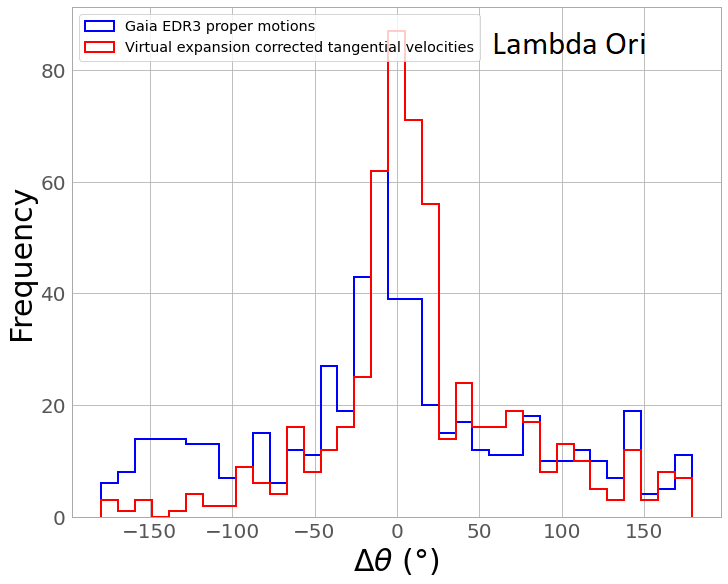}}}%
\qquad
\subfloat{{\includegraphics[width=245pt]{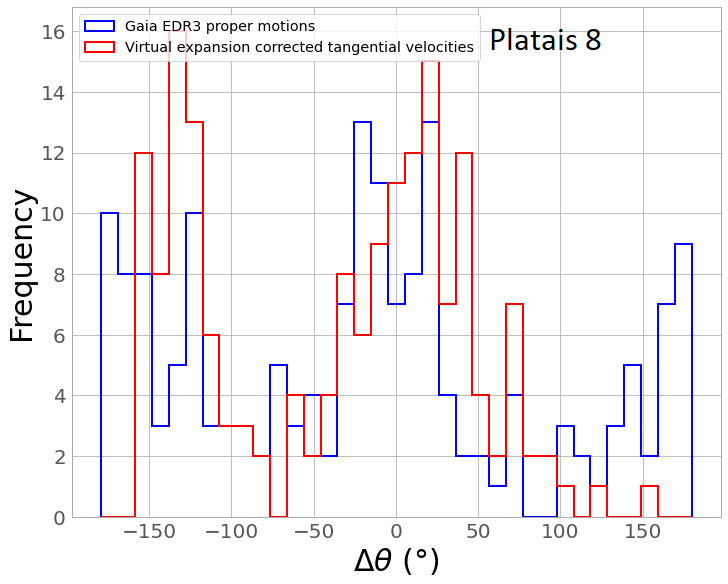}}}%
\qquad
\subfloat{{\includegraphics[width=245pt]{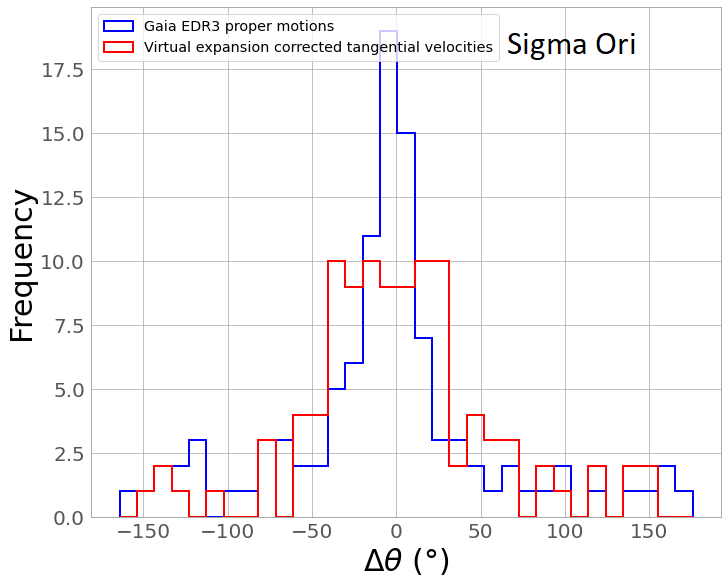}}}%
\qquad
\subfloat{{\includegraphics[width=245pt]{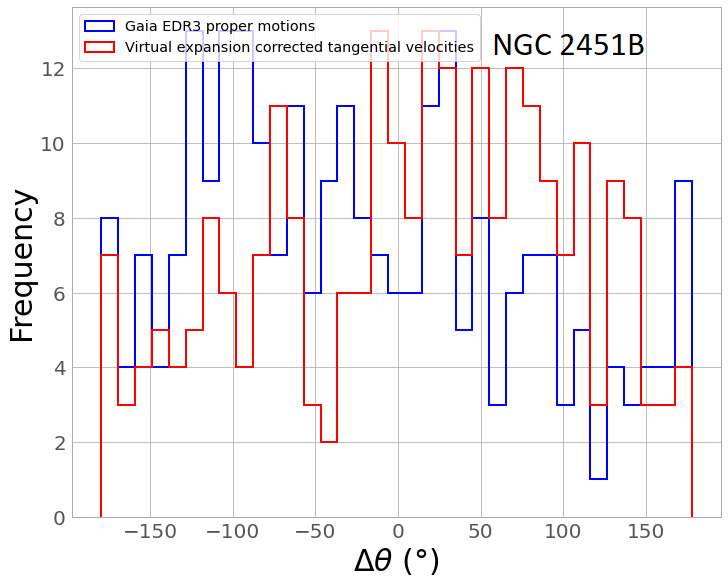}}}%
\qquad
\subfloat{{\includegraphics[width=245pt]{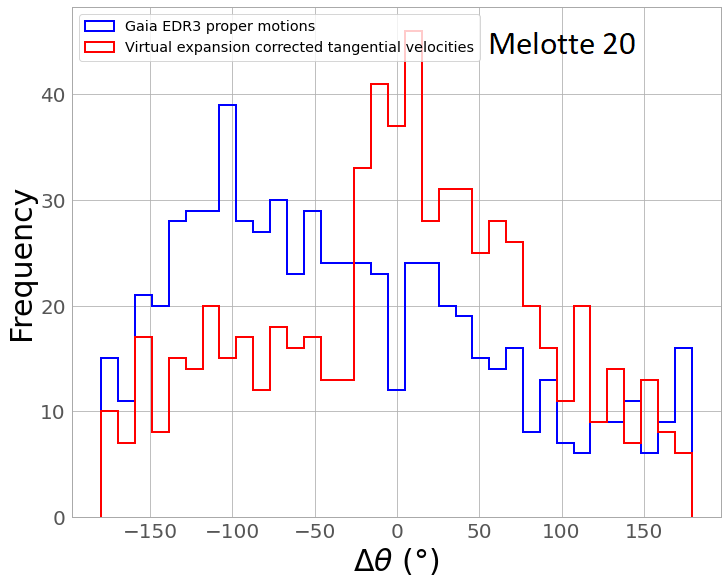}}}%
\qquad
\subfloat{{\includegraphics[width=245pt]{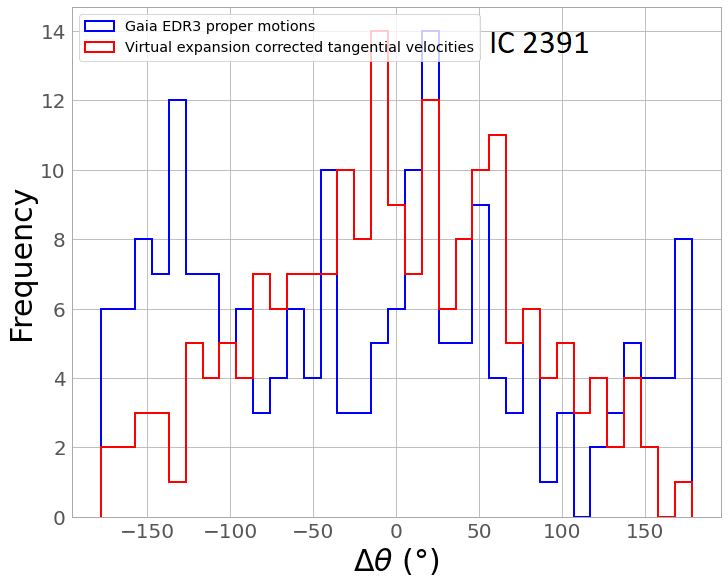}}}%
\caption{Histogram of $\Delta\theta$ calculated using proper motion vectors (blue) or virtual expansion corrected tangential velocity vectors (red) relative to the cluster center for members of $\lambda$ Ori, Platais 8, $\sigma$ Ori, NGC 2451B, Melotte 20 \& IC 2391 (\textit{right to left, descending}).  }
\label{VangleA1}
\end{figure*}

\begin{figure*}
\subfloat{{\includegraphics[width=245pt]{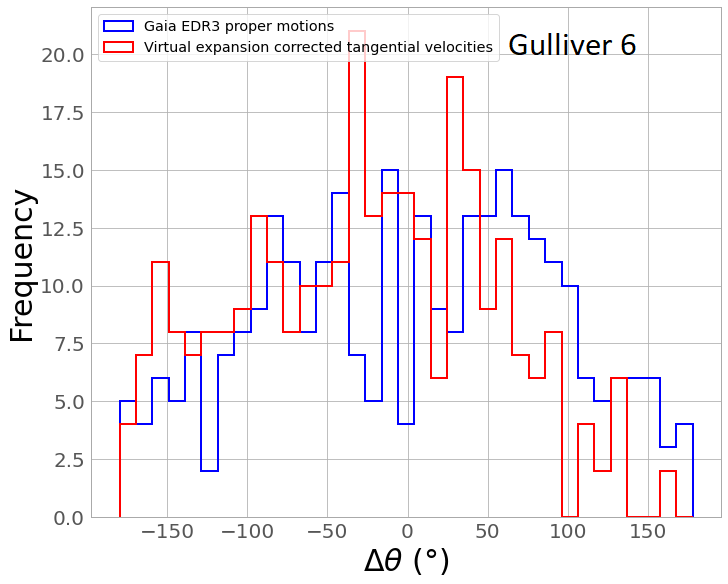}}}%
\qquad
\subfloat{{\includegraphics[width=245pt]{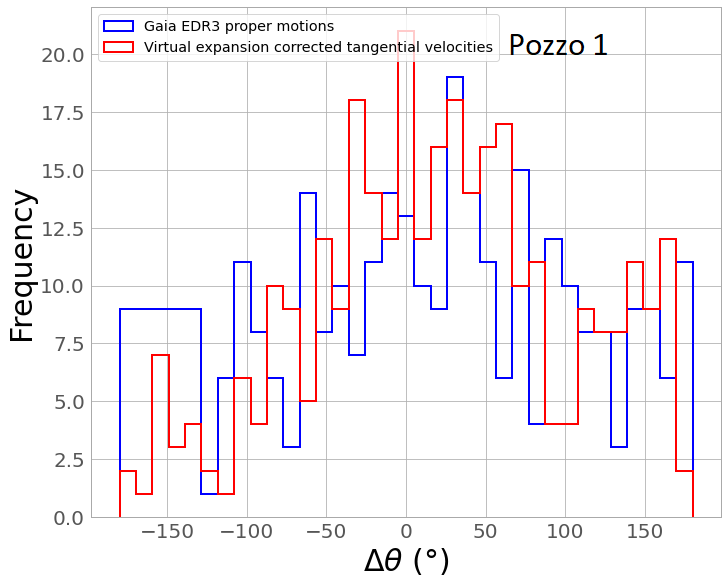}}}%
\qquad
\subfloat{{\includegraphics[width=245pt]{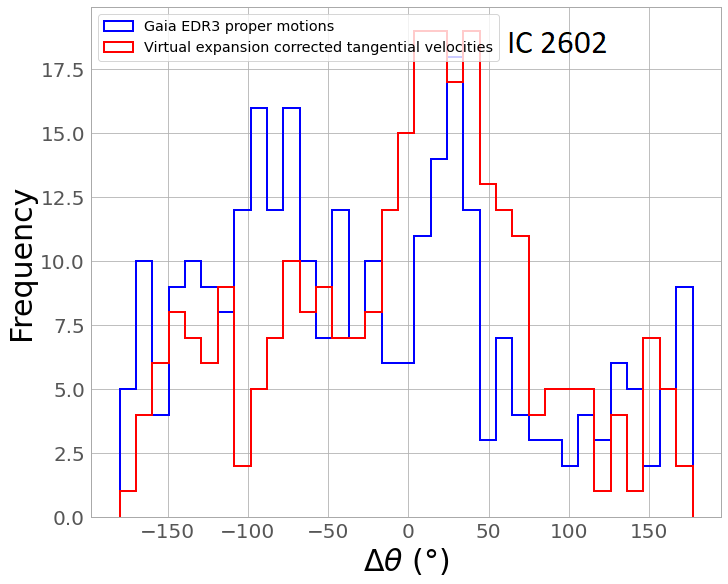}}}%
\qquad
\subfloat{{\includegraphics[width=245pt]{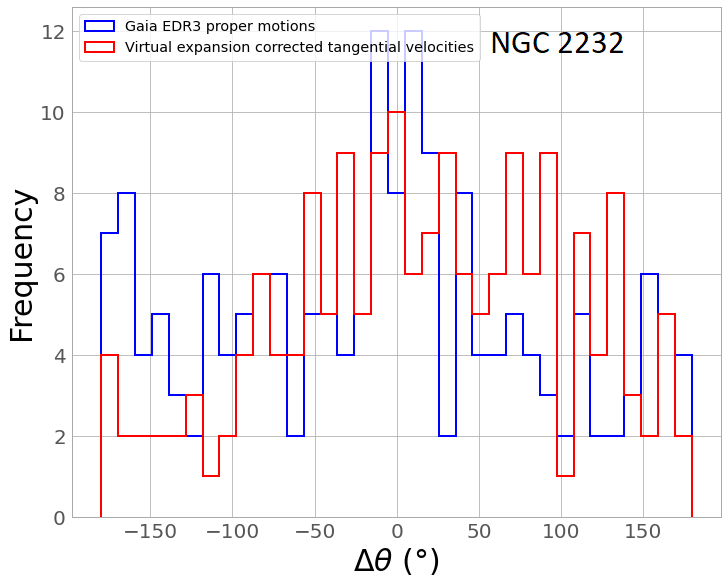}}}%
\qquad
\subfloat{{\includegraphics[width=245pt]{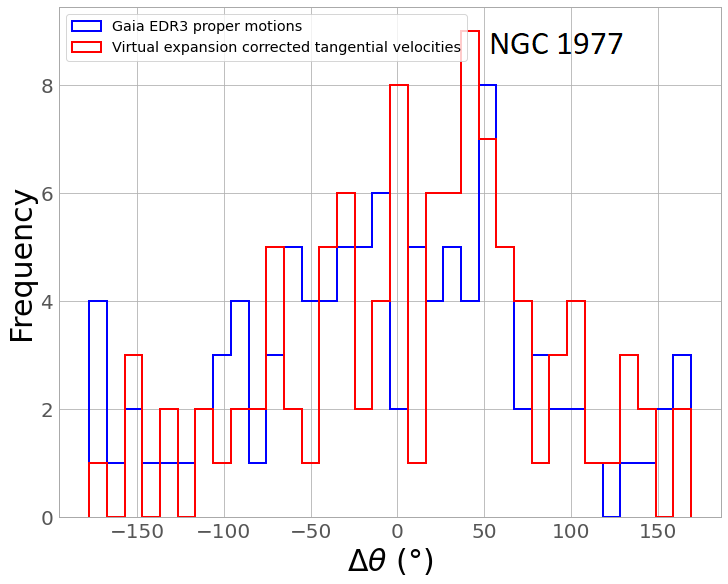}}}%
\qquad
\subfloat{{\includegraphics[width=245pt]{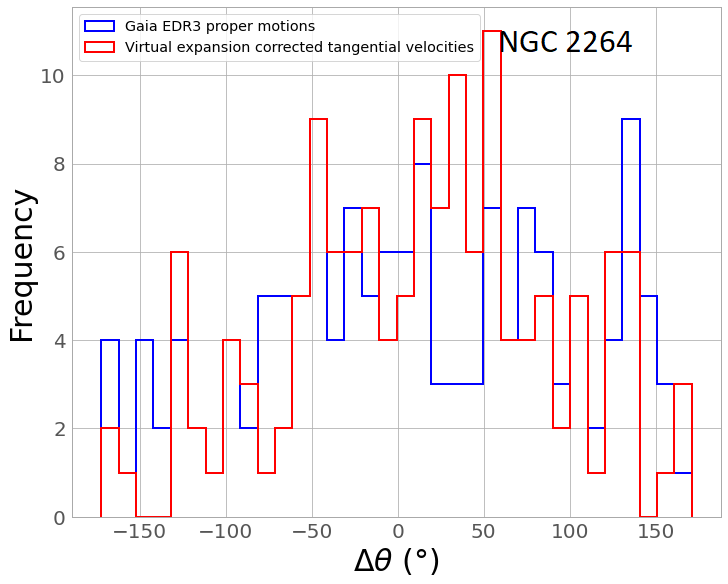}}}%
\caption{Histogram of $\Delta\theta$ calculated using proper motion vectors (blue) or virtual expansion corrected tangential velocity vectors (red) relative to the cluster center for members of Gulliver 6, Pozzo 1, IC 2602, NGC 2232, NGC 1977 \& NGC 2264 (\textit{right to left, descending}).  }
\label{VangleA2}
\end{figure*}

\begin{figure*}
\subfloat{{\includegraphics[width=245pt]{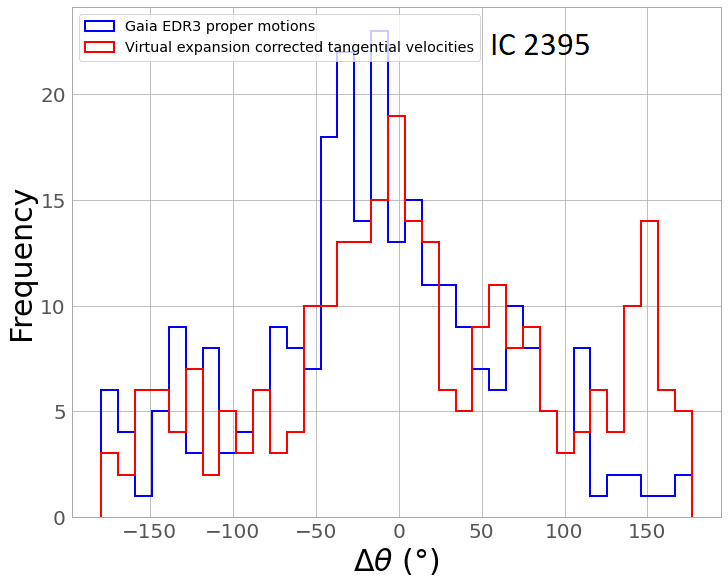}}}%
\qquad
\subfloat{{\includegraphics[width=245pt]{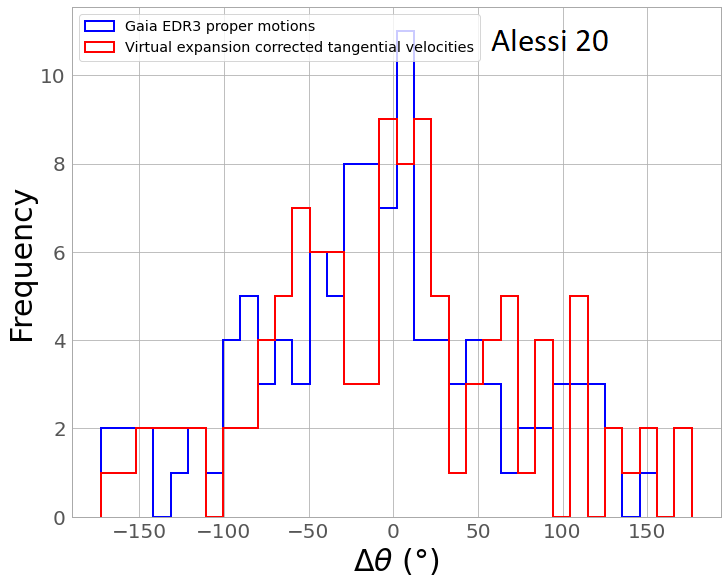}}}%
\qquad
\subfloat{{\includegraphics[width=245pt]{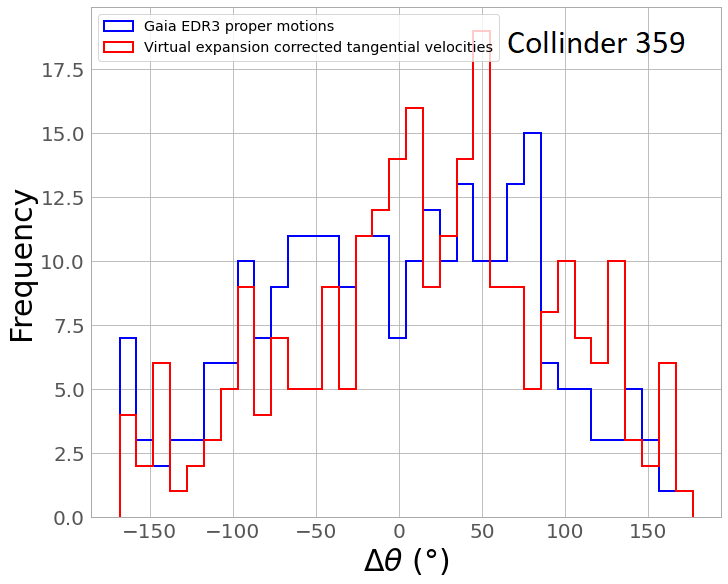}}}%
\qquad
\subfloat{{\includegraphics[width=245pt]{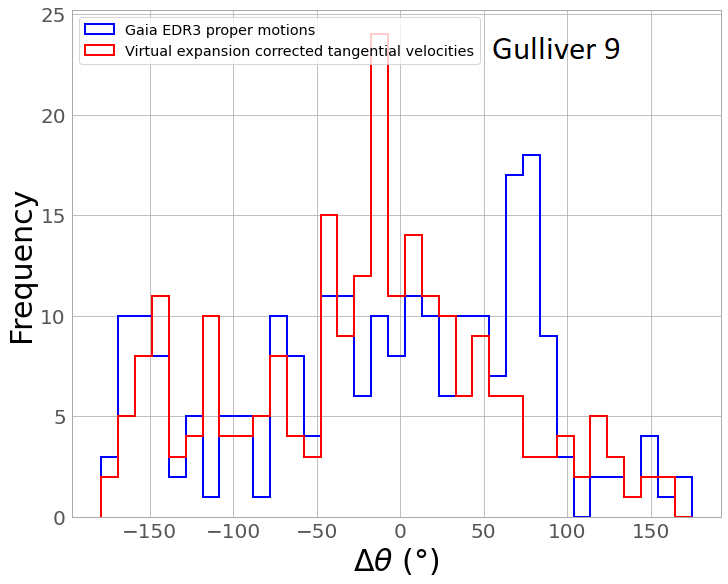}}}%
\qquad
\subfloat{{\includegraphics[width=245pt]{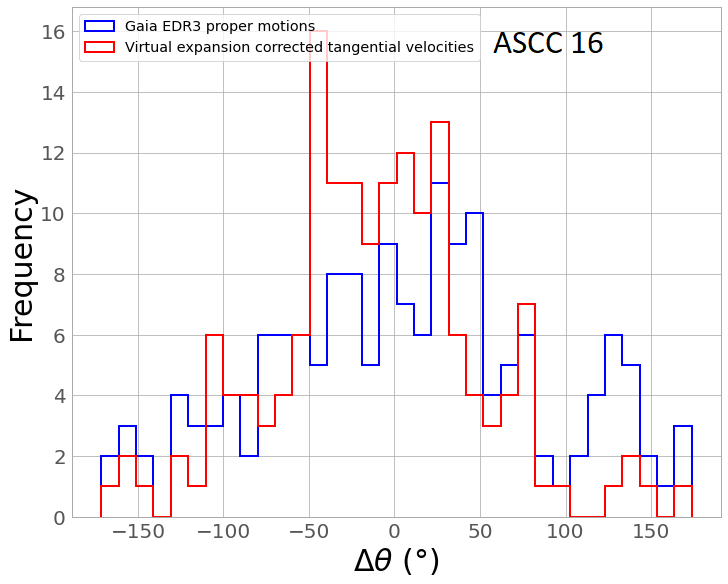}}}%
\qquad
\subfloat{{\includegraphics[width=245pt]{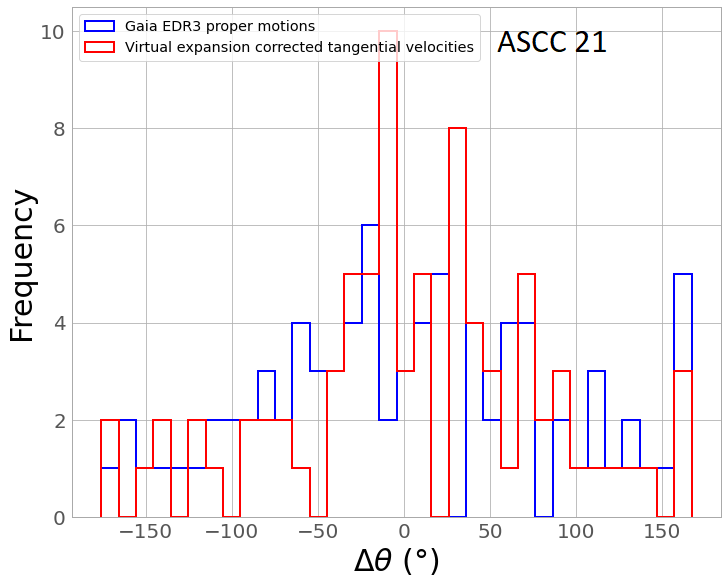}}}%
\caption{Histogram of $\Delta\theta$ calculated using proper motion vectors (blue) or virtual expansion corrected tangential velocity vectors (red) relative to the cluster center for members of IC 2395, Alessi 20, Collinder 359, Gulliver 9, ASCC 16 \& ASCC 21 (\textit{right to left, descending}).  }
\label{VangleA3}
\end{figure*}

\begin{figure*}
\subfloat{{\includegraphics[width=245pt]{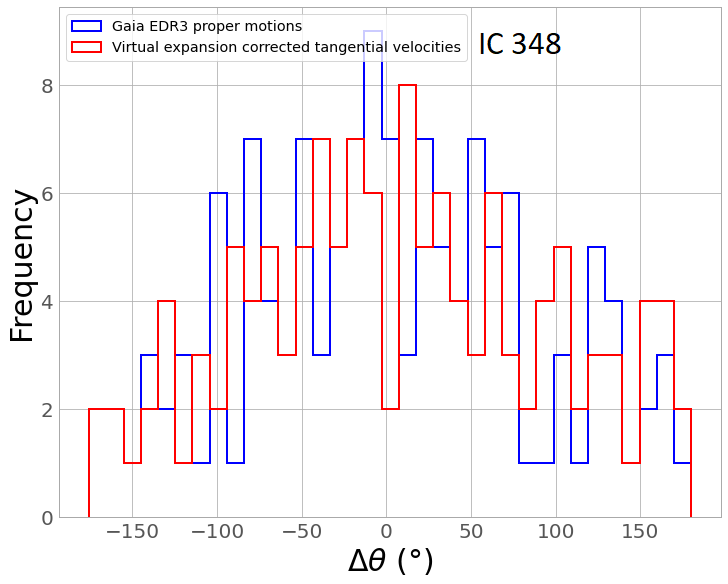}}}%
\qquad
\subfloat{{\includegraphics[width=245pt]{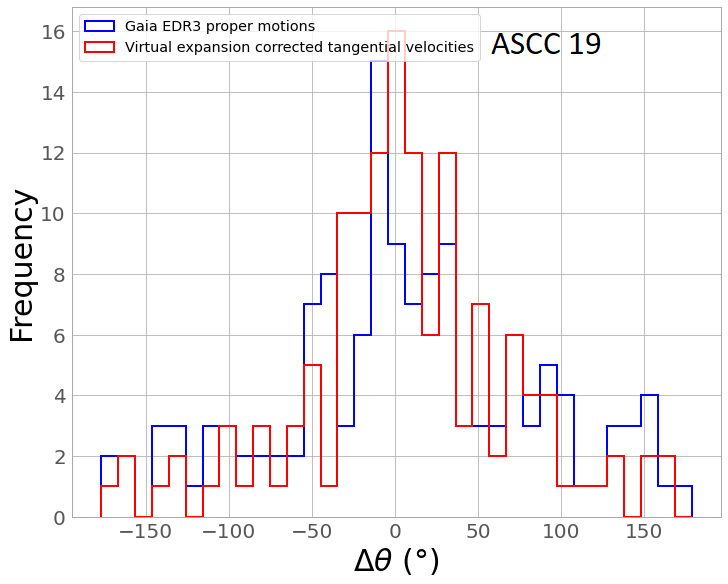}}}%
\qquad
\subfloat{{\includegraphics[width=245pt]{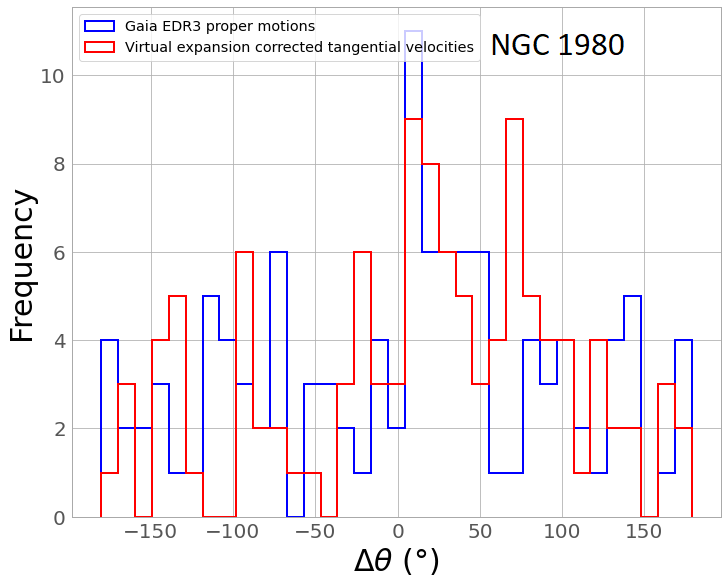}}}%
\qquad
\subfloat{{\includegraphics[width=245pt]{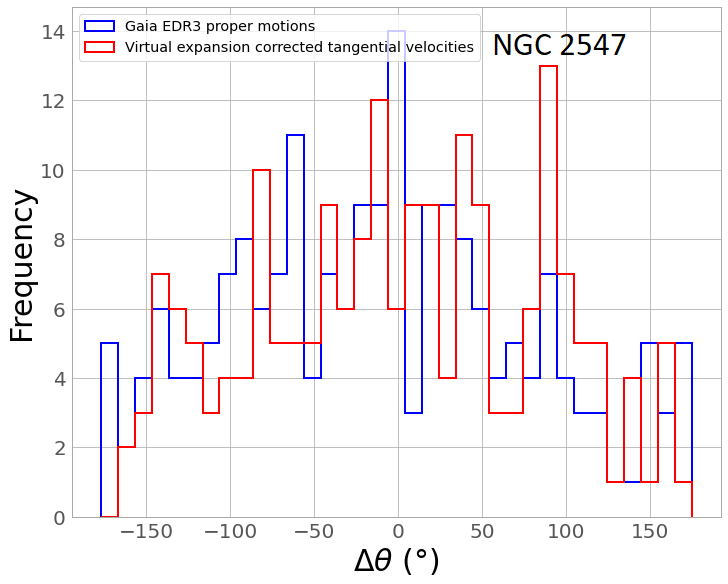}}}%
\qquad
\subfloat{{\includegraphics[width=245pt]{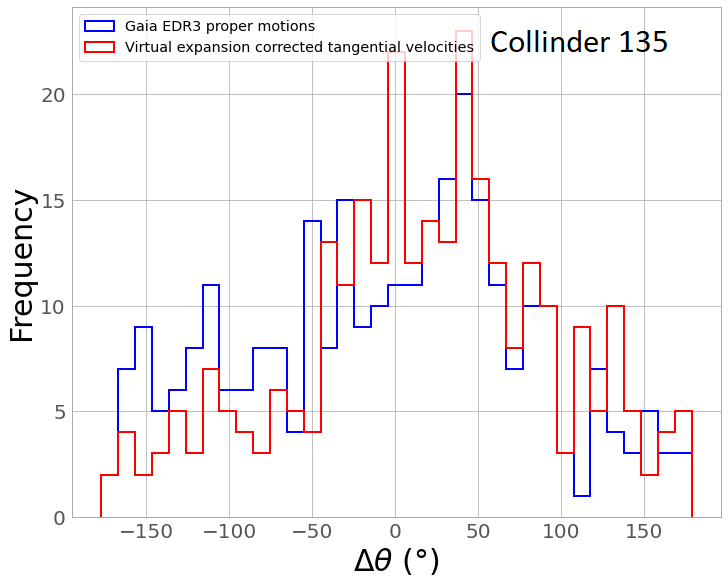}}}%
\caption{Histogram of $\Delta\theta$ calculated using proper motion vectors (blue) or virtual expansion corrected tangential velocity vectors (red) relative to the cluster center for members of IC 348, ASCC 19, NGC 1980, NGC 2547 \& Collinder 135 (\textit{right to left, descending}).  }
\label{VangleA4}
\end{figure*}

\end{document}